\documentclass[11pt,fleqn]{article}
\pdfoutput=0

\usepackage{a4}
\usepackage{hyperref}
\usepackage{amstext}
\usepackage{amsfonts}
\usepackage{amssymb}
\usepackage{amsmath}
\usepackage{color}
\usepackage{subfig}
\usepackage{epsfig}
\usepackage{booktabs}
\usepackage{caption}
\usepackage{multirow}
\usepackage{footnote}
\usepackage{dsfont}
\usepackage{longtable}
\usepackage{xcolor,colortbl}
\usepackage{bbm}

\newcommand{\ltapprox}{\raisebox{-0.5ex}{$\,\stackrel{<}{\scriptstyle\sim}\,$}}

\begin{document}


\begin{center}

{\Large {\bf{} Continuum limit of the }$D${\bf{} meson, }$D_s${\bf{} meson and charmonium}}

{\Large {\bf{} spectrum from $N_f=2+1+1$ twisted mass lattice QCD}}

\vspace{0.5cm}

\textbf{Krzysztof Cichy$^{1,2}$, Martin Kalinowski$^1$, Marc Wagner$^1$}

$^1$~Goethe-Universit\"at Frankfurt am Main, Institut f\"ur Theoretische Physik, \\ Max-von-Laue-Stra{\ss}e 1, D-60438 Frankfurt am Main, Germany

$^2$~Adam Mickiewicz University, Faculty of Physics, Umultowska 85, 61-614 Pozna\'n, Poland

\begin{center}
\vspace*{0.4cm}
\includegraphics
[width=0.2\textwidth,angle=0]
{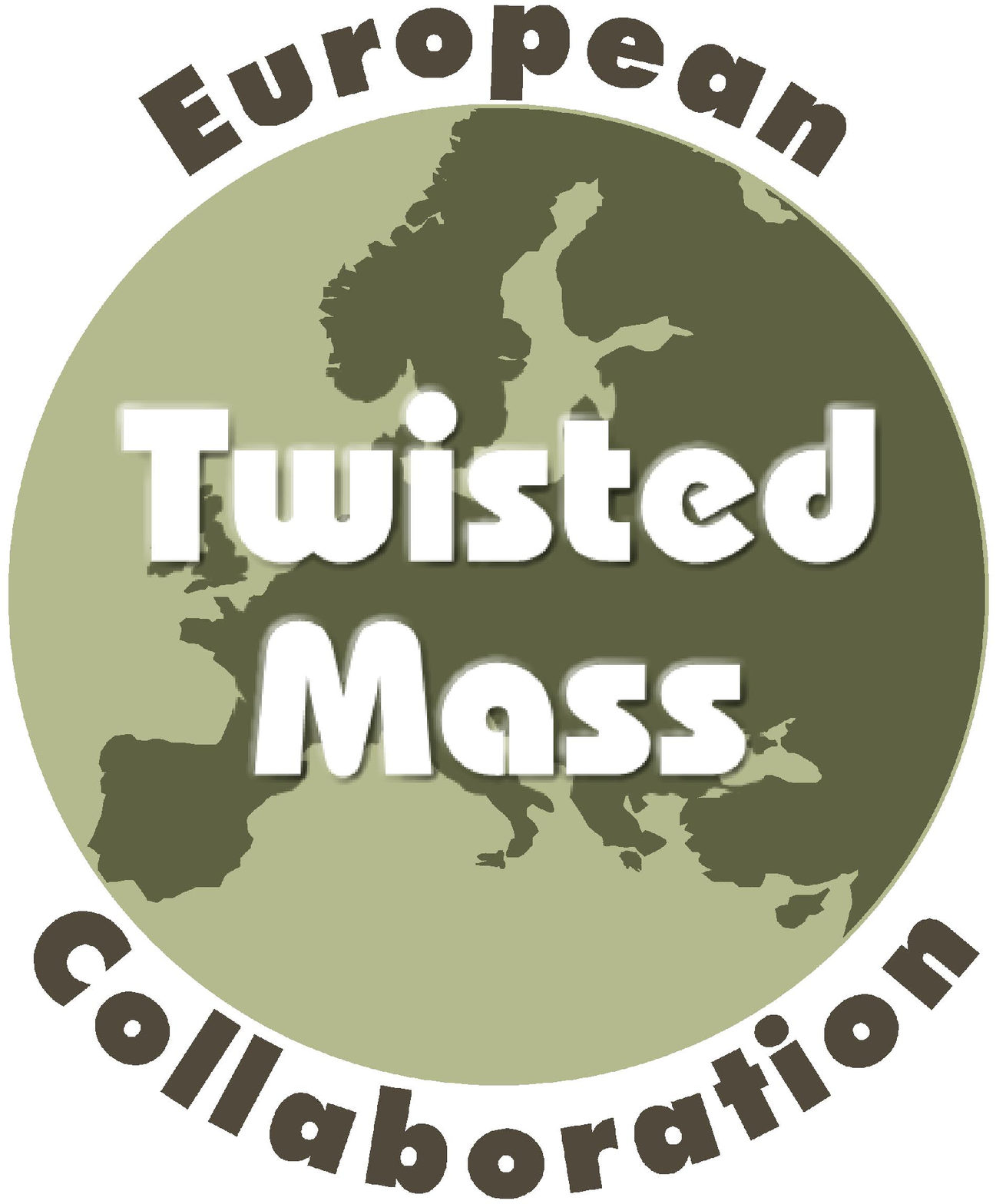}
\end{center}

\vspace{0.4cm}

November 22, 2016

\end{center}

\vspace{0.1cm}

\begin{tabular*}{16cm}{l@{\extracolsep{\fill}}r} \hline \end{tabular*}

\vspace{-0.40cm}
\begin{center} \textbf{Abstract} \end{center}
\vspace{-0.40cm}
We compute masses of $D$ meson, $D_s$ meson and charmonium states using $N_f=2+1+1$ Wilson twisted mass lattice QCD. 
All results are extrapolated to physical light quark masses, physical strange and charm quark masses and to the continuum.
Our analysis includes states with spin $J = 0,1,2$, parity $\mathcal{P} = -,+$ and in case of charmonium also charge conjugation $\mathcal{C} = -,+$. 
Computations are based on a large set of quark-antiquark meson creation operators. 
We investigate and quantify all sources of systematic errors, including fitting range uncertainties, finite volume effects, isospin breaking effects and the choice of the fitting ansatz for the combined chiral and continuum extrapolation such that the resulting meson masses can be compared directly and in a meaningful way to experimental results.
Within combined statistical and systematic errors, which are between below two per mille and three percent, our results agree with available experimental results for most of the states. 
In the few cases where we observe discrepancies, we discuss possible reasons.

\begin{tabular*}{16cm}{l@{\extracolsep{\fill}}r} \hline \end{tabular*}

\thispagestyle{empty}


\newpage

\setcounter{page}{1}

\section{Introduction}
\label{sec:intro}

Quite a number of $D$ meson, $D_s$ meson and charmonium states have been observed in experiments \cite{PDG}. Several of them are both experimentally and theoretically well-understood, i.e.\ their quantum numbers and their structures are known. Examples include, in particular, the pseudoscalar ground state mesons $D$, $D_s$ and $\eta_c(1S)$ and the vector ground state mesons $D^\ast$, $D_s^\ast$ and $J/\psi(1S)$. There are, however, open questions regarding some of the more recently found excitations. Examples are the positive parity mesons $D_{s0}^\ast(2317)$ and $D_{s1}(2460)$, first reported by BaBar \cite{Aubert:2003fg} and CLEO \cite{Besson:2003cp}, respectively, which are unexpectedly light compared to expectations from quark model calculations. This could be an indication that these states are not just quark-antiquark pairs, but have a more complicated structure, e.g.\ are composed of two quarks and two antiquarks, a scenario at the moment neither established nor ruled out. The situation is similar for some of the charmonium-like $X$ states, e.g.\ $X(3872)$ first observed by Belle \cite{Choi:2003ue}.

There are many interesting approaches to study $D$ mesons, $D_s$ mesons, and charmonium states theoretically, e.g.\ quark models \cite{Ebert:2009ua}, effective theories respecting QCD symmetries \cite{Eshraim:2014eka}, or Dyson-Schwinger and Bethe-Salpeter equations \cite{Fischer:2014cfa}. Of course, it would be highly desirable to understand these mesons and their properties starting from first principles, i.e.\ the QCD Lagrangian, without any assumptions, model simplifications or truncations. The corresponding method is lattice QCD, a numerical technique to compute QCD observables, which allows one to investigate and quantify all sources of systematic error. Computing the spectrum and investigating the structure of mesons using lattice QCD is, however, a challenging task. Several problems have only partly been solved or require investing a rather large amount of high performance computing resources, for example simulations with physically light $u/d$ quarks. Similarly, to remove discretization errors, one has to study the continuum limit, which necessitates time consuming simulations at several different lattice spacings. Particularly problematic is the investigation of mesons, which readily decay into lighter multi-particle states. Such states should theoretically be treated as resonances and not as stable quark-antiquark states, which is technically extremely difficult, even for simple cases, where only a single decay channel exists. Examples are $D_0^\ast(2400)$ and $D_1(2430)$ with quantum numbers $J^P = 0^+$ and $J^P = 1^+$. Similarly, it is very challenging to study mesons which might have a structure more complicated than a simple quark-antiquark pair, e.g.\ candidates for tetraquarks or hybrid mesons. While there has been a lot of impressive progress regarding lattice QCD hadron spectroscopy within the last couple of years, there is certainly still a lot of room for improvement. Simple states, in particular pseudoscalar and vector ground state mesons, have, meanwhile, been studied very accurately, including simulations at or extrapolations to physically light $u/d$ quark masses and the continuum limit. On the other hand, the majority of studies concerned with parity, radial and orbital excitations are still at a more exploratory stage, i.e.\ have quite often been performed at unphysically heavy quark masses or at a single finite lattice spacing. Recent reviews discussing the status of lattice QCD computations of $D$ and $D_s$ mesons and of charmonium are Refs.\ \cite{Mohler:2015zsa} and \cite{DeTar:2011nn,Prelovsek:2015fra}, respectively.

The most common approach to compute meson masses using lattice QCD is to employ meson creation operators, which are composed of a quark and an antiquark, and to extract meson masses from the exponential decay of corresponding correlation functions\footnote{For a basic introduction to lattice hadron spectroscopy, cf.\ Ref.\ \cite{Weber:2013eba}.}. This strategy yields accurate and solid results for mesons which resemble quark-antiquark pairs and which are quite stable, i.e.\ many of the low-lying states in the $D$ meson, $D_s$ meson and charmonium sector. Recent lattice QCD papers following this strategy to compute masses and spectra of $D$ and $D_s$ mesons and of charmonium are Refs.\ \cite{Dudek:2007wv,Dong:2009wk,Burch:2009az,Dudek:2010wm,Mohler:2011ke,Namekawa:2011wt,
Bali:2011dc,Bali:2011rd,Liu:2012ze,Yang:2012mya,Dowdall:2012ab,Bali:2012ua,
Moir:2013ub,Galloway:2014tta,Bali:2015lka,Cheung:2016bym}. Rigorous treatments of more complicated mesonic systems like the previously mentioned unstable $D_0^\ast(2400)$ and $D_1(2430)$ mesons or the tetraquark candidates $D_{s0}^\ast(2317)$ and $D_{s1}(2460)$, require more advanced techniques, including the implementation of meson creation operators composed of two quarks and two antiquarks and possibly studies of the volume dependence of the masses of corresponding scattering states\footnote{For a basic introduction on how to study resonances using lattice QCD, cf.\ Ref.\ \cite{Prelovsek:2011nk}.}. Examples of recent lattice QCD papers exploring and using such techniques to study specific $D$, $D_s$, or charmonium states are Refs.\ \cite{Gong:2011nr,Mohler:2012na,Liu:2012zya,Prelovsek:2013cra,
Prelovsek:2013xba,Mohler:2013rwa,Ikeda:2013vwa,Lang:2014yfa,Prelovsek:2014swa,Guerrieri:2014nxa,Padmanath:2015era,Lang:2015sba,Moir:2016srx}.

The goal of this paper is to compute the masses of several low-lying $D$ meson, $D_s$ meson and charmonium states using Wilson twisted mass lattice QCD with 2+1+1 dynamical quark flavors. One of the main advantages of this particular discretization of QCD is automatic $\mathcal{O}(a)$ improvement, i.e.\ discretization errors appear only quadratically in the small lattice spacing $a$ and are, hence, strongly suppressed. From a technical point of view, we employ a large variety of quark-antiquark meson creation operators and are, hence, able to address total angular momentum $J = 0,1,2$, parity $\mathcal{P} = -,+$ and in case of charmonium charge conjugation $\mathcal{C} = -,+$.
Computations have been performed for ten different gauge link ensembles with unphysically heavy $u/d$ quark masses corresponding to pion masses $m_\pi \approx 225 \ldots 470 \, \textrm{MeV}$ and with lattice spacings $a \approx 0.0619(18) \, \textrm{fm} \, , \, 0.0815(30) \, \textrm{fm} \, , \, 0.0885(36) \, \textrm{fm}$ \cite{Carrasco:2014cwa}. The uncertainty in the scale setting has been propagated to the final results. Moreover, each meson mass has been computed twice using two different valence Wilson twisted mass quark discretizations. This rather large amount of lattice data allows solid extrapolations both to physically light $u/d$ quark masses and to the continuum. What is more, finite volume effects have been investigated and found to be negligible.

As mentioned above, some $D$ meson, $D_s$ meson and charmonium states are quite unstable or might have a structure significantly different from a quark-antiquark pair. Even though we present results for these states in this work, a rigorous treatment will require more advanced techniques, in particular the inclusion of four-quark creation operators as discussed above. We are in the process of developing such techniques using a similar lattice QCD setup \cite{Alexandrou:2012rm,Abdel-Rehim:2014zwa,Berlin:2015faa}. The techniques and results presented in this paper are an important prerequisite for such more advanced computations.

Parts of this work have been presented at recent conferences \cite{Kalinowski:2012re,Kalinowski:2013wsa,Wagner:2013laa,Cichy:2015tma} and certain technical aspects are discussed in detail in Ref.\ \cite{Kalinowski:2015bwa}.

This paper is structured as follows. In Section~\ref{sec:setup}, we summarize the 2+1+1 flavor Wilson twisted mass lattice QCD setup and the meson creation operators we use. We also explain in detail how we tune the strange and charm quark masses, and how we extrapolate the meson masses to physically light $u/d$ quark masses and to the continuum. In Section~\ref{sec:results}, we present our results for $D$ and $D_s$ mesons and for charmonium states and discuss and quantify all possible error sources. These results are summarized in plots and tables in Section~\ref{sec:summary}, where we also give a brief outlook.


\section{Computational setup}
\label{sec:setup}

In the following, we summarize our lattice QCD setup and the technical steps by which we determine masses of $D$ mesons, of $D_s$ mesons and of charmonium states. For further details, we refer to Ref.\ \cite{Kalinowski:2015bwa}, where some aspects have already been discussed extensively.


\subsection{\label{SEC459}Gauge link ensembles, sea quarks, valence quarks}

We use gauge link configurations generated with 2+1+1 dynamical quark flavors by the European Twisted Mass Collaboration (ETMC) \cite{Baron:2008xa,Jansen:2009xp,Baron:2009zq,Baron:2010bv,Baron:2011sf}. The gluonic action is the Iwasaki gauge action \cite{Iwasaki:1985we}. For the light degenerate $(u,d)$ quark doublet, the standard Wilson twisted mass action
\begin{eqnarray}
\label{EQN001} S_{\textrm{light}}[\chi^{(l)},\bar{\chi}^{(l)},U] \ \ = \ \ \sum_x \bar{\chi}^{(l)}(x) \Big(D_\textrm{W}(m_0) + i \mu \gamma_5 \tau_3\Big) \chi^{(l)}(x)
\end{eqnarray}
has been used \cite{Frezzotti:2000nk}, for the heavy $(c,s)$ sea quark doublet, the Wilson twisted mass formulation for non-degenerate quarks
\begin{eqnarray}
\label{EQN002} S_{\textrm{heavy}}[\chi^{(h)},\bar{\chi}^{(h)},U] \ \ = \ \ \sum_x \bar{\chi}^{(h)}(x) \Big(D_\textrm{W}(m_0) + i \mu_\sigma \gamma_5 \tau_1 + \mu_\delta \tau_3\Big) \chi^{(h)}(x)
\end{eqnarray}
\cite{Frezzotti:2003xj}. $D_\mathrm{W}$ denotes the Wilson Dirac operator,
\begin{eqnarray}
\label{EQN302} D_\mathrm{W}(m_0) \ \ = \ \ \frac{1}{2} \Big(\gamma_\mu \Big(\nabla_\mu + \nabla^\ast_\mu\Big) - a\nabla^\ast_\mu \nabla_\mu\Big) + m_0 ,
\end{eqnarray}
$\chi^{(l)} = (\chi^{(u)},\chi^{(d)})$ and $\chi^{(h)} = (\chi^{(c)},\chi^{(s)})$ are the quark fields in the so-called twisted basis and $\tau_1$ and $\tau_3$ denote the first and third Pauli matrices acting in flavor space. At maximal twist, physical quantities, e.g.\ meson masses, are automatically $\mathcal{O}(a)$ improved \cite{Frezzotti:2003ni,Frezzotti:2004wz,Frezzotti:2005gi,Chiarappa:2006ae}. The tuning has been done by adjusting $m_0$ such that the PCAC quark mass in the light quark sector vanishes (cf.\ \cite{Baron:2010bv} for details). For a review on Wilson twisted mass lattice QCD, we refer to Ref.\ \cite{Shindler:2007vp}.

In this work, we use ten ensembles, which differ in the light $u/d$ quark mass (corresponding pion masses $225 \, \textrm{MeV} \ltapprox m_\pi \ltapprox 470 \, \textrm{MeV}$), the lattice spacing $0.0619 \, \textrm{fm} \ltapprox a \ltapprox 0.0885 \, \textrm{fm}$ and the spacetime volume (scale setting via the pion mass and the pion decay constant \cite{Carrasco:2014cwa}). The $s$ and the $c$ quark masses are represented by $\mu_\sigma$ and $\mu_\delta$. These values have been chosen such that the lattice QCD results for $2 m_K^2 - m_\pi^2$ and for $m_D$, quantities which depend only weakly on the light $u/d$ quark mass, are close to the corresponding physical values \cite{Baron:2010bv,Baron:2010th,Baron:2010vp}. Details of these gauge link ensembles are collected in Table~\ref{tab.ensembles}. Ensemble A40.24 is not used to generate final results for meson masses, but only to confirm the absence of finite volume effects (cf.\ Section\ \ref{sec:FVE}).

\begin{table}[htb]
\centering
\begin{tabular}{|c|c|c|c|c|c|c|c|c|}
      \hline
      \multirow{2}*{ensemble} & \multirow{2}*{$\beta$} & \multirow{2}*{$(L/a)^3 \times T/a$} & \multirow{2}*{$a\mu$} & \multirow{2}*{$a\mu_\sigma$} & \multirow{2}*{$a\mu_\delta$} & $a$ & $m_\pi$ & \# of \\
      & & & & & & $[\textrm{fm}]$ & $[\textrm{MeV}]$ & config. \\
      \hline
      A30.32 &$1.90$& $32^3 \times 64$ & $0.0030$ & $0.150$ & $0.190$ & $0.0885(36)$  & $276.5(8)\phantom{0.}$ & $1530$          \\
      A40.32 &	    & $32^3 \times 64$ & $0.0040$ &         &         & 	  & $314.9(7)\phantom{0.}$ & $\phantom{0}846$\\
      A40.24 &	    & $24^3 \times 48$ & $0.0040$ &         &         & 	  & $321.1(1.1)$   & $1302$          \\
      A80.24 &	    & $24^3 \times 48$ & $0.0080$ &         &         &	          & $443.3(8)\phantom{0.}$ & $1850$          \\
      B25.32 &$1.95$& $32^3 \times 64$ & $0.0025$ & $0.135$ & $0.170$ &	$0.0815(30)$  & $260.0(1.1)$   & $\phantom{0}615$\\
      B55.32 &	    & $32^3 \times 64$ & $0.0055$ &         &         &	          & $375.2(5)\phantom{0.}$   & $1173$          \\
      B85.24 &	    & $24^3 \times 48$ & $0.0085$ &         &         &	          & $468.4(1.0)$   & $1176$          \\
      
      D15.48 &$2.10$& $48^3 \times 96$ & $0.0015$ & $0.120$ & $0.1385$&	$0.0619(18)$  & $224.1(1.1)$             & $\phantom{0}300$\\
      D20.48 &	    & $48^3 \times 96$ & $0.0020$ &         &         &	          & $257.0(1.0)$             & $\phantom{0}132$\\
      D30.48 &	    & $48^3 \times 96$ & $0.0030$ &         &         &	          & $310.8(1.0)$             & $\phantom{0}172$\\
      \hline
\end{tabular}
\caption{\label{tab.ensembles}Ensembles of gauge link configurations (ensemble name, inverse gauge coupling $\beta$, lattice volume $(L/a)^3 \times T/a$, light quark mass $a\mu$, mass parameters $a\mu_\sigma$ and $a\mu_\delta$ for the heavy doublet, lattice spacing $a$ in fm, pion mass in MeV, number of gauge field configurations).}
\end{table}

For the light degenerate $(u,d)$ valence quark doublet, we use the same action which was used to simulate the corresponding sea quarks, i.e.\ the action (\ref{EQN001}).

For the heavy $s$ and $c$ valence quarks, we use twisted mass doublets of degenerate quarks, i.e.\ a different discretization than for the corresponding sea quarks. We use the action (\ref{EQN001}) with the replacements $\chi^{(l)} \rightarrow \chi^{(s)} = (\chi^{(s^+)} , \chi^{(s^-)})$, $\mu \rightarrow \mu_s$ and $\chi^{(l)} \rightarrow \chi^{(c)} = (\chi^{(c^+)} , \chi^{(c^-)})$, $\mu \rightarrow \mu_c$, respectively. We do this to avoid mixing of strange and charm quarks, which inevitably takes place in a unitary non-degenerate Wilson twisted mass setup, and which is particularly problematic for observables containing charm quarks, e.g.\ masses of $D$ and $D_s$ mesons and of charmonium (cf.\ \cite{Baron:2010th,Baron:2010vp} for a detailed discussion of these problems). These degenerate valence doublets allow two realizations for strange as well as for charm quarks, either with a twisted mass term $+i \mu_{s,c} \gamma_5$ (i.e.\ $\chi^{(s^+)}$ or $\chi^{(c^+)}$) or $-i \mu_{s,c} \gamma_5$ (i.e.\ $\chi^{(s^-)}$ or $\chi^{(c^-)}$). For a quark-antiquark meson creation operator, e.g.\ $\bar{\chi}^{(1)} \gamma_5 \chi^{(2)}$, the sign combinations $(+,-)$ and $(-,+)$ for the antiquark $\bar{\chi}^{(1)}$ and the quark $\chi^{(2)}$ are related by symmetry, i.e.\ the corresponding correlation functions are identical. These correlation functions differ, however, from their counterparts with sign combinations $(+,+)$ and $(-,-)$ due to different discretization errors. We have performed computations for both types of sign combinations and refer to them as $(+,-)\equiv(\pm,\mp)$ discretization and $(+,+)\equiv(\pm,\pm)$ discretization, respectively (cf.\ the following subsection for details).


\subsection{\label{SEC599}Meson creation operators and trial states}

In the continuum, a quark-antiquark operator creating a meson trial state with definite quantum numbers $J^{\mathcal{P}\mathcal{C}}$ (total angular momentum $J$, parity $\mathcal{P}$, charge conjugation $\mathcal{C}$), when applied to the vacuum $| \Omega \rangle$, is
\begin{eqnarray}
\label{EQN101} O_{\Gamma,\bar{\psi}^{(1)} \psi^{(2)}}^\textrm{physical} \ \ \equiv \ \ \frac{1}{\sqrt{V}} \int d^3r \, \bar{\psi}^{(1)}(\mathbf{r}) \int_{|\Delta \mathbf{r}| = R} d^3\Delta r \, U(\mathbf{r};\mathbf{r} + \Delta \mathbf{r}) \Gamma(\Delta \mathbf{r}) \psi^{(2)}(\mathbf{r} + \Delta \mathbf{r}) .
\end{eqnarray}
$(1/\sqrt{V}) \int d^3r$ projects to vanishing total momentum ($V$ is the spatial volume), i.e.\ realizes a meson at rest. $\int_{|\Delta \mathbf{r}| = R} d^3\Delta r$ denotes an integration over a sphere of radius $R$, which is the distance between the antiquark and the quark. $\Gamma(\Delta \mathbf{r})$ is a suitable combination of spherical harmonics and $\gamma$ matrices (cf.\ Table~\ref{tab.operators}, column ``$\Gamma(\mathbf{n})$, pb''), which determines total angular momentum $J$, parity $\mathcal{P}$ and, in the case of identical quark flavors, charge conjugation $\mathcal{C}$. $U(\mathbf{r};\mathbf{r} + \Delta \mathbf{r})$ is a straight gluonic parallel transporter connecting the antiquark and the quark in a gauge invariant way. For $D$ mesons, e.g.\ $\bar{\psi}^{(1)} \psi^{(2)} = \bar{u} c$, for $D_s$ mesons, e.g.\ $\bar{\psi}^{(1)} \psi^{(2)} = \bar{s} c$ and for charmonium, $\bar{\psi}^{(1)} \psi^{(2)} = \bar{c} c$.

\begin{table}[p]
\centering
\begin{tabular}{|c|c|c|c|c|c|c|c|c|c|c|}
\hline
\multicolumn{1}{|c|}{}&  \multicolumn{3}{c|}{continuum}        &  \multicolumn{4}{c|}{twisted mass lattice QCD} \\
\cline{2-8} 
 \multicolumn{1}{|c|}{index}&  \multicolumn{1}{ c|}{\multirow{2}{*}{$\Gamma(\mathbf{n})$, pb}}        &   \multicolumn{1}{c|}{\multirow{2}{*}{$J$}}       &   \multicolumn{1}{c|}{\multirow{2}{*}{$\mathcal{P}\mathcal{C}$}} &  \multicolumn{1}{c|}{\multirow{2}{*}{tb, $(\pm,\mp)$}} &  \multicolumn{1}{c|}{\multirow{2}{*}{tb, $(\pm,\pm)$}} &\multicolumn{2}{c|}{\multirow{2}{*}{$\mathrm{O}^S \otimes  \mathrm{O}^L \rightarrow \mathrm{O}^J$}} \\
 \multicolumn{1}{|c|}{}&                                                                        &                                                    &                                                                  &                                                        &                                                        &\multicolumn{2}{c|}{}\\
\hline
\hline
1&$\gamma_5$					&\multirow{8}{*}{0}&$- +$&pb&$\pm i\gamma_5\times$&\multirow{4}{*}{$A_1\otimes A_1$ }&\multirow{8}{*}{$A_1$}  \\  
2&$\gamma_0\gamma_5$				&&$- +$&$\pm i\gamma_5\times$&pb&&  \\  
3&$\mathds{1}$					&&$+ +$&pb&$\pm i\gamma_5\times$&&  \\  
4&$\gamma_0$					&&$+ -$&$\pm i\gamma_5\times$&pb&&  \\ 
\cline{2-2}\cline{4-7} 
5&$        \gamma_5\gamma_j\mathbf{n}_j$&&$- -$&$\pm i\gamma_5\times$&pb&\multirow{4}{*}{$T_1\otimes T_1$}&  \\  
6&$\gamma_0\gamma_5\gamma_j\mathbf{n}_j$&&$- +$&pb&$\pm i\gamma_5\times$&&  \\  
7&$                \gamma_j\mathbf{n}_j$&&$+ +$&$\pm i\gamma_5\times$&pb&&  \\  
8&$\gamma_0        \gamma_j\mathbf{n}_j$&&$+ +$&pb&$\pm i\gamma_5\times$&&  \\ 
\hline\hline
1&$                \gamma_1$&\multirow{16}{*}{1}&$- -$&$\pm i\gamma_5\times$&pb&\multirow{4}{*}{$T_1\otimes A_1$}&\multirow{16}{*}{$T_1$}  \\
2&$\gamma_0        \gamma_1$&&$- -$&pb&$\pm i\gamma_5\times$&&  \\ 
3&$        \gamma_5\gamma_1$&&$+ +$&$\pm i\gamma_5\times$&pb&&  \\ 
4&$\gamma_0\gamma_5\gamma_1$&&$+ -$&pb&$\pm i\gamma_5\times$&&  \\
\cline{2-2}\cline{4-7} 
5&$\mathbf{n}_1$			&&$- -$&pb&$\pm i\gamma_5\times$&\multirow{4}{*}{$A_1\otimes T_1$}&  \\
6&$\gamma_0\mathbf{n}_1$		&&$- +$&$\pm i\gamma_5\times$&pb&&  \\ 
7&$\gamma_5\mathbf{n}_1$		&&$+ -$&pb&$\pm i\gamma_5\times$&&  \\ 
8&$\gamma_0\gamma_5\mathbf{n}_1$	&&$+ -$&$\pm i\gamma_5\times$&pb&&  \\
\cline{2-2}\cline{4-7} 
 9&$                (\mathbf{n}\times\vec{\gamma})_1$&&$+ +$&$\pm i\gamma_5\times$&pb&\multirow{4}{*}{$T_1\otimes T_1$}&  \\
10&$\gamma_0        (\mathbf{n}\times\vec{\gamma})_1$&&$+ +$&pb&$\pm i\gamma_5\times$&&  \\ 
11&$\gamma_5        (\mathbf{n}\times\vec{\gamma})_1$&&$- -$&$\pm i\gamma_5\times$&pb&&  \\ 
12&$\gamma_0\gamma_5(\mathbf{n}\times\vec{\gamma})_1$&&$- +$&pb&$\pm i\gamma_5\times$&&  \\
\cline{2-2}\cline{4-7} 
13&$                \gamma_1(2\mathbf{n}^2_1-\mathbf{n}^2_2-\mathbf{n}^2_3)$&&$- -$&$\pm i\gamma_5\times$&pb&\multirow{4}{*}{$T_1\otimes E$}&  \\
14&$\gamma_0        \gamma_1(2\mathbf{n}^2_1-\mathbf{n}^2_2-\mathbf{n}^2_3)$&&$- -$&pb&$\pm i\gamma_5\times$&&  \\ 
15&$        \gamma_5\gamma_1(2\mathbf{n}^2_1-\mathbf{n}^2_2-\mathbf{n}^2_3)$&&$+ +$&$\pm i\gamma_5\times$&pb&&  \\ 
16&$\gamma_0\gamma_5\gamma_1(2\mathbf{n}^2_1-\mathbf{n}^2_2-\mathbf{n}^2_3)$&&$+ -$&pb&$\pm i\gamma_5\times$&&  \\
\hline\hline
1&$                (\mathbf{n}_1^2+\mathbf{n}_2^2-2\mathbf{n}_3^2)$&\multirow{8}{*}{2}&$+ +$&pb&$\pm i\gamma_5\times$&\multirow{4}{*}{$A_1\otimes E$}&\multirow{8}{*}{$E$} \\
2&$\gamma_0        (\mathbf{n}_1^2+\mathbf{n}_2^2-2\mathbf{n}_3^2)$&&$+ -$&$\pm i\gamma_5\times$&pb&&  \\ 
3&$        \gamma_5(\mathbf{n}_1^2+\mathbf{n}_2^2-2\mathbf{n}_3^2)$&&$- +$&pb&$\pm i\gamma_5\times$&&  \\ 
4&$\gamma_0\gamma_5(\mathbf{n}_1^2+\mathbf{n}_2^2-2\mathbf{n}_3^2)$&&$- +$&$\pm i\gamma_5\times$&pb&&  \\
\cline{2-2}\cline{4-7}
5&$                (\gamma_1\mathbf{n}_1+\gamma_2\mathbf{n}_2-2\gamma_3\mathbf{n}_3)$&&$+ +$&$\pm i\gamma_5\times$&pb&\multirow{4}{*}{$T_1\otimes T_1$}&  \\
6&$\gamma_0        (\gamma_1\mathbf{n}_1+\gamma_2\mathbf{n}_2-2\gamma_3\mathbf{n}_3)$&&$+ +$&pb&$\pm i\gamma_5\times$&&  \\ 
7&$        \gamma_5(\gamma_1\mathbf{n}_1+\gamma_2\mathbf{n}_2-2\gamma_3\mathbf{n}_3)$&&$- -$&$\pm i\gamma_5\times$&pb&&  \\ 
8&$\gamma_0\gamma_5(\gamma_1\mathbf{n}_1+\gamma_2\mathbf{n}_2-2\gamma_3\mathbf{n}_3)$&&$- +$&pb&$\pm i\gamma_5\times$&&  \\
\hline\hline
1&$                (\gamma_3\mathbf{n}_2+\gamma_2\mathbf{n}_3)$&\multirow{8}{*}{2}&$+ +$&$\pm i\gamma_5\times$&pb&\multirow{4}{*}{$T_1\otimes T_1$}&\multirow{8}{*}{$T_2$}  \\
2&$\gamma_0        (\gamma_3\mathbf{n}_2+\gamma_2\mathbf{n}_3)$&&$+ +$&pb&$\pm i\gamma_5\times$&&  \\ 
3&$        \gamma_5(\gamma_3\mathbf{n}_2+\gamma_2\mathbf{n}_3)$&&$- -$&$\pm i\gamma_5\times$&pb&&  \\ 
4&$\gamma_0\gamma_5(\gamma_3\mathbf{n}_2+\gamma_2\mathbf{n}_3)$&&$- +$&pb&$\pm i\gamma_5\times$&&  \\
\cline{2-2}\cline{4-7}
5&$                \gamma_1(\mathbf{n}_2^2-\mathbf{n}_3^2)$&&$- -$&$\pm i\gamma_5\times$&pb&\multirow{4}{*}{$T_1\otimes E$}&  \\
6&$\gamma_0        \gamma_1(\mathbf{n}_2^2-\mathbf{n}_3^2)$&&$- -$&pb&$\pm i\gamma_5\times$&&  \\ 
7&$        \gamma_5\gamma_1(\mathbf{n}_2^2-\mathbf{n}_3^2)$&&$+ +$&$\pm i\gamma_5\times$&pb&&  \\ 
8&$\gamma_0\gamma_5\gamma_1(\mathbf{n}_2^2-\mathbf{n}_3^2)$&&$+ -$&pb&$\pm i\gamma_5\times$&&  \\ 
\hline
\end{tabular}
\caption{\label{tab.operators}Meson creation operators.}
\end{table}

Our lattice meson creation operators are of similar form,
\begin{eqnarray}
\label{EQN507} O_{\Gamma,\bar{\chi}^{(1)} \chi^{(2)}}^\textrm{twisted} \ \ \equiv \ \ \frac{1}{\sqrt{V/a^3}} \sum_\mathbf{n} \bar{\chi}^{(1)}(\mathbf{n}) \sum_{\Delta \mathbf{n} = \pm \mathbf{e}_x , \pm \mathbf{e}_y , \pm \mathbf{e}_z} U(\mathbf{n};\mathbf{n} + \Delta \mathbf{n}) \Gamma(\Delta \mathbf{n}) \chi^{(2)}(\mathbf{n} + \Delta \mathbf{n}) ,
\end{eqnarray}
where the integration over a sphere with the center at $\mathbf{r}$ has been replaced by the sum over the six neighboring lattice sites of $\mathbf{n}$ and $U(\mathbf{n};\mathbf{n} + \Delta \mathbf{n})$ denotes the link between $\mathbf{n}$ and $\mathbf{n} + \Delta \mathbf{n}$. Moreover, physical basis quark operators $\bar{\psi}^{(1)}$, $\psi^{(2)}$ have been replaced by their twisted basis counterparts $\bar{\chi}^{(1)}$, $\chi^{(2)}$.

In the continuum, the relation between the physical and the twisted basis is given by the twist rotation,
\begin{eqnarray}
\label{EQN589} \psi^{(f)} \ \ = \ \ \exp\Big(i \gamma_5 \tau_3 \omega / 2\Big) \chi^{(f)} \quad , \quad \bar{\psi}^{(f)} \ \ = \ \ \bar{\chi}^{(f)} \exp\Big(i \gamma_5 \tau_3 \omega / 2\Big)
\end{eqnarray}
with the twist angle $\omega$, where $\omega = \pi / 2$ at maximal twist. $\chi^{(f)}$ denotes either the light doublet $\chi^{(l)} = (\chi^{(u)} , \chi^{(d)})$, the strange doublet $\chi^{(s)} = (\chi^{(s^+)} , \chi^{(s^-)})$ or the charm doublet $\chi^{(c)} = (\chi^{(c^+)} , \chi^{(c^-)})$.

When transforming a twisted basis quark bilinear $\bar{\chi}^{(1)} \Gamma \chi^{(2)}$ as e.g.\ appearing in (\ref{EQN507}) to the physical basis or vice versa, the result depends not only on $\Gamma$, but also on the flavor combination, i.e.\ whether $\bar{\chi}^{(1)}$ and $\chi^{(2)}$ are upper components (twisted mass term $+i \mu \gamma_5$) or lower components (twisted mass term $-i \mu \gamma_5$) of twisted basis doublets. In the columns ``tb, $(\pm,\mp)$'' and ``tb, $(\pm,\pm)$'' of Table~\ref{tab.operators}, we list for all flavor combinations ($+$ and $-$ denote the signs in front of the twisted mass terms for $\bar{\chi}^{(1)}$ and $\chi^{(2)}$) and all $\Gamma$ combinations of our meson creation operators, how physical and twisted bases are related. ``pb'' indicates that the twisted basis $\Gamma$ is the same as the physical basis $\Gamma$, while ``$\pm i \gamma_5 \times$'' denotes that the physical basis $\Gamma$ has to be multiplied from the left with $\pm i \gamma_5$ to obtain the corresponding twisted $\Gamma$.

Isospin $I$ and parity $\mathcal{P}$ are symmetries of QCD. While in Wilson twisted mass lattice QCD the $z$ component of isospin $I_z$ is still a quantum number, $I$ and $\mathcal{P}$ are broken by $\mathcal{O}(a)$ due to the Wilson term $-\bar{\chi}^{(f)} (a/2) \nabla^\ast_\mu \nabla_\mu \chi^{(f)}$ appearing in the twisted mass actions (\ref{EQN001}) and (\ref{EQN302}).

For $D$ mesons, we use trial states $O_{\Gamma,\bar{\chi}^{(1)} \chi^{(2)}}^\textrm{twisted} | \Omega \rangle$ with defined $I_z$, e.g.\ $\bar{\chi}^{(1)} \chi^{(2)} = \bar{\chi}^{(d)} \chi^{(c^+)}$ is suited for $D$ mesons with $I_z = +1/2$. There are eight appropriate flavor combinations for $D$ mesons, where the four with opposite signs in front of the twisted mass terms ($(+,-)$ discretization),
\begin{eqnarray}
\label{EQN348} \bar{\chi}^{(d)} \chi^{(c^+)} \ \ , \ \ \bar{\chi}^{(u)} \chi^{(c^-)} \ \ , \ \ \bar{\chi}^{(c^-)} \chi^{(u)} \ \ , \ \ \bar{\chi}^{(c^+)} \chi^{(d)}
\end{eqnarray}
are related by symmetry and yield identical correlation functions. Similarly, the four flavor combinations with identical signs in front of the twisted mass terms ($(+,+)$ discretization),
\begin{eqnarray}
\label{EQN349} \bar{\chi}^{(u)} \chi^{(c^+)} \ \ , \ \ \bar{\chi}^{(d)} \chi^{(c^-)} \ \ , \ \ \bar{\chi}^{(c^+)} \chi^{(u)} \ \ , \ \ \bar{\chi}^{(c^-)} \chi^{(d)}
\end{eqnarray}
also yield identical correlation functions. At finite lattice spacing, $(+,-)$ and $(+,+)$ correlation functions slightly differ due to discretization errors. As a consequence, $D$ meson masses computed on the one hand with a $(+,+)$ and on the other hand with a $(+,-)$ flavor combination, but which are otherwise identical, will differ in mass. Because of automatic $\mathcal{O}(a)$ improvement of Wilson twisted mass lattice QCD at maximal twist, this mass splitting will be proportional to $a^2$, i.e.\ is expected to be rather small and will vanish quadratically when approaching the continuum limit. This theoretical expectation is confirmed by our numerical results shown in Section\ \ref{sec:results}. We use both $(+,-)$ results and $(+,+)$ results when performing continuum extrapolations.

Note that parity is not a symmetry, i.e.\ there is no rigorous separation between $\mathcal{P} = +$ and $\mathcal{P} = -$ states in Wilson twisted mass lattice QCD. Nevertheless, it is possible to assign parity quantum numbers to the extracted meson masses in a clean and unambiguous way (cf.\ \cite{Kalinowski:2015bwa}, Section~4.2 for a detailed numerical example).

Identical considerations apply for $D_s$ mesons, when replacing $(u,d) \rightarrow (s^+,s^-)$.

For charmonium creation operators, there are two appropriate flavor combinations,
\begin{eqnarray}
\bar{\chi}^{(c^+)} \chi^{(c^+)} \ \ , \ \ \bar{\chi}^{(c^-)} \chi^{(c^-)} ,
\end{eqnarray}
which are again related by symmetry. Since we ignore disconnected contributions to correlation functions throughout this work, we can also consider flavor combinations with opposite signs in front of the twisted mass terms,
\begin{eqnarray}
\label{EQN350} \bar{\chi}^{(c^+)} \chi^{(c^-)} \ \ , \ \ \bar{\chi}^{(c^-)} \chi^{(c^+)} .
\end{eqnarray}

On a cubic lattice, rotational symmetry is reduced to symmetry with respect to cubic rotations. There are only five different irreducible representations of the cubic group $\mathrm{O}$ (labeled by $A_1$, $T_1$, $E$, $T_2$, $A_2$). For our creation operators, we list the corresponding representations for spin, orbital angular momentum and total angular momentum in Table~\ref{tab.operators}, column ``$\mathrm{O}^S \otimes \mathrm{O}^L \rightarrow \mathrm{O}^J$''.
Assignment of continuum angular momentum $J$ to our resulting meson masses is discussed in Section~\ref{sec:results}.

To enhance the overlap of trial states $O_{\Gamma,\bar{\chi}^{(1)} \chi^{(2)}}^\textrm{twisted} | \Omega \rangle$ to low lying meson states, we use standard smearing techniques. This allows one to read off meson masses from the exponential decay of correlation functions at rather small temporal separations, where the signal-to-noise ratio is favorable. Smearing is done in two steps. First, we replace spatial gauge links by their APE smeared counterparts \cite{Albanese:1987ds}. Then, we use Gaussian smearing on the quark fields $\chi^{(l)}$, $\chi^{(s)}$ and $\chi^{(c)}$, which resorts to the APE smeared spatial links. The parameters we have chosen are $N_\textrm{APE} = 10$, $\alpha_\textrm{APE} = 0.5$, $N_\textrm{Gauss} = 30, 36, 65$ (for A, B and D ensembles, respectively) and $\kappa_\textrm{Gauss} = 0.5$. This corresponds to a Gaussian width of the smeared quark fields of approximately $0.24 \, \textrm{fm}$; cf.\ Ref.\ \cite{Jansen:2008si} for detailed equations.


\subsection{Determination of meson masses}


\subsubsection{\label{SEC609}Computation and analysis of correlation matrices}

For each twisted mass sector characterized by flavor $\bar{\chi}^{(1)} \chi^{(2)}$, the cubic representation $\mathrm{O}^J$ and, in the case of charmonium, either $\mathcal{C}$ for twisted mass signs $(+,+)$ or $\mathcal{C} \circ \mathcal{P}^{(\textrm{tm})}\,$\footnote{Twisted mass parity $\mathcal{P}^{(\textrm{tm})}$ is parity combined with flipping the sign in front of the twisted mass term, e.g.\ $u\leftrightarrow d$.} for twisted mass signs $(+,-)$ (for a detailed discussion, cf.\ Ref.\ \cite{Kalinowski:2015bwa}), we compute temporal correlation matrices of meson creation operators
\begin{eqnarray}
\label{EQN698} C_{\Gamma_j;\Gamma_k;\bar{\chi}^{(1)} \chi^{(2)}}(t) \ \ \equiv \ \ \langle \Omega | \Big(S(O_{\Gamma_j,\bar{\chi}^{(1)} \chi^{(2)}}^\textrm{twisted})\Big)^\dagger(t) \Big(S(O_{\Gamma_k,\bar{\chi}^{(1)} \chi^{(2)}}^\textrm{twisted})\Big)(0) | \Omega \rangle .
\end{eqnarray}
$j$ and $k$ label the rows and columns of these correlation matrices or, equivalently, are indices of the meson creation operators entering a correlation matrix (cf.\ Table~\ref{tab.operators}, column ``index''). $S(\ldots)$ indicates that APE smeared gauge links and Gaussian smeared quark fields are used for the meson creation operators (cf.\ the discussion in Section~\ref{SEC599}). For the computations, we use a generalization of the one-end trick, which is explained in detail in Ref.\ \cite{Kalinowski:2015bwa}. Since parity is only an approximate symmetry in twisted mass lattice QCD, we consider correlation matrices of meson creation operators with both $\mathcal{P} = +$ and $\mathcal{P} = -$.

In Table~\ref{tab.ensembles}, we list for each ensemble the number of gauge link configurations used for the computation of the correlation matrices $C_{\Gamma_j;\Gamma_k;\bar{\chi}^{(1)} \chi^{(2)}}$. The four stochastic sources needed for the one-end trick are located on a timeslice, which is randomly chosen for every gauge link configuration. We use a single set of four stochastic timeslice sources, i.e.\ a single sample for each gauge link configuration.

We determine meson masses from the correlation matrices $C_{\Gamma_j;\Gamma_k;\bar{\chi}^{(1)} \chi^{(2)}}$ 
by solving standard generalized eigenvalue problems (cf.\ e.g.\ Ref.\ \cite{Blossier:2009kd} and references therein). From the resulting eigenvalues, we obtain the masses, while the resulting eigenvectors provide information about the operator content, from which one can read off the parity quantum number as well as information regarding the structure of the meson (e.g.\ which spin or orbital angular momentum is dominant). For a detailed discussion, we refer to Ref.\ \cite{Kalinowski:2015bwa}, Section\ 4.2.


\subsubsection{\label{SEC595}Tuning $s$ and $c$ valence quark masses to their physical values}
\label{sec:tuning}

We tune the $s$ and $c$ valence quark masses via $2 m_K^2 - m_\pi^2$ (which does not depend on the $u/d$ quark mass at leading order of chiral perturbation theory) and $m_D$ (which does only weakly depend on the $u/d$ quark mass). For that purpose, we use experimental results for the masses of the electrically neutral mesons $\pi^0$, $K^0$ and $D^0$ in the case of neutral charm-light and charm-charm mesons, whereas for the charged charm-light and charm-strange mesons, we instead use the electrically charged mesons $\pi^\pm$, $K^\pm$ and $D^\pm$.
The idea behind this choice is that for neutral (charged) mesons, the electrical charges are opposite (combine to $\pm1$) both for the input meson masses used for quark mass tuning as well as for all the meson masses predicted by our lattice QCD computation.
Even though electromagnetism is not part of our lattice setup, we expect that by using this procedure, electromagnetic effects are to a sizable extent incorporated in the quark mass tuning. 
As we will discuss in Section\ \ref{sec:results}, when analyzing our results, this procedure seems to work very well, i.e.\ for states, which can be computed with small statistical errors of $\ltapprox 5 \, \textrm{MeV}$ (e.g.\ the $D_s$ meson, $\eta_c(1S)$ and $J/\psi(1S)$), we obtain perfect agreement with experimental results. 
On the other hand, when performing the tuning in the non-ideal opposite way, e.g.\ using neutral mesons for quark mass tuning to predict the $D_s$ meson mass, the lattice result differs from its experimental counterpart by $\approx 5 \, \textrm{MeV}$, which is the typical order of magnitude of isospin breaking effects in meson masses.

We also note that to compute $2 m_K^2 - m_\pi^2$ and $m_D$ for the $s$ and $c$ valence quark mass tuning, we always use the $(+,-)$ discretization, which is known to yield smaller discretization errors for these pseudoscalar mesons\footnote{Moreover, the calculation of the pion mass using the $(+,+)$ discretization requires the computation of disconnected diagrams, which would lead to significantly larger statistical errors and hence less precise $s$ and $c$ valence quark mass tuning.} \cite{Urbach:2007rt,Frezzotti:2007qv}.

The technical aspects of the tuning procedure are explained in more detail in Ref.\ \cite{Kalinowski:2015bwa}. For each ensemble (characterized by $a$ and $m_\pi \equiv m_\pi^{(+,-)}$), we perform computations of $m_K$ for two valence $s$ quark masses $\mu_{s,1}$ and $\mu_{s,2}$, different by around $10 \%$ and both in the region of the physical $s$ quark mass. The physical valence $s$ quark mass can then be determined according to
\begin{eqnarray}
\nonumber & & \hspace{-0.7cm} \mu_{s,\textrm{phys}}(a,m_\pi) \ \ = \ \ \mu_{s,2}(a,m_\pi) \\
\label{EQN753} & & \hspace{0.675cm} + \Big(\mu_{s,1}(a,m_\pi) - \mu_{s,2}(a,m_\pi)\Big) \frac{X_{\textrm{exp}} - X^{(+,-)}(a,m_\pi,\mu_{s,2})}{X^{(+,-)}(a,m_\pi,\mu_{s,1}) - X^{(+,-)}(a,m_\pi,\mu_{s,2})} , 
\end{eqnarray}
where $X^{(+,-)}(a,m_\pi,\mu_s) \equiv 2 (m_K^{(+,-)}(a,m_\pi,\mu_s))^2 - m_\pi)^2$ and $X_{\textrm{exp}} = 2 m_{K^0}^2 - m_{\pi^0}^2 \approx 0.477 \, \textrm{GeV}^2$ (for neutral mesons) or $X_{\textrm{exp}} = 2 m_{K^\pm}^2 - m_{\pi^\pm}^2 \approx 0.467 \, \textrm{GeV}^2$ (for charged mesons). Analogously, the physical valence $c$ quark mass is
\begin{eqnarray}
\nonumber & & \hspace{-0.7cm} \mu_{c,\textrm{phys}}(a,m_\pi) \ \ = \ \ \mu_{c,2}(a,m_\pi) \\
\label{EQN754} & & \hspace{0.675cm} + \Big(\mu_{c,1}(a,m_\pi) - \mu_{c,2}(a,m_\pi)\Big) \frac{m_{D_\textrm{exp}} - m_D^{(+,-)}(a,m_\pi,\mu_{c,2})}{m_D^{(+,-)}(a,m_\pi,\mu_{c,1}) - m_D^{(+,-)}(a,m_\pi,\mu_{c,2})} ,
\end{eqnarray}
where $m_{D_\textrm{exp}}\approx1.865$ GeV (for neutral mesons) or $m_{D_\textrm{exp}}\approx1.870$ GeV (for charged mesons).

\begin{figure}[t!]
\begin{center}
\includegraphics[width=0.345\textwidth,angle=-90]{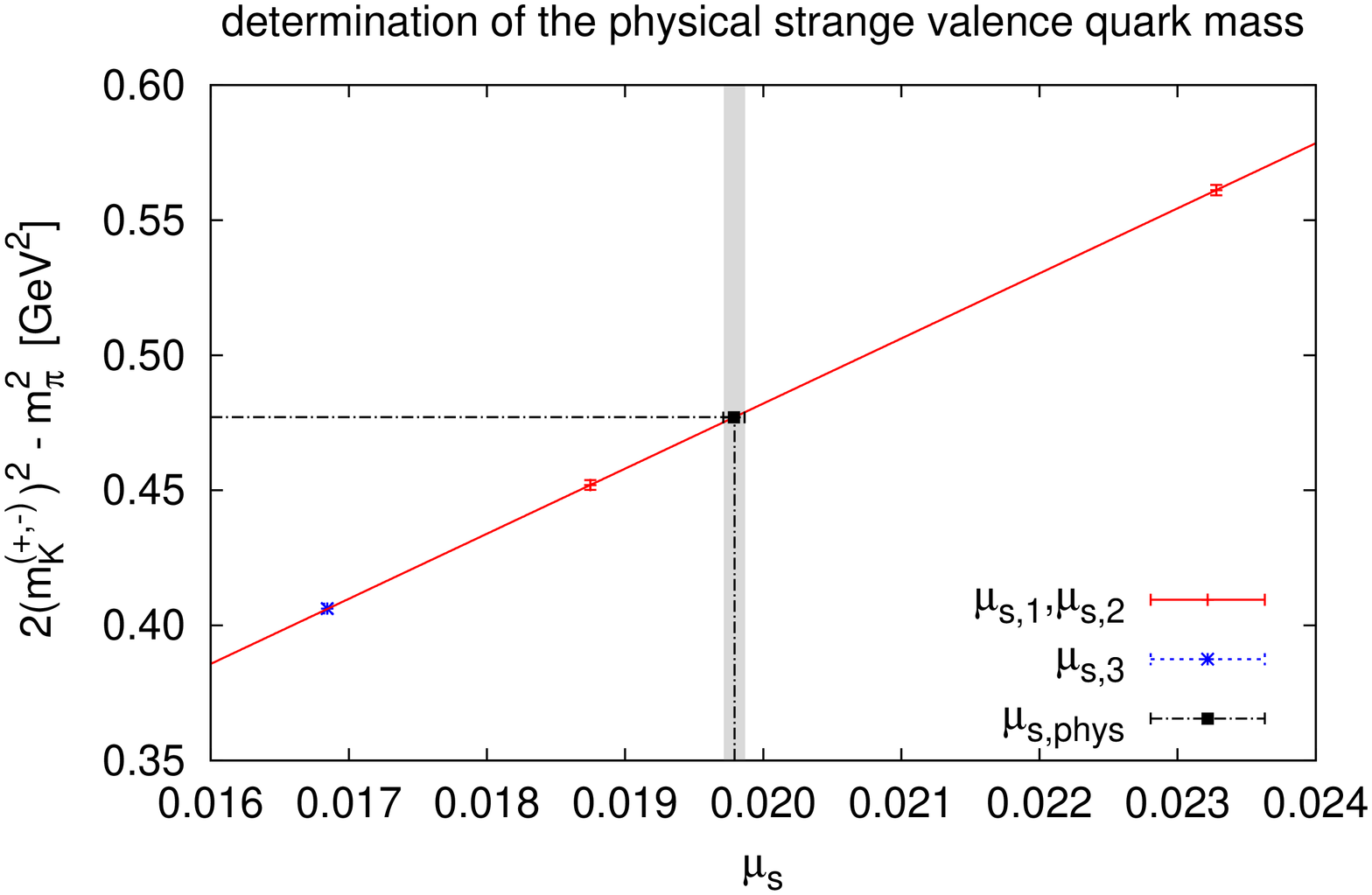}
\includegraphics[width=0.345\textwidth,angle=-90]{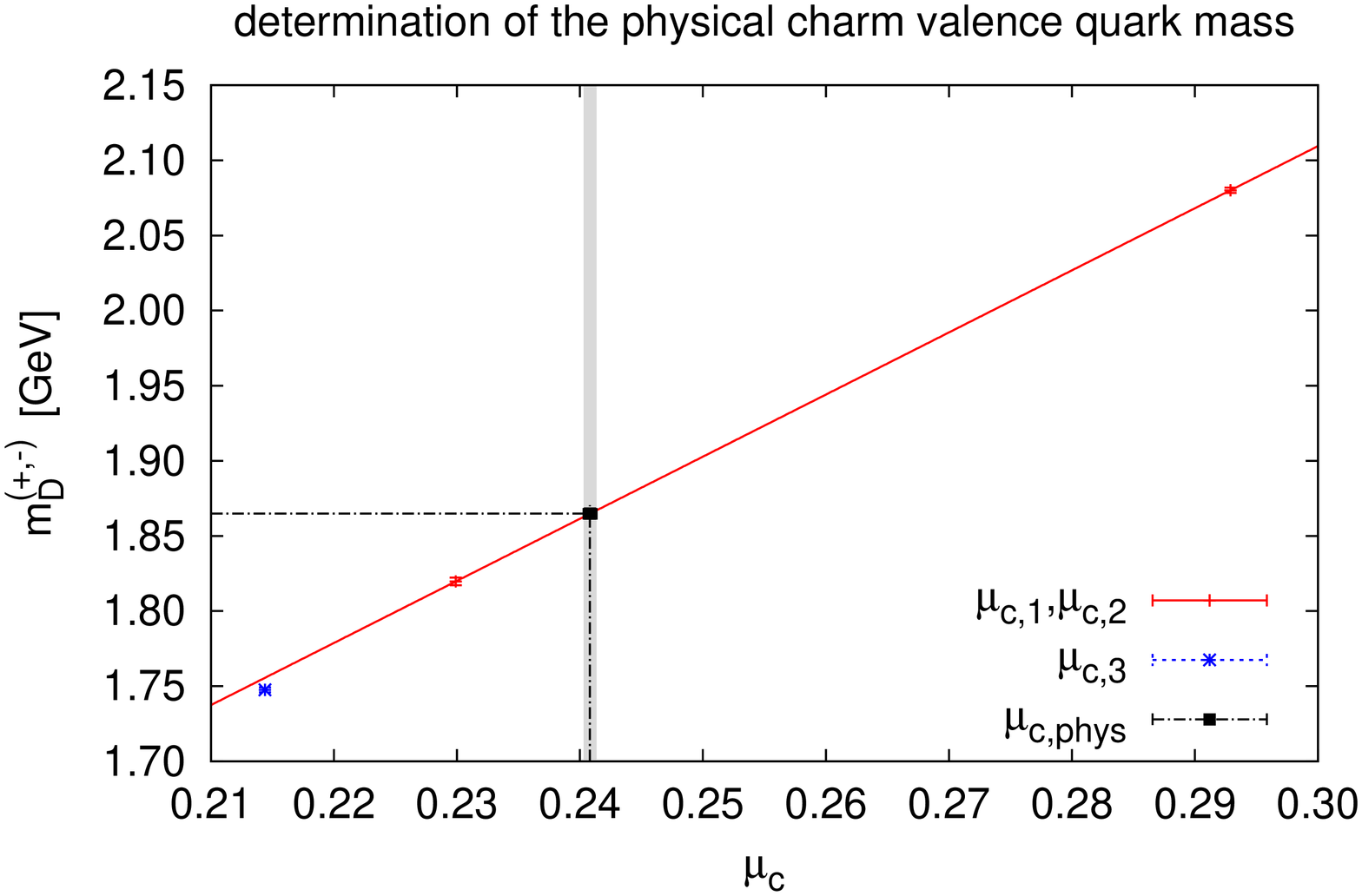}
\caption{\label{fig:tuning}An example of the tuning of $s$ and $c$ valence quark masses (for neutral mesons) to their physical values (ensemble A80.24).}
\end{center}
\end{figure}

An example of the tuning of $s$ and $c$ valence quark masses (for neutral mesons) to their physical values for ensemble A80.24 is shown in Figure~\ref{fig:tuning} (the red dots correspond to $\mu_{s,1}$, $\mu_{s,2}$, $\mu_{c,1}$ and $\mu_{c,2}$, the black dots to $\mu_{s,\textrm{phys}}$ and $\mu_{c,\textrm{phys}}$). For this specific case, we also performed computations for a third $s$ and $c$ valence quark mass, $\mu_{s,3}$ and $\mu_{c,3}$ (represented by the blue points in Figure~\ref{fig:tuning}). This shows that the linear dependence of $2 (m_K^{(+,-)})^2 - m_\pi^2$ and $m_D^{(+,-)}$ on $\mu_s$ and $\mu_c$ assumed in (\ref{EQN753}) and (\ref{EQN754}), respectively, is well satisfied for our numerical data with only a tiny deviation from the linear behavior observed in the case of $m_D^{(+,-)}$.

Having once determined the physical $s$ and $c$ valence quark masses for an ensemble, we can linearly  interpolate/extrapolate all $D$ meson and charmonium masses for that ensemble according to
\begin{eqnarray}
\nonumber & & \hspace{-0.7cm} m^{(+,\pm)}(a,m_\pi) \ \ = \ \ m^{(+,\pm)}(a,m_\pi,\mu_{c,2}) \\
 & & \hspace{0.675cm} + \Big(m^{(+,\pm)}(a,m_\pi,\mu_{c,1}) - m^{(+,\pm)}(a,m_\pi,\mu_{c,2})\Big) \frac{\mu_{c,\textrm{phys}}(a,m_\pi) - \mu_{c,2}(a,m_\pi)}{\mu_{c,1}(a,m_\pi) - \mu_{c,2}(a,m_\pi)}
\end{eqnarray}
and, similarly, all $D_s$ meson masses according to
\begin{eqnarray}
\nonumber & & \hspace{-0.7cm} m^{(+,\pm)}(a,m_\pi) \ \ = \ \ m^{(+,\pm)}(a,m_\pi,\mu_{s,2},\mu_{c,2}) \\
\nonumber & & \hspace{0.675cm} + \Big(m^{(+,\pm)}(a,m_\pi,\mu_{s,1},\mu_{c,s}) - m^{(+,\pm)}(a,m_\pi,\mu_{s,2},\mu_{c,2})\Big) \frac{\mu_{s,\textrm{phys}}(a,m_\pi) - \mu_{s,2}(a,m_\pi)}{\mu_{s,1}(a,m_\pi) - \mu_{s,2}(a,m_\pi)} \\
 & & \hspace{0.675cm} + \Big(m^{(+,\pm)}(a,m_\pi,\mu_{s,2},\mu_{c,1}) - m^{(+,\pm)}(a,m_\pi,\mu_{s,2},\mu_{c,2})\Big) \frac{\mu_{c,\textrm{phys}}(a,m_\pi) - \mu_{c,2}(a,m_\pi)}{\mu_{c,1}(a,m_\pi) - \mu_{c,2}(a,m_\pi)} .
\end{eqnarray}


\subsubsection{\label{SEC594}Extrapolating meson masses to physically light $u/d$ quark masses and to the continuum}
\label{sec:fits}

We use the procedure described in the previous subsection to compute meson masses for nine of the ten ensembles listed in Table~\ref{tab.ensembles} using both the $(+,-)$ and the $(+,+)$ valence quark discretizations (ensemble A40.24 is only used to exclude finite volume effects; cf.\ Section\ \ref{sec:FVE}). Then, we extrapolate these meson masses to physically light $u/d$ quark masses (for simplicity, also denoted as the chiral extrapolation in the following) and to the continuum. For each meson state, we perform an independent fit to the corresponding 18 lattice QCD masses $m^{(+,\pm)}(a,m_\pi)$ (nine ensembles characterized by $a$ and $m_\pi$, two discretizations $(+,-)$ and $(+,+)$).

Our most general fitting ansatz is first order in $a^2$ (discretization errors proportional to $a$ are excluded due to automatic $\mathcal{O}(a)$ improvement of Wilson twisted mass lattice QCD) and linear in $m_\pi^2$ (typically the leading order in chiral perturbation theory):
\begin{eqnarray}
\label{eq:TM} & & \hspace{-0.7cm} m^{(+,-)}(a,m_\pi) \ \ \equiv \ \ m^{(+,-)} + c^{(+,-)} a^2 + \alpha^{(+,-)} \Big(m_\pi^2 - m_{\pi^0,\textrm{exp}}^2\Big) , \\
\label{eq:OS} & & \hspace{-0.7cm} m^{(+,+)}(a,m_\pi) \ \ \equiv \ \ m^{(+,+)} + c^{(+,+)} a^2 + \alpha^{(+,+)} \Big(m_\pi^2 - m_{\pi^0,\textrm{exp}}^2\Big) ,
\end{eqnarray}
where $m^{(+,-)}$, $m^{(+,+)}$, $c^{(+,-)}$, $c^{(+,+)}$, $\alpha^{(+,-)}$ and $\alpha^{(+,+)}$ are fitting parameters and $m_{\pi^0,\textrm{exp}} \approx 135 \, \textrm{MeV}$\,\footnote{For both the neutral and charged meson masses, we extrapolate to the mass of the neutral pion. The difference in meson masses if the charged pion mass is taken for extrapolating to the physical light quark mass is negligible with respect to the total error, in contrast to the effect of taking neutral/charged $\pi$, $K$ and $D$ meson masses for the $s$ and $c$ quark masses tuning described in Section\ \ref{sec:tuning}.}. We adopt three slightly different strategies to obtain and cross-check $u/d$ quark mass and continuum extrapolated results for meson masses.
\begin{itemize}
\item \textbf{Strategy 1} \\
We take $m^{(+,-)}$, $m^{(+,+)}$, $c^{(+,-)}$, $c^{(+,+)}$, $\alpha^{(+,-)}$ and $\alpha^{(+,+)}$ as six independent fitting parameters. In this case, the fits of (\ref{eq:TM}) to $(+,-)$ lattice QCD results and of (\ref{eq:OS}) to $(+,+)$ lattice QCD results decouple and we obtain for each meson mass two independent estimates $m^{(+,-)}$ and $m^{(+,+)}$, respectively. Since these estimates correspond to different discretizations, this strategy allows one to test their universality ($m^{(+,-)}$ and $m^{(+,+)}$ should coincide).

\item \textbf{Strategy 2} \\
Alternatively, we can take this universality for granted and set $m^{(+,-)} \equiv m^{(+,+)} \equiv m$, leaving five fitting parameters, $m$, $c^{(+,-)}$, $c^{(+,+)}$, $\alpha^{(+,-)}$ and $\alpha^{(+,+)}$. This allows one to perform a single combined $u/d$ quark mass and continuum extrapolation using both discretizations simultaneously.

\item \textbf{Strategy 3} \\
We can additionally assume that the slope of the chiral extrapolation is discretization independent, i.e.\ set $\alpha^{(+,-)} = \alpha^{(+,+)} \equiv \alpha$ and perform fits with four fitting parameters, $m$, $c^{(+,-)}$, $c^{(+,+)}$ and $\alpha$. This assumption seems reasonable, because any discretization dependence of the $u/d$ quark mass dependence is expected to be of higher order, i.e.\ $\mathcal{O}(a^2 (m_\pi^2 - m_{\pi^0,\textrm{exp}}^2))$.
\end{itemize}
In the following, our preferred fitting strategy is \textbf{Strategy 2} and all results presented in the following have been generated accordingly. Note, however, that we have cross-checked these results by also using the other two strategies, thereby confirming the universality of our two valence quark discretizations as well as discretization independence of the $u/d$ quark mass dependence (see Section\ \ref{sec:ansatz} for more details).

Note that we have also investigated higher order discretization effects by adding corresponding terms to our fitting ansatz (\ref{eq:TM}) and (\ref{eq:OS}). In the twisted mass formulation, all odd powers of the lattice spacing are excluded at maximal twist \cite{Frezzotti:2003ni,Frezzotti:2004wz,Frezzotti:2005gi,Chiarappa:2006ae}. From the typical size of $\mathcal{O}(a^2)$ effects, one expects the $\mathcal{O}(a^4)$ corrections to be on the 0.05-0.2\% level, and indeed we have found no statistical significance of such higher order terms in any of the computed meson masses. Similarly, we have performed fits including higher order terms in $m_\pi^2 - m_{\pi^0,\textrm{exp}}^2$. We have as well found that these are negligible at the current level of statistical precision.


\subsubsection{\label{sec:analysis}Determination of statistical and systematic errors}
\label{sec:syst}

The input meson masses $m^{(+,\pm)}(a,m_\pi,\mu_s,\mu_c)$ for the quark mass and continuum extrapolations are extracted from correlation matrices (\ref{EQN698}) by solving generalized eigenvalue problems, as discussed in Section~\ref{SEC609}. For each meson mass $m^{(+,\pm)}(a,m_\pi,\mu_s,\mu_c)$, one needs to decide for a temporal fitting interval, where a constant (representing $m^{(+,\pm)}(a,m_\pi,\mu_s,\mu_c)$) is fitted to the corresponding effective mass. Therefore, the obtained input meson masses as well as the subsequently generated extrapolations depend to some extent on the chosen fitting intervals. To arrive at results which are quite independent of a possibly somewhat arbitrary single specific choice of such fitting intervals, we adopt a systematic procedure. This procedure combines results obtained by choosing a variety of fitting intervals and provides an estimate of the related uncertainty by taking the spread of these results into account.

Our procedure is applied independently to each meson state. It consists of the following steps:
\begin{enumerate}
\item For each ensemble, each valence quark discretization and each $s$ and $c$ valence quark mass, i.e.\ for each meson mass $m^{(+,\pm)}(a,m_\pi,\mu_s,\mu_c)$, choose the minimal Euclidean time for extracting the meson mass from the corresponding effective mass, $t_{\rm min}(a,m_\pi,\mu_s,\mu_c)/a$, as the smallest $t/a$ for which a fit in the interval $[t/a,\,t/a+2]$ yields a $\chi^2/{\rm dof} \leq 1.3$. (This we take as an indication that the contamination by excited states is of the same order of magnitude or smaller than statistical errors.)

\item For each ensemble, each valence quark discretization and each $s$ and $c$ valence quark mass, i.e.\ for each meson mass $m^{(+,\pm)}(a,m_\pi,\mu_s,\mu_c)$, choose the maximal time for extracting the meson mass from the corresponding effective mass, $t_{\rm max}(a,m_\pi,\mu_s,\mu_c)/a$, as the largest $t/a$ for which the statistical error of the effective mass is still smaller than five times the error of the effective mass at time $t_{\rm min}(a,m_\pi,\mu_s,\mu_c)/a$. (In this way data points with rather large statistical errors, which usually do not contain any useful information, are excluded.)

\item Given $t_{\rm min}(a,m_\pi,\mu_s,\mu_c)/a$ and $t_{\rm max}(a,m_\pi,\mu_s,\mu_c)/a$, generate 13 sets of input meson masses $m^{(+,\pm)}(a,m_\pi,\mu_s,\mu_c)$ obtained from fitting intervals (with minimal length of three timeslices)
\begin{eqnarray}
\nonumber & & \hspace{-0.7cm} \Big[t_{\rm min}(a,m_\pi,\mu_s,\mu_c)/a + \textrm{floor}\Big(n \delta(a,m_\pi,\mu_s,\mu_c)\Big) \ , \ t_{\rm max}(a,m_\pi,\mu_s,\mu_c)/a\Big] \quad , \\
 & & \hspace{0.675cm} \quad \delta(a,m_\pi,\mu_s,\mu_c) \ \ \equiv \ \ \frac{t_{\rm max}(a,m_\pi,\mu_s,\mu_c)/a - t_{\rm min}(a,m_\pi,\mu_s,\mu_c)/a-2}{12} ,
\end{eqnarray}
$n=0,\ldots,12$ (we will later refer to this as the shift parameter), and perform the corresponding 13 quark mass and continuum extrapolations as explained in Section~\ref{SEC595} and Section~\ref{SEC594}. These resulting input meson masses are denoted by $m_n$. For convenience, we also define $\tilde{m}_n$, which are the same values, but sorted in ascending order, i.e.\ $\{ m_0, m_1, \ldots , m_{12} \} = \{ \tilde{m}_0, \tilde{m}_1, \ldots , \tilde{m}_{12} \}$ and $\tilde{m}_0 < \tilde{m}_1 < \ldots < \tilde{m}_{12}$.

\item From the set of 13 resulting $u/d$, $s$ and $c$ quark mass and continuum extrapolated meson masses $\{ \tilde{m}_0, \tilde{m}_1, \ldots , \tilde{m}_{12} \}$, we discard the two smallest and the two largest outliers, i.e.\ $\tilde{m}_0$, $\tilde{m}_1$, $\tilde{m}_{11}$ and $\tilde{m}_{12}$. The 9 remaining estimates $\tilde{m}_2 < \tilde{m}_3 < \ldots < \tilde{m}_{10}$ cover roughly a $1 \, \sigma$ region assuming a Gaussian distribution ($9 / 13 \approx 69 \%$).

\item The final result for the meson mass is generated from $\{ \tilde{m}_2, \tilde{m}_3, \ldots , \tilde{m}_{10} \}$ as follows:
\begin{itemize}
\item \textbf{Central value }$\overline{m}$: \\ $\overline{m}$ is that $\tilde{m}_n \in \{ \tilde{m}_2, \tilde{m}_3, \ldots , \tilde{m}_{10} \}$ which is the closest to $(\tilde{m}_{10} + \tilde{m}_2) / 2$, i.e.\ closest to the mean of the largest and the smallest $\tilde{m}_n$.

\item \textbf{Statistical uncertainty }$\Delta m_\textrm{stat}$: \\ $\Delta m_\textrm{stat}$ is the statistical error associated with that $\tilde{m}_n$ taken as the central value $\overline{m}$ \footnote{The statistical errors of all estimates $m_n$ are determined via evolved jackknife analyses starting at the level of the correlation matrices (\ref{EQN698}). To exclude statistical correlations between gauge link configurations which are close in Monte Carlo simulation time, we performed a suitable binning of these configurations.}. 

\item \textbf{Systematic uncertainty }$\Delta m_\textrm{syst}$: \\ $\Delta m_\textrm{syst} = \textrm{max}(|\overline{m} - \tilde{m}_2| \, , \, |\overline{m} - \tilde{m}_{10}|)$, i.e.\ the larger of the two usually roughly equal differences between the central value and the largest and the smallest values in $\{ \tilde{m}_2, \tilde{m}_3, \ldots , \tilde{m}_{10} \}$, respectively.
\end{itemize}
\end{enumerate}

\begin{figure}[t!]
\begin{center}
\includegraphics[width=0.30\textwidth,angle=-90]{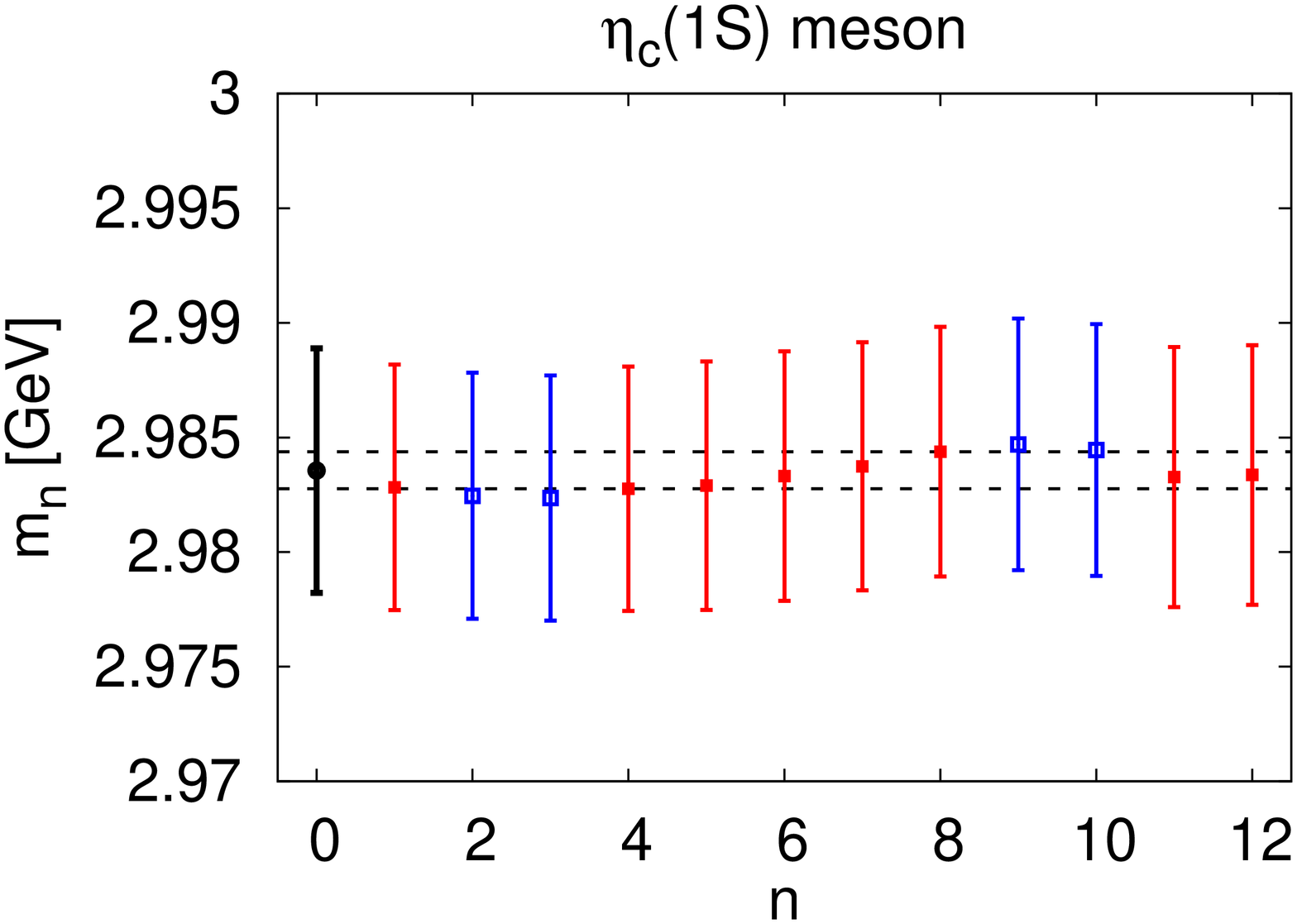}
\includegraphics[width=0.30\textwidth,angle=-90]{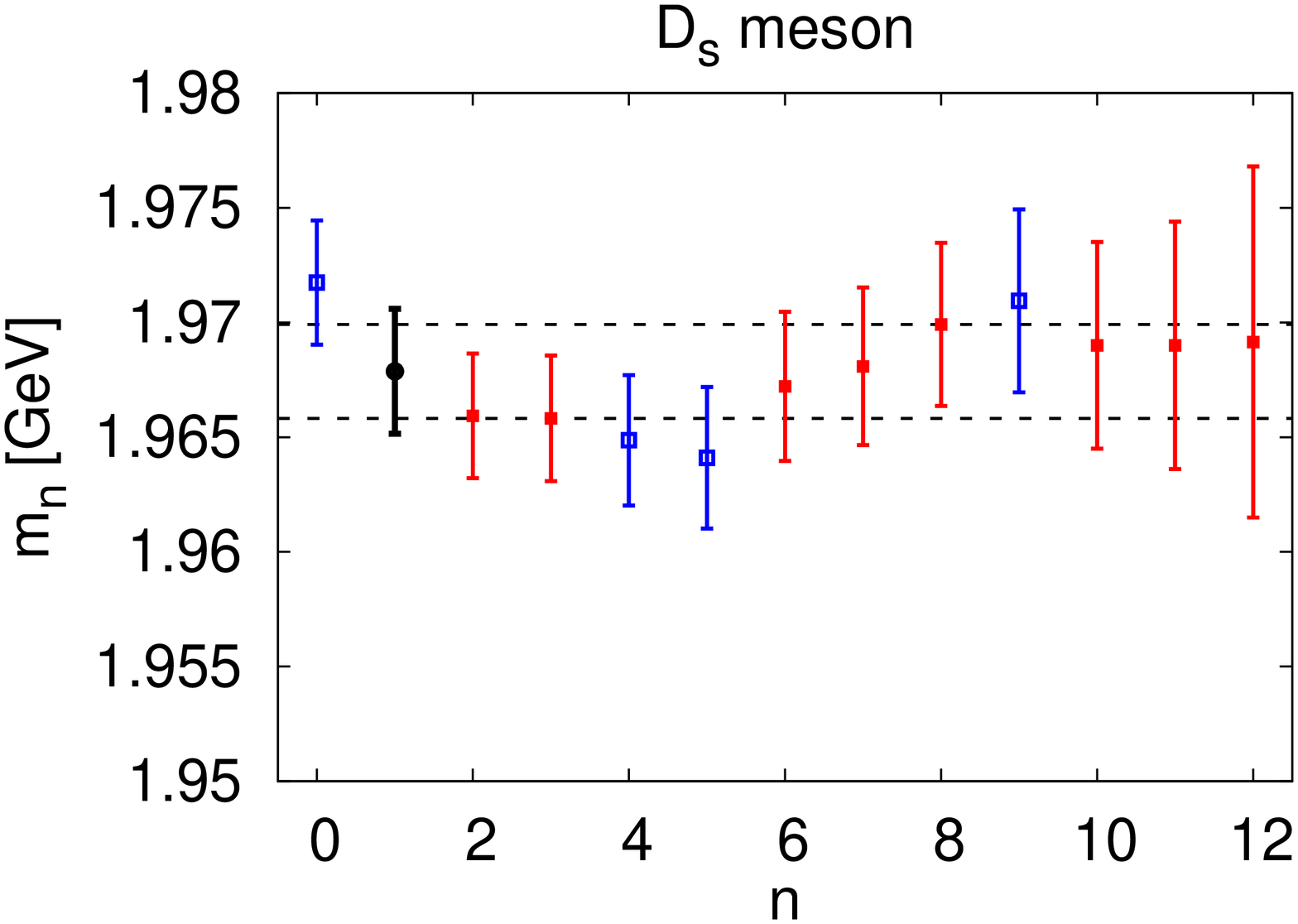}
\includegraphics[width=0.30\textwidth,angle=-90]{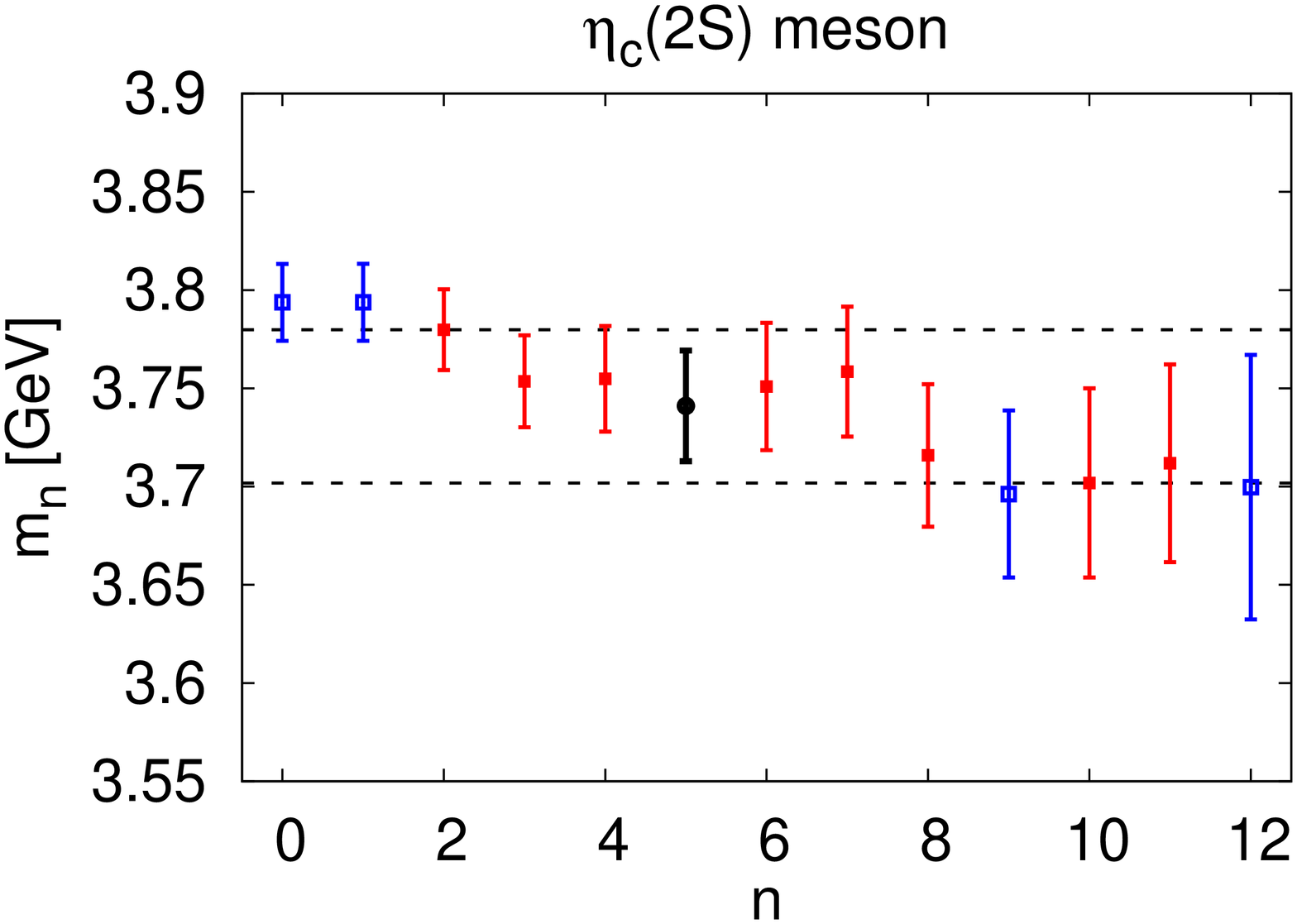}
\includegraphics[width=0.30\textwidth,angle=-90]{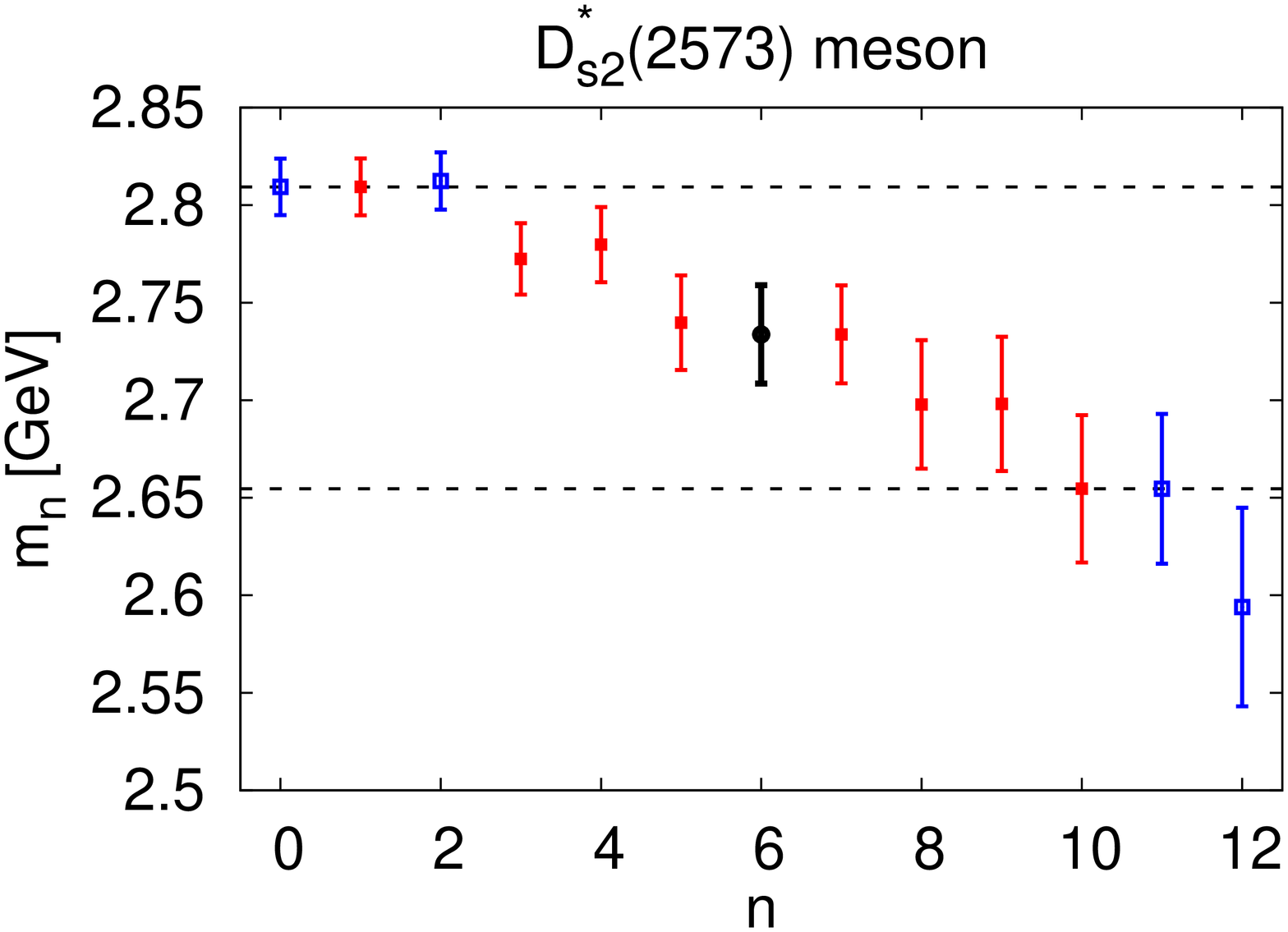}
\caption{\label{fig:syst}Illustration of our procedure to determine the systematic uncertainty. Each plot shows the 13 estimates $m_n$ (generated with \textbf{Strategy 2}). The four outliers, which are removed in step~4, are colored blue. That $m_n$ taken as the central value $\overline{m}$ and defining the statistical error $\Delta m_\textrm{stat}$ is colored black. All remaining $m_n$ are colored red. The resulting systematic uncertainty is represented by dashed black lines.
\textbf{(top):}~$\eta_c(1S)$ and $D_s$, cases, where the fitting range uncertainty is smaller than the statistical error (the vertical scale in both plots is the same for better comparison).
\textbf{(bottom):}~$\eta_c(2S)$ and $D_{s2}^*(2573)$, cases, where the fitting range uncertainty is larger than the statistical error (the vertical scale in both plots is the same for better comparison).
}
\end{center}
\end{figure}

This procedure, in particular step~4 and step~5, is illustrated in Figure~\ref{fig:syst} for four mesons, $\eta_c(1S)$, $D_s$, $\eta_c(2S)$ and $D_{s2}^*(2573)$, ordered ascendingly according to $\Delta m_\textrm{syst} / \Delta m_\textrm{stat}$ (i.e.\ straightforward cases with very clear effective mass plateaus first, more delicate cases last). The ratio $\Delta m_\textrm{syst} / \Delta m_\textrm{stat}$ assumes values roughly between $0.2$ and $3.0$, which clearly shows that for some states, the statistical uncertainty dominates, while for other states, the systematic uncertainty is dominant. Each of the plots shows the 13 estimates $m_n$ (generated with \textbf{Strategy 2}). The four outliers, which are removed in step~4, are colored blue. That $m_n$ taken as the central value $\overline{m}$ and defining the statistical error $\Delta m_\textrm{stat}$ is colored black. All remaining $m_n$ are colored red.

In the case of $\eta_c(1S)$, the extracted estimates $m_n$ are essentially independent of the choice of the fitting range for effective mass plateaus. Consequently, the systematic uncertainty is small and the total error is close to the statistical error, which is nearly the same for all 13 plateau choices. Hence, our procedure generates essentially the same result as one would obtain by a less evolved analysis of choosing a single specific effective mass plateau.

In the case of $D_{s2}^*(2573)$, the situation is rather opposite. The choice of the effective mass plateau has a strong influence on the resulting meson mass, i.e.\ the extracted estimates $m_n$ differ within statistical errors by a few $\sigma$. Correspondingly, our systematic procedure generates a rather large systematic uncertainty, which represents the spread of the estimates $m_n$. This systematic uncertainty is significantly larger than any of the statistical errors of the $m_n$ and, hence, dominates the total error. For such cases, we consider our systematic procedure more conservative and the corresponding results hence superior compared to results obtained by just deciding for a single specific fitting interval for each effective mass.
Note, however, that if the meson mass systematically drops down when the fit starts at larger Euclidean times, as for $D_{s2}^*(2573)$, the resulting fitting range uncertainty may still be underestimated.
In such a case, the proper identification of this systematic error would require going to even larger $t/a$, which is impossible due to the strongly decreasing signal-to-noise ratio at large $t/a$.
In our discussion of specific states in Section\ \ref{sec:results}, we will always emphasize this whenever this is the case.


\subsubsection{\label{SEC811}Differences of meson masses}

In addition to individual meson masses, we also determine the following differences of meson masses:
\begin{itemize}
\item All $D$ mesons, e.g.\ flavor combinations $\bar{u} c$ in (\ref{EQN101}), with respect to the $D$ meson ground state ($J^{\mathcal{P}} = 0^-$).

\item All $D_s$ mesons, e.g.\ flavor combinations $\bar{s} c$ in (\ref{EQN101}), with respect to the $D_s$ meson ground state ($J^{\mathcal{P}} = 0^-$).

\item All charmonium states, e.g.\ flavor combinations $\bar{c} c$ in (\ref{EQN101}), with respect to the $\eta_c$ charmonium ground state ($J^{\mathcal{P} \mathcal{C}} = 0^{-+}$).

\item The difference between the $D_s$ meson ground state ($J^{\mathcal{P}} = 0^-$) and the $D$ meson ground state ($J^{\mathcal{P}} = 0^-$).
\end{itemize}

Performing a statistical and systematic error analysis for a mass difference (exactly as detailed in the previous subsection for absolute masses) should give a somewhat reduced error compared to just subtracting the two mean values of the individual meson masses and adding the corresponding errors in quadrature. The reason is that correlations between the two effective masses of the mesons are in that way eliminated.

These mass differences are collected at the end of our paper in Table~\ref{tab:all_diff}. Even though we prefer individual meson masses, such a table of meson mass differences might be particularly helpful to compare with work from other authors focusing exclusively on mass differences and not on individual masses.


\section{Results}
\label{sec:results}
In this section, we discuss our results, in particular the chiral and continuum extrapolations for the meson masses considered in this work.
We begin with some general remarks valid for all states.

For each state, we provide a plot showing all 18 data points corresponding to nine of the ten ensembles listed in Table\ \ref{tab.ensembles} and the two different discretizations of valence quarks, $(+,-)$ and $(+,+)$.
Each such plot shows data from fitting intervals corresponding to the central value according to our systematic procedure (cf. Section~\ref{sec:syst}), e.g.\ the black points and error bars in  Figure\ \ref{fig:syst}.
The straight lines in every plot are fits of equations (\ref{eq:TM}) and (\ref{eq:OS}).
We remark that the 18 meson masses at a fixed lattice spacing and pion mass are plotted only with statistical errors.
Hence, in cases when the systematic error dominates over the statistical one, the $\chi^2/{\rm dof}$ can be significantly larger than 1.
However, when the systematic error is taken into account in the definition of the $\chi^2$ function, we always obtain values of $\chi^2/{\rm dof}$ indicating good fits, i.e.\ $\chi^2/{\rm dof}\approx1$.

The obtained continuum result at the physical pion mass is compared to the experimental value given in the latest PDG review \cite{PDG}, with updates from the online version of PDG, available at \url{http://pdglive.lbl.gov}.

The correlation matrices, in particular their size and operator content, is the same as in Ref.\ \cite{Kalinowski:2015bwa} (see Sections 4 and 5 of this reference for details).

\subsection{$D$ mesons}
We start with mesons containing a charm quark and a light quark, i.e.\ with $D$ mesons.
The plots refer to our computations of neutral $D$ mesons, which differ from their charged counterparts only in the strange and charm quark masses tuning (cf. Sections\ \ref{sec:tuning} and \ref{sec:fits}).
Plots for charged $D$ mesons are qualitatively identical and hence not shown.
The final results for both cases are collected in Table\ \ref{tab:all}.
Note that our estimates of electromagnetic effects and effects from different $u$ and $d$ quark masses are very crude (for a more detailed discussion, cf.\ Section~\ref{sec:tuning} and~\ref{sec:isospin}).

\begin{figure}[t!]
\begin{center}
\includegraphics[width=0.52\textwidth,angle=-90]{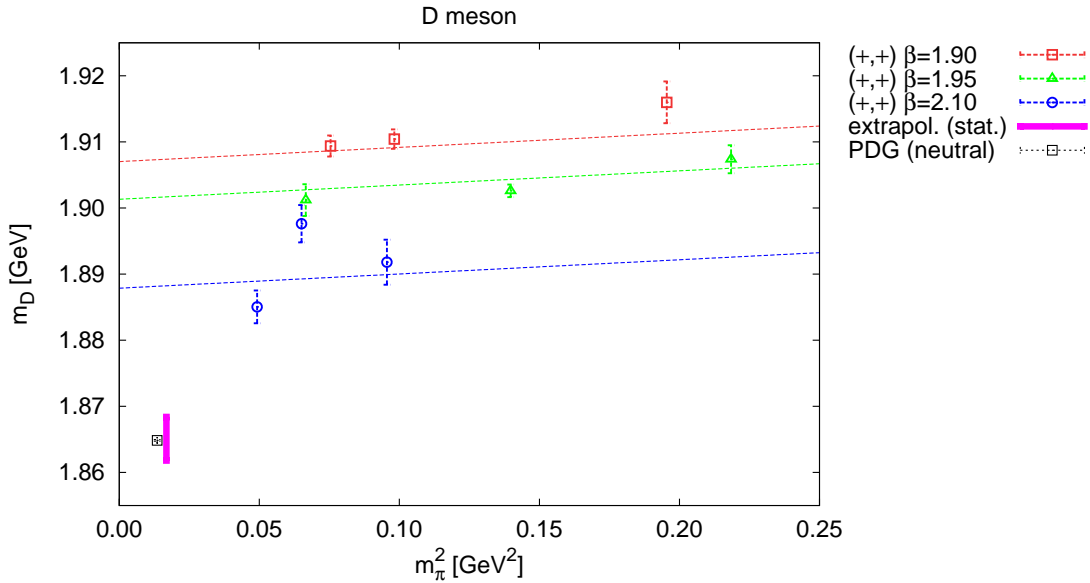}
\caption{\label{fig:D} Chiral and continuum extrapolation for the neutral $D$ meson ($J^{\mathcal{P}}=0^-$). The $(+,-)$ results are used for tuning of the charm quark mass -- hence only $(+,+)$ results are used. PDG value of the mass: 1.86484(5) GeV. Our lattice QCD result: 1.8651(33) GeV (only statistical error, magenta, identical to the total error). }
\end{center}
\end{figure}

\subsubsection{$A_1$ representation (spin $J=0$)}
\label{sec:DA1}
The ground state corresponds to the $D$ meson ($J^{\mathcal{P}}=0^-$) and the first excited state to the $D_0^*$ meson ($J^{\mathcal{P}}=0^+$). 
As mentioned above, these states have to be extracted from a single correlation matrix due to twisted mass parity breaking -- even though these states are in different channels in the continuum (different parity), they belong to the same sector in twisted mass lattice QCD.
The $A_1$ representation corresponds to continuum spins $J=0,4,\ldots$, but the $J=4$ and higher states are expected to be much heavier than the $J=0$ ones.
Hence, it should be appropriate to assume that these low-lying states have $J=0$.

The mass of the $D$ meson computed with the $(+,-)$ discretization is used for the tuning of the charm quark. Hence, for this case we only perform a combined chiral and continuum extrapolation using the $(+,+)$ discretization (see Figure\ \ref{fig:D}).
We find a result compatible with the PDG value, which confirms that the $(+,+)$ discretization indeed yields the same continuum results, as guaranteed by universality (see also Section\ \ref{sec:ansatz} for a more systematic comparison of results obtained from both discretizations independently).
Of course, when using the experimental charged $\pi$, $K$ and $D$ meson masses for the tuning of the strange and charm quark, our $(+,+)$ result is consistent with the PDG mass of the charged $D$ meson, as expected.

\begin{figure}[t!]
\begin{center}
\includegraphics[width=0.52\textwidth,angle=-90]{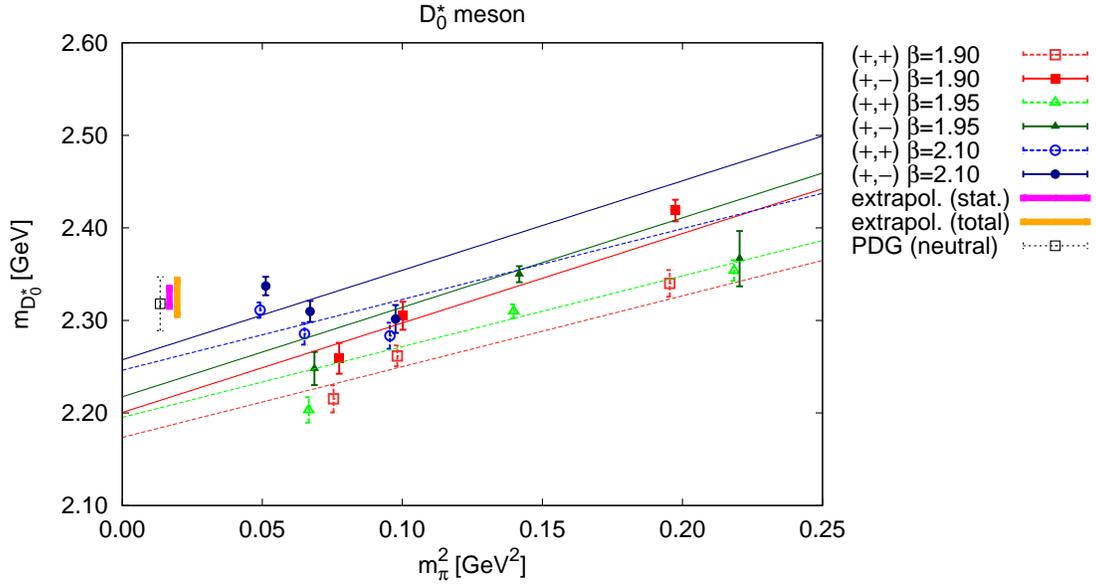}
\caption{\label{fig:D0star} Chiral and continuum extrapolation for the neutral $D_0^*$ meson ($J^{\mathcal{P}}=0^+$). PDG value of the mass: 2.318(29) GeV (neutral). Our lattice QCD result: 2.325(10) GeV (only statistical error, magenta), 2.325(19) GeV (total error, orange).}
\end{center}
\end{figure}

The first excitation (Figure\ \ref{fig:D0star}) in this sector is the $D_0^*$ meson (with $J^{\mathcal{P}}=0^+$).
The eigenvectors resulting from the generalized eigenvalue problem provide information about the structure of the $D_0^*$. They suggest that it is a roughly equal mixture of four $\Gamma$ structures ($\Gamma=\mathbbm{1},\gamma_0,\gamma_j\textbf{n}_j,\gamma_0\gamma_j\textbf{n}_j$), leading to the conclusion that this meson is a roughly equal superposition of $S$ and $P$ waves.
This is interesting, because in quark model calculations this state is often considered to be a pure $P$ wave state.
Our continuum value of the $D_0^*$ meson mass is compatible with the experimental value.
However, the behavior with changing light quark mass is somewhat untypical -- the meson mass increases significantly with increasing light quark mass for A and B ensembles, whereas the D ensembles, where the pion masses are smaller, indicate an opposite trend for pion masses below $250 \, \textrm{MeV}$. Indeed, excluding the D ensembles from the extrapolation fits, leads to a somewhat smaller continuum value, however still compatible with the experimental result.
This behavior might be related to the fact that this state can decay to a $D$ meson and a pion and has a rather large width of $\Gamma=267(40)$ MeV.
In other words, the state that we extract may contain a two-meson contribution with the same quantum numbers, $D+\pi$, where the presence of the pion could be the reason for the dependence on the light quark mass.
Therefore, this state may require a more sophisticated lattice treatment, taking the possibility of such a decay explicitly into account, as explored e.g.\ in Ref.\ \cite{Mohler:2012na}.

\begin{figure}[t!]
\begin{center}
\includegraphics[width=0.52\textwidth,angle=-90]{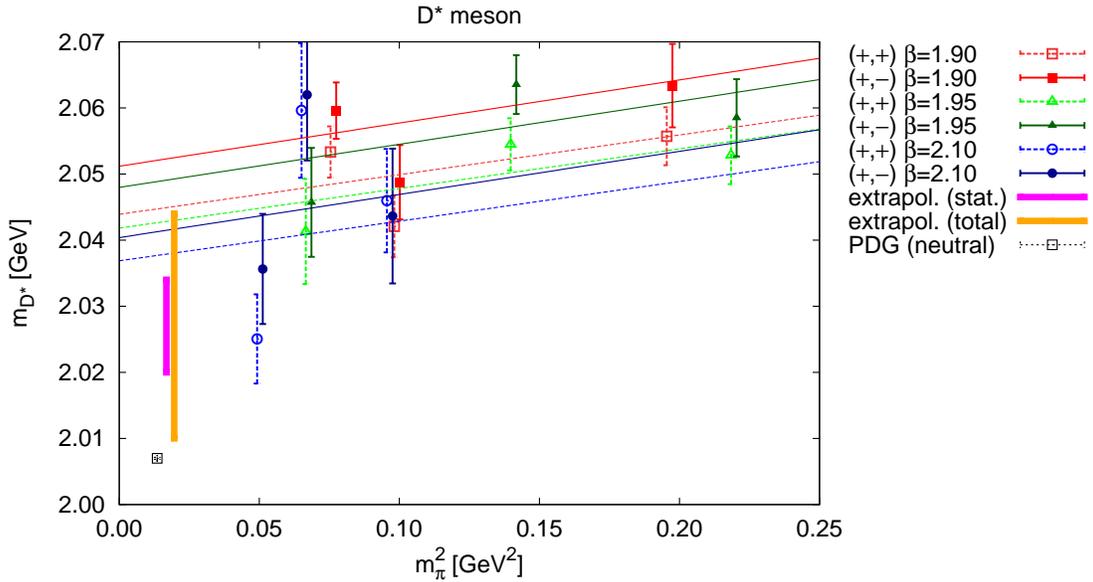}
\caption{\label{fig:Dstar} Chiral and continuum extrapolation for the neutral $D^*$ meson ($J^{\mathcal{P}}=1^-$). PDG value of the mass: 2.00697(8) GeV (neutral). Our lattice QCD result: 2.027(7) GeV (only statistical error, magenta), 2.027(17) GeV (total error, orange).}
\end{center}
\end{figure}
\subsubsection{$T_1$ representation (spin $J=1$)}
\label{sec:DT1}
The $T_1$ representation corresponds to continuum spins $J=1,3,4,\ldots$ and again it seems reasonable to assume that the low-lying states that we extract are all $J=1$ states. Note that the $J=3$ states appear also in the $T_2$ representation and finding a state at a similar mass in both $T_1$ and $T_2$ is hence an indication that it is $J=3$. On the other hand, $T_1$ representation states with no counterpart in the $T_2$ representation are most likely $J=1$.

The lowest $T_1$ charm-light state is the $D^*$ meson ($J^{\mathcal{P}}=1^-$), see Figure\ \ref{fig:Dstar}.
The $D^*$ meson cannot decay to $D+\pi$ at our values of the pion mass, hence our computation should not be affected by two-meson contributions with the same quantum numbers.
The value we get is slightly above the PDG result when only the statistical error is taken into account.
However, after considering the fitting range uncertainty, we get agreement with the experiment within the total error (represented by orange boxes in our plots).

\begin{figure}[t!]
\begin{center}
\includegraphics[width=0.52\textwidth,angle=-90]{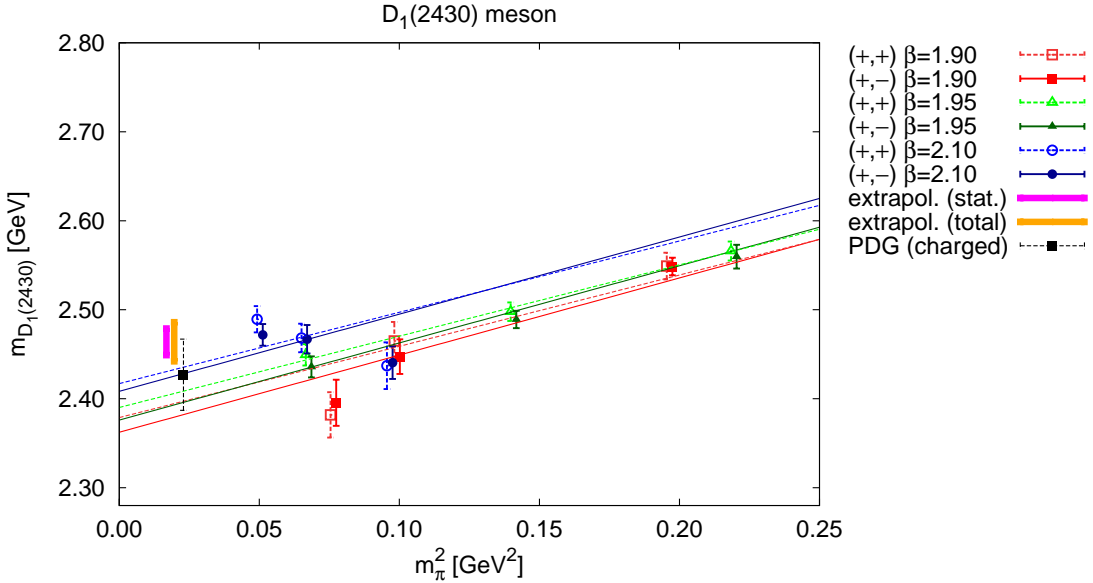}
\caption{\label{fig:D1} Chiral and continuum extrapolation for the neutral $D_1(2430)$ meson ($J^{\mathcal{P}}=1^+$, $j=1/2$). PDG value of the mass: 2.427(40) GeV (charged, no neutral experimental result available). Our lattice QCD result: 2.464(15) GeV (only statistical error, magenta), 2.464(22) GeV (total error, orange).}
\end{center}
\end{figure}

\begin{figure}[t!]
\begin{center}
\includegraphics[width=0.52\textwidth,angle=-90]{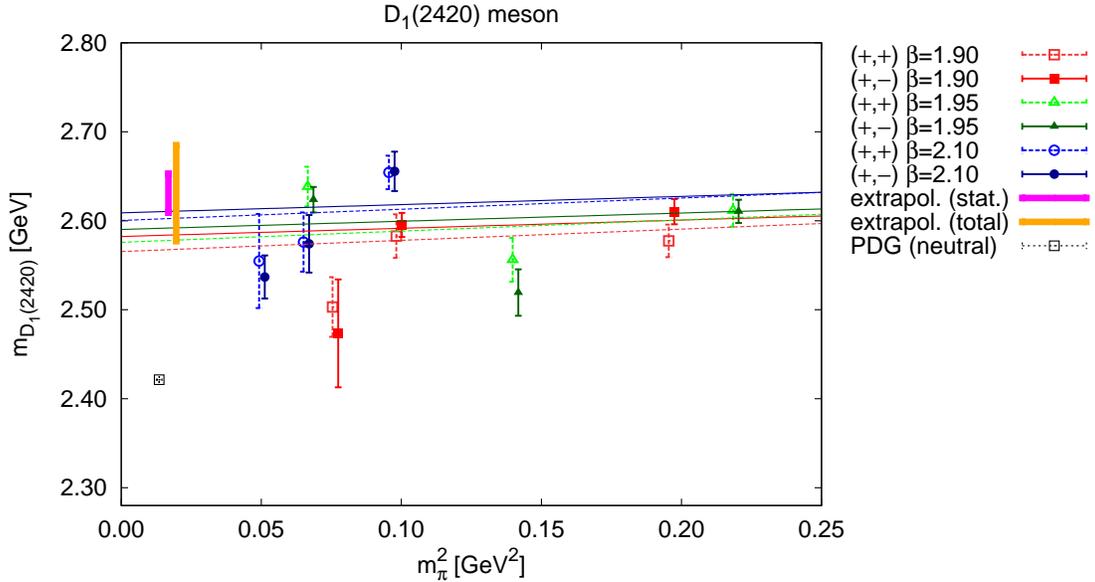}
\caption{\label{fig:D1a} Chiral and continuum extrapolation for the neutral $D_1(2420)$ meson ($J^{\mathcal{P}}=1^+$, $j=3/2$). PDG value of the mass: 2.4214(6) GeV (neutral). Our lattice QCD result: 2.631(22) GeV (only statistical error, magenta), 2.631(54) GeV (total error, orange).}
\end{center}
\end{figure}

For the two lowest excitations in this channel, both with $J^{\mathcal{P}}=1^+$, we again use the eigenvector components obtained from the generalized eigenvalue problem to resolve the structure. 
This is particularly important for these states, since their experimental masses are very close and hence the assignment of lattice results to the two states characterized by the spin of the light degrees of freedom $j\approx1/2$ and $j\approx3/2$ is not straightforward.
Nevertheless, it can be done in an unambiguous way, which we described in detail in Ref.\ \cite{Kalinowski:2015bwa}, Section\ 5.1.2.
According to the arguments presented there, the $D_1(2430)$ has $j\approx1/2$, while the $D_1(2420)$ has $j\approx3/2$.
This implies that only the former can decay via an $S$ wave (to $D^*+\pi$) and our lattice treatment of this state in terms of quark-antiquark operators may again be too simple.
On the other hand, the latter is protected by angular momentum $j\approx3/2$ and hence this decay is strongly suppressed.
Our combined chiral and continuum extrapolation is, however, in good agreement with the experimental value for $D_1(2430)$ (Figure\ \ref{fig:D1}), while for $D_1(2420)$ (Figure\ \ref{fig:D1a}), we obtain a result which is almost $4\,\sigma$ away, even after considering the uncertainty from the fitting range.
It is interesting to note that $j\approx3/2$ static-light mesons came out too heavy using a similar lattice QCD setup, with a similar discrepancy with respect to the experimental result \cite{Michael:2010aa}.
Hence, it is not too surprising that also in the present computation, we encounter the same problem.
A possible explanation of this fact is that the creation operator used to extract the meson might not generate a trial state similar enough to the $D_1(2420)$.
In practice, this would lead to a contamination of the considered correlation function with higher excited states.
Thus, one would need to go to later Euclidean times to disentangle the $D_1(2420)$ from the next excited state.
Indeed, this explanation is favored by our present data, since the meson mass decreases when it is extracted from plateau fitting intervals starting at higher $t/a$ (the plot is qualitatively similar to the lower right plot of Figure\ \ref{fig:syst}).
However, before a definite conclusion can be reached, the signal-to-noise ratio drops too much, i.e.\ it is not excluded that fits starting at even higher $t/a$ would give even smaller values of the mass, thus lowering our preferred central value and increasing the fitting range uncertainty.
Another hint for this interpretation is provided by the results from our finest lattice spacing (D ensembles, blue points in Figure\ \ref{fig:D1a}).
For these ensembles, the temporal extent of the lattice is $T/a=48$ (as compared to $T/a=32$ or $T/a=24$ for the other ensembles) and hence larger Euclidean times in lattice units can be reached.
Consequently, if the chiral extrapolation is performed using these ensembles alone, the obtained result is much closer to the experimental value and at this lattice spacing the cut-off effects should be rather small.
The situation is similar for some of our higher orbital or radial excitations, but mostly with smaller discrepancies to experimental results of around $2\,\sigma$. In such cases, more precise data at larger temporal separations or better optimized creation operators might be needed to clarify the situation and to obtain more robust results.

\begin{figure}[t!]
\begin{center}
\includegraphics[width=0.52\textwidth,angle=-90]{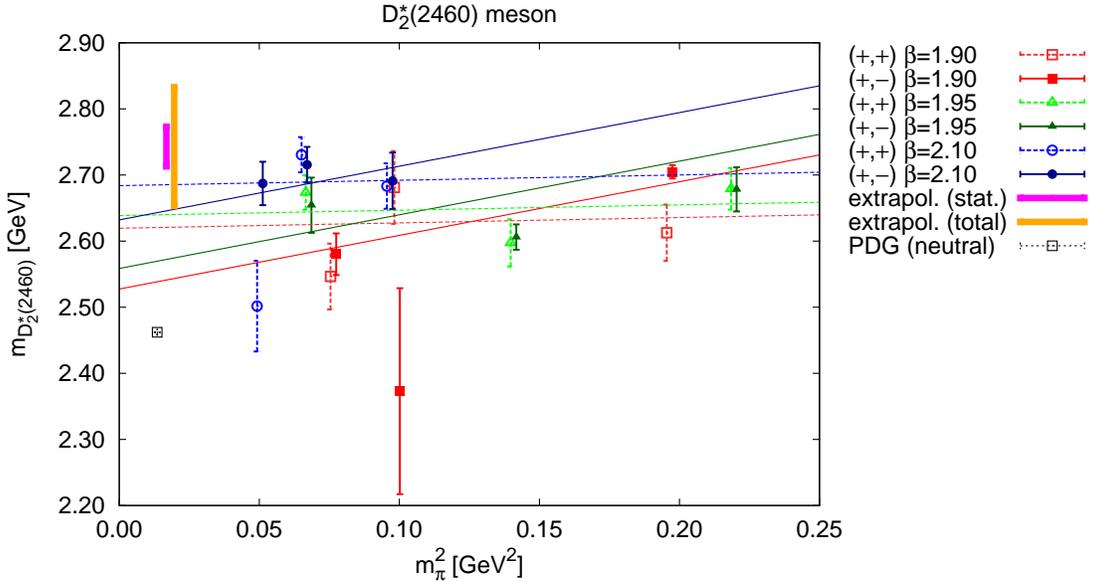}
\caption{\label{fig:D2} Chiral and continuum extrapolation for the neutral $D_2^*(2460)$ meson ($J^{\mathcal{P}}=2^+$). PDG value of the mass: 2.4626(6) GeV (neutral). Our lattice QCD result ($E$ representation): 2.743(30) GeV (only statistical error, magenta), 2.743(90) GeV (total error, orange).}
\end{center}
\end{figure}

\subsubsection{$T_2$ and $E$ representations (spin $J=2$)}
\label{sec:DT2E}
The $T_2$ representation contains continuum spins $J=2,3,4,\ldots$, while the $E$ representation contains $J=2,4,\ldots$.
Thus, if a state with similar mass is present in both $T_2$ and $E$, it should corresponds to spin $J=2$ or $J=4$ (although the latter is most likely excluded for low lying states).

We extracted three states in the $J=2$ channel (i.e.\ from both representations $T_2$ and $E$), the $D_2^*(2460)$ with $J^{\mathcal{P}}=2^+$ (Figure\ \ref{fig:D2} for the $E$ representation) and two other states of opposite parity, with unknown experimental counterparts. 

For the former, the statistical error is quite large and the plateau quality is not very good, which is reflected in the size of the systematic uncertainty related to the choice of the fitting interval.
Moreover, our systematic procedure of extracting the fitting range uncertainty again points to a possible contamination by excited states, since the results for increasing shift parameter $n$ (again, the plot for different fitting ranges resembles the lower right plot of Figure\ \ref{fig:syst}) go down systematically.
However, again the signal-to-noise ratio decreases too quickly to reach a definite conclusion.
Moreover, the mass value from chiral fits using only our D ensembles is much closer to the experimental value than the one from the chiral and continuum extrapolation taking all ensembles into account.
Hence, we interpret the observed discrepancy as a sign that this state is among those for which the lattice techniques need to be improved for a reliable extraction, e.g.\ by using operators with better overlap with the desired state.

The other two extracted states in this channel ($J^{\mathcal{P}}=2^-$) suffer even more from the aforementioned problem.
Therefore, we do not quote any value for them.
We only remark that the chiral extrapolation from only the D ensembles leads to a value consistent with the experimental one for the $D(2750)$ meson.
The presence of another state with these quantum numbers points to the possible existence of another, rather closely lying state, which can be expected to be discovered experimentally (two close lying states with $J^{\mathcal{P}}=2^-$ were also observed in Ref.\ \cite{Mohler:2011ke}).

\begin{figure}[t!]
\begin{center}
\includegraphics[width=0.52\textwidth,angle=-90]{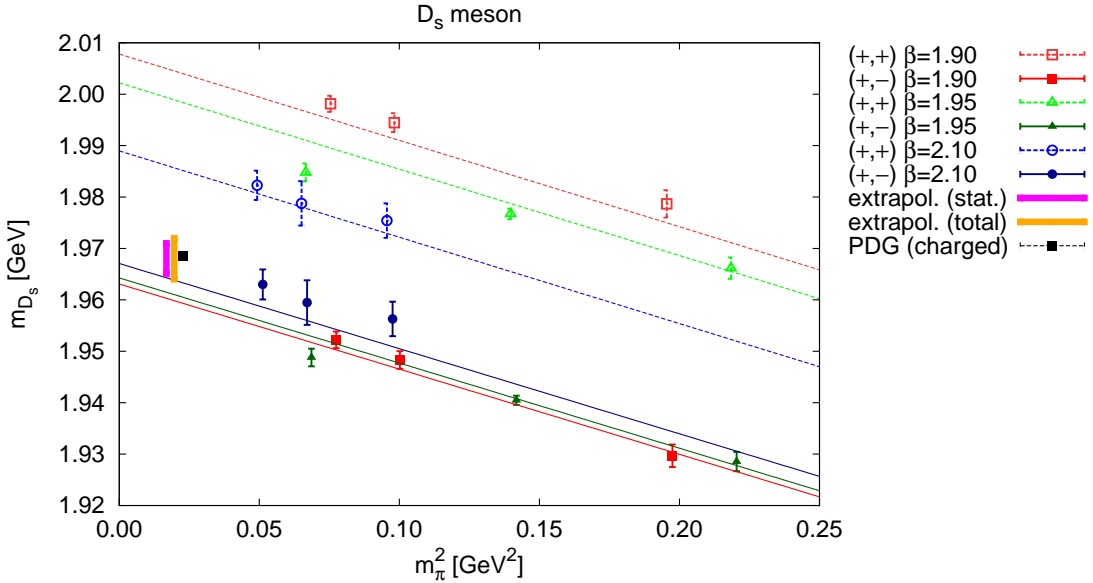}
\caption{\label{fig:Ds} Chiral and continuum extrapolation for the $D_s$ meson ($J^{\mathcal{P}}=0^-$). PDG value of the mass: 1.96830(10) GeV. Our lattice QCD result: 1.9679(27) GeV (only statistical error, magenta), 1.9679(40) GeV (total error, orange).}
\end{center}
\end{figure}
\subsection{$D_s$ mesons}
\subsubsection{$A_1$ representation (spin $J=0$)}
We treat the two lowest lying charm-strange mesons in the $A_1$ representation very similarly to the charm-light ones.
The ground state $D_s$ meson ($J^{\mathcal{P}}=0^-$) can be extracted from lattice results rather precisely (Figure\ \ref{fig:Ds}), i.e.\ with a relative error smaller than two per mille, which is the best precision from among all our states.
We find agreement with experiment and the effective mass plateau quality is very good such that the uncertainty from the choice of the fitting range is slightly smaller than the statistical error (see the upper right plot of Figure\ \ref{fig:syst}).
The precision of the data allows one also to obtain the slopes of the chiral extrapolation ($\alpha^{(+,-)}$ and $\alpha^{(+,+)}$) with small statistical errors and we find that they are compatible: $-0.166(12)$ and $-0.168(13)$, respectively. 
This confirms the expectation that discretization effects in the $\alpha$ parameters are quite small, i.e.\ that higher order terms proportional to $a^2(m_\pi^2 - m_{\pi,\textrm{exp}}^2)$ are negligible in equations (\ref{eq:TM}) and (\ref{eq:OS}).
We also note that the discretization effects in the $(+,-)$ setup are much smaller ($c^{(+,-)}=-1.0(4)$) than those in the $(+,+)$ setup ($c^{(+,+)}=+4.7(4)$).
This again agrees with our expectations -- as mentioned above, the $(+,-)$ setup is known to usually give smaller cut-off effects in comparison with the $(+,+)$ valence quark discretization for light-light and heavy-light pseudoscalar mesons.
However, this is not the case for all investigated mesons and we will comment more about this when discussing other states and in Section\ \ref{sec:ansatz}.

\begin{figure}[t!]
\begin{center}
\includegraphics[width=0.52\textwidth,angle=-90]{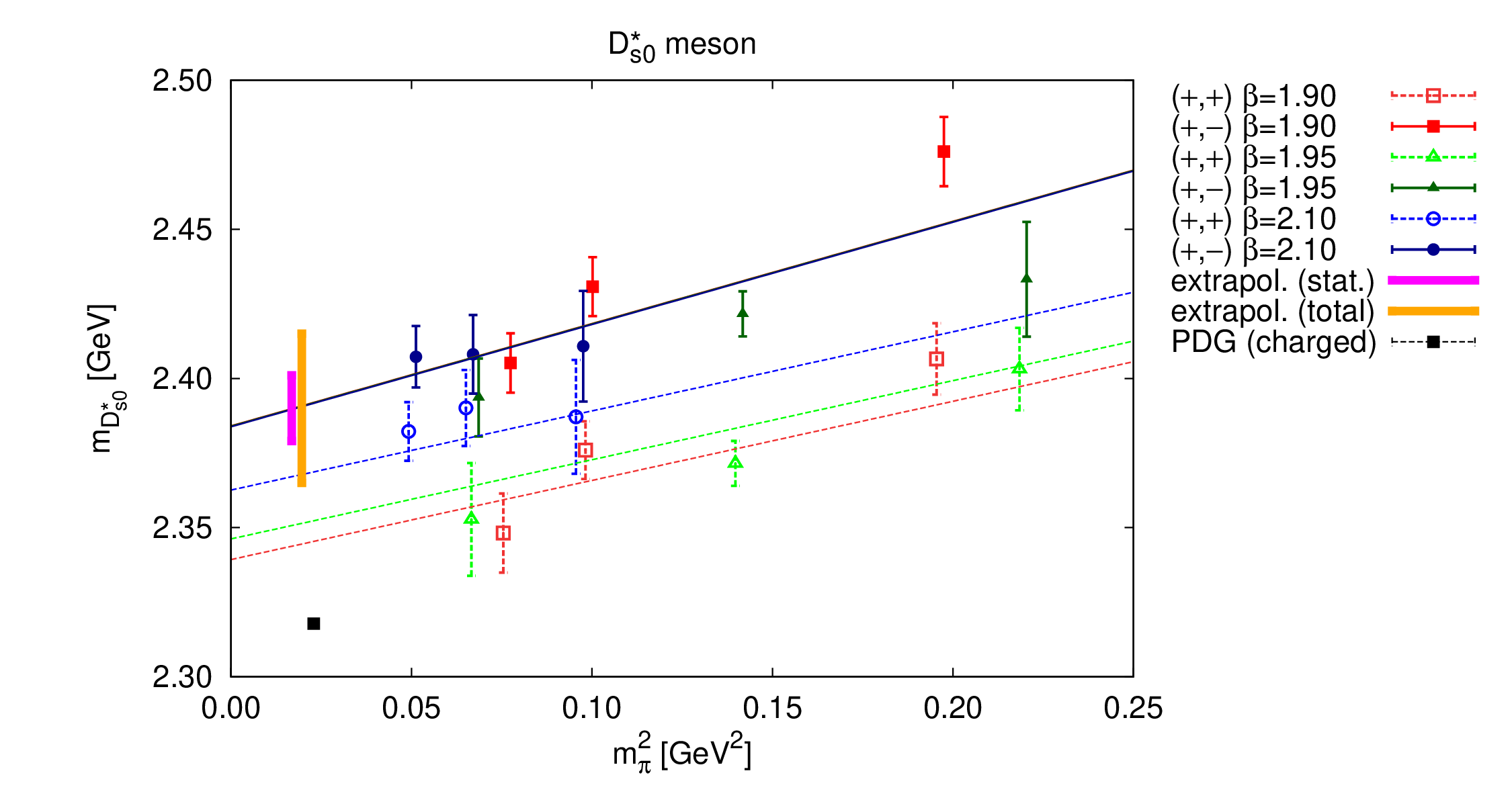}
\caption{\label{fig:Ds0star} Chiral and continuum extrapolation for the $D_{s0}^*$ meson ($J^{\mathcal{P}}=0^+$). PDG value of the mass: 2.3177(6) GeV. Our lattice QCD result: 2.390(11) GeV (only statistical error, magenta), 2.390(25) GeV (total error, orange).}
\end{center}
\end{figure}

The first excited state in this channel is the $D_{s0}^*$ meson ($J^{\mathcal{P}}=0^+$), where we arrive at very similar conclusions regarding its structure as for the $D_0^*$ meson, i.e.\ that it is a roughly equal superposition of quarks in an $S$ wave and a $P$ wave (cf.\ the discussion in Section\ \ref{sec:DA1}).
We obtain rather good statistical precision. The fitting range uncertainty is twice larger, but still we observe an almost $3\,\sigma$ discrepancy with respect to experiment.
Note that this meson is frequently considered to be a tetraquark candidate and many previous studies based on quark models (e.g.\ \cite{Ebert:2009ua}) and lattice QCD (e.g.\ \cite{Mohler:2011ke}) found its mass to be in contradiction with experiment when assuming the standard quark-antiquark structure. The latter was also used in our setup and hence we are not in a position to present unambiguos evidence for or against a tetraquark or some other exotic structure, but our result is indicative of some non-standard behavior.

\begin{figure}[t!]
\begin{center}
\includegraphics[width=0.52\textwidth,angle=-90]{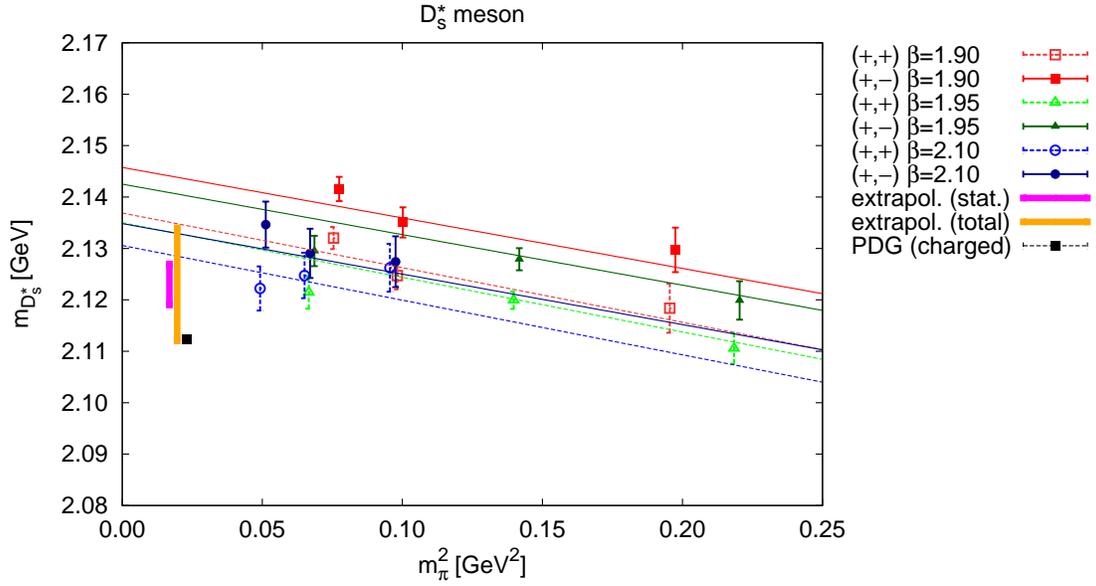}
\caption{\label{fig:Dsstar} Chiral and continuum extrapolation for the $D_{s}^*$ meson ($J^{\mathcal{P}}=1^-$). PDG value of the mass: 2.1121(4) GeV. Our lattice QCD result: 2.1226(38) GeV (only statistical error, magenta), 2.123(11) GeV (total error, orange).}
\end{center}
\end{figure}

\subsubsection{$T_1$ representation (spin $J=1$)}
The ground state spin-1 $D_s$ meson is the $D_s^*$ ($J^{\mathcal{P}}=1^-$). Our result (see Figure\ \ref{fig:Dsstar}) is consistent with experiment when both the statistical and the systematic uncertainties are taken into account.

\begin{figure}[t!]
\begin{center}
\includegraphics[width=0.52\textwidth,angle=-90]{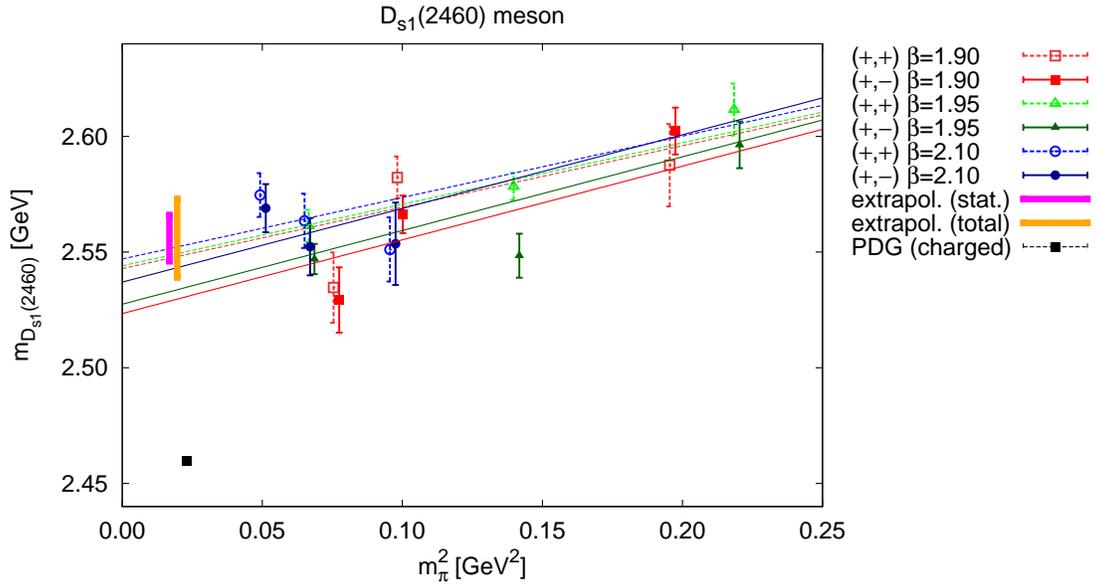}
\caption{\label{fig:Ds1} Chiral and continuum extrapolation for the $D_{s1}(2460)$ meson ($J^{\mathcal{P}}=1^+$, $j=1/2$). PDG value of the mass: 2.4595(6) GeV. Our lattice QCD result: 2.556(10) GeV (only statistical error, magenta), 2.556(17) GeV (total error, orange).}
\end{center}
\end{figure}
\begin{figure}[t!]
\begin{center}
\includegraphics[width=0.52\textwidth,angle=-90]{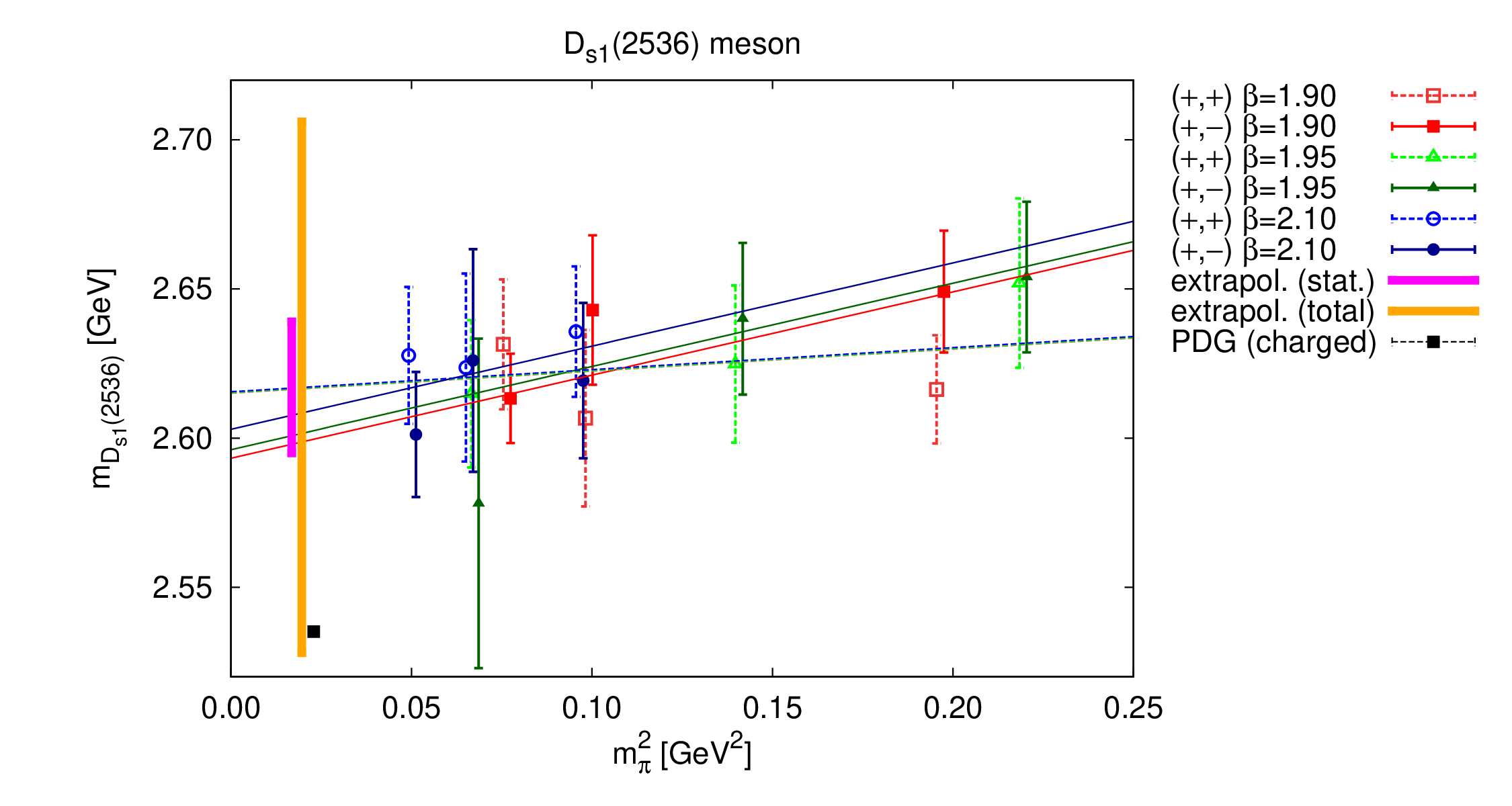}
\caption{\label{fig:Ds1a} Chiral and continuum extrapolation for the $D_{s1}(2536)$ meson ($J^{\mathcal{P}}=1^+$, $j=3/2$). PDG value of the mass: 2.53511(6) GeV. Our lattice QCD result: 2.617(22) GeV (only statistical error, magenta), 2.617(89) GeV (total error, orange).}
\end{center}
\end{figure}

We also extract the two lowest excitations in this channel -- the $J^{\mathcal{P}}=1^+$ states with $j\approx1/2$ ($D_{s1}(2460)$, Figure\ \ref{fig:Ds1}) and $j\approx3/2$ ($D_{s1}(2536)$, Figure\ \ref{fig:Ds1a}), where $j$ again denotes the spin of light degrees of freedom. In both cases, our result is above the PDG value.
For $D_{s1}(2460)$, the discrepancy remains at the level of $8\,\sigma$ even with systematic uncertainties taken into account.
As for the $D_{s0}^*$, this might indicate that this state is not a standard meson, but rather has some exotic structure.
Similar conclusions have been obtained using quark models (e.g.\ \cite{Ebert:2009ua}), which is why this meson is often considered as a tetraquark candidate.
The $D_{s1}(2536)$ is the charm-strange analogue of $D_1(2420)$.
Having $j\approx3/2$, it is expected to share the same problems as $D_1(2420)$.
Indeed, the effective mass plateau quality is rather low and we obtain a mass which is slightly too large with respect to experiment.
However, for this case, our systematic procedure points to a large fitting range uncertainty of around 85 MeV.
Within this uncertainty, the mass agrees with experiment.
As for the corresponding $D_1(2420)$, an optimization of the creation operator might help to obtain a more precise result.

\begin{figure}[t!]
\begin{center}
\includegraphics[width=0.52\textwidth,angle=-90]{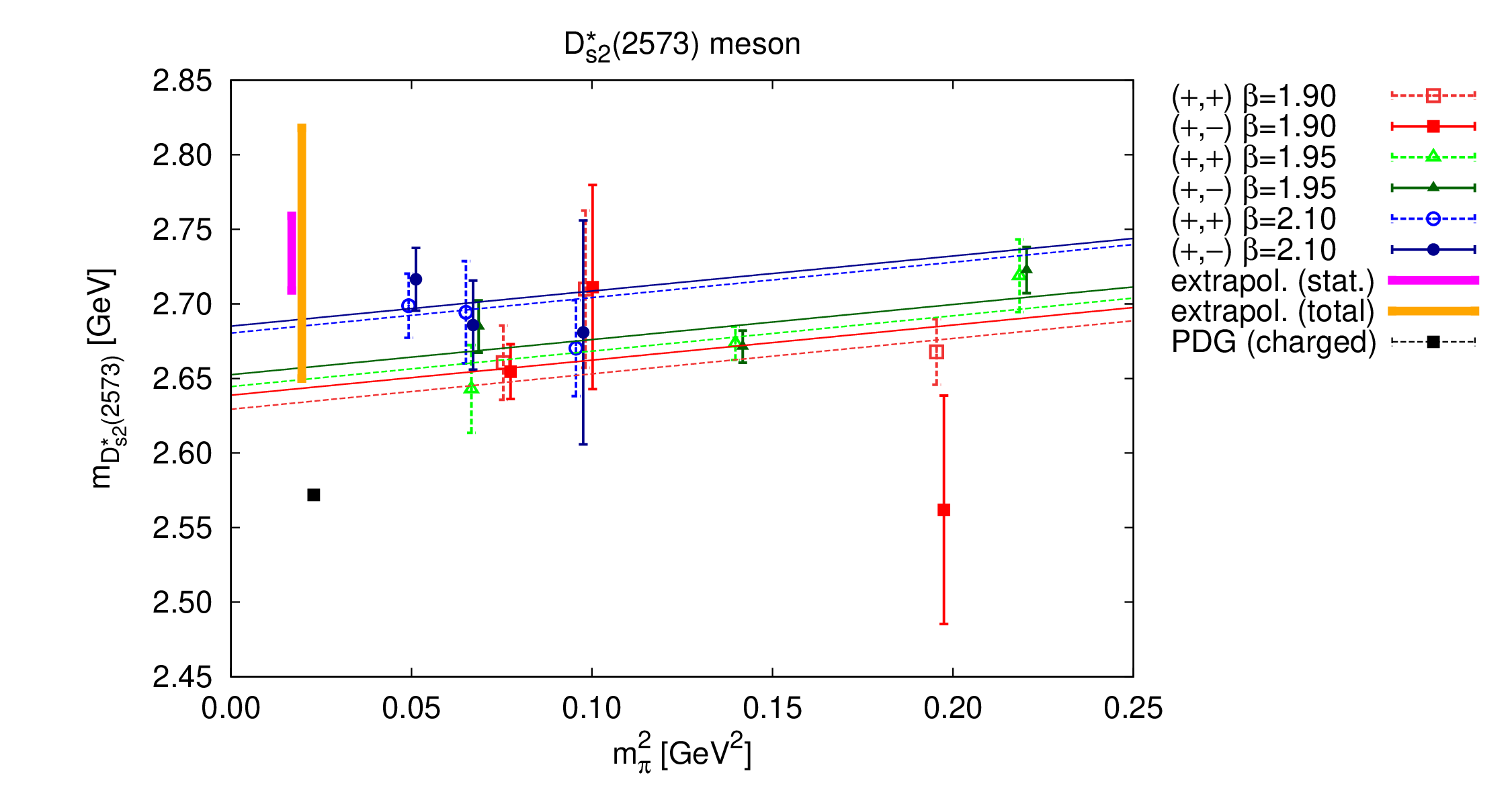}
\caption{\label{fig:Ds2} Chiral and continuum extrapolation for the $D_{s2}^*(2573)$ meson ($J^{\mathcal{P}}=2^+$). PDG value of the mass: 2.5719(8) GeV. Our lattice QCD result ($E$ representation): 2.734(25) GeV (only statistical error, magenta), 2.734(84) GeV (total error, orange).}
\end{center}
\end{figure}

\subsubsection{$T_2$ and $E$ representations (spin $J=2$)}
We could extract one state in the spin-2 channel (Figure\ \ref{fig:Ds2}), using both the $E$ and the $T_2$ representations -- the $D_{s2}^*(2573)$ meson ($J^{\mathcal{P}}=2^+$).
As for the analogous charm-light state ($D_2^*(2460)$), we observe a rather poor effective mass plateau quality, with no obvious choice of the fitting interval.
Our systematic procedure to capture such effects hence generates a large uncertainty (70..80 MeV, see the lower right plot of Figure\ \ref{fig:syst} for the $E$ representation).
Within the combined statistical and systematic uncertainty, our result is consistent with the experimental value.

We also observe two states with $J^{\mathcal{P}}=2^-$, but the quality of the data is not good enough to quote a meaningful and trustworthy value (cf.\ also Section\ \ref{sec:DT2E} for a discussion of the corresponding charm-light states).

\begin{figure}[t!]
\begin{center}
\includegraphics[width=0.52\textwidth,angle=-90]{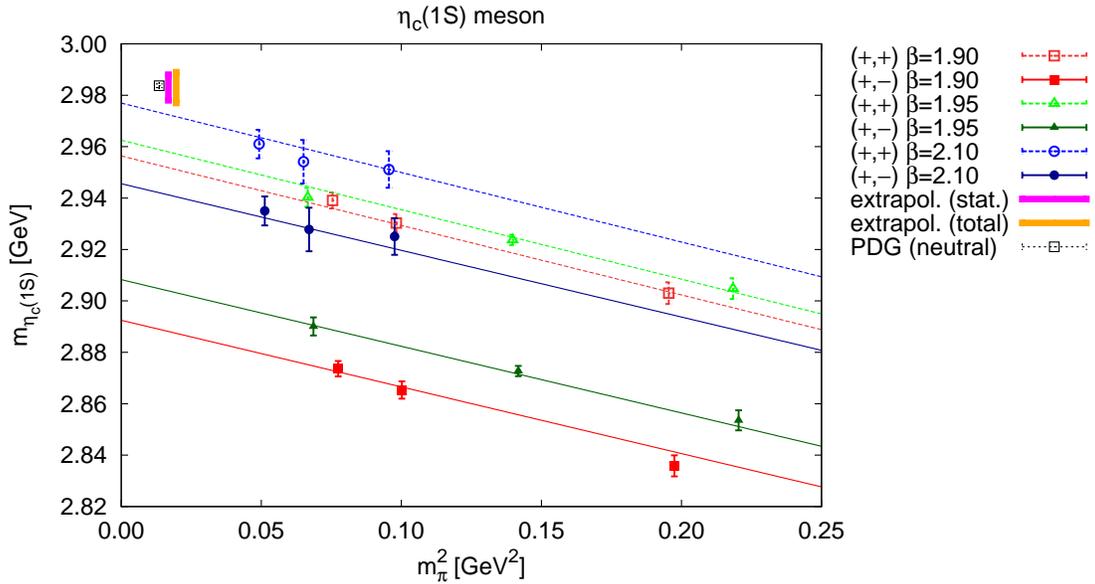}
\caption{\label{fig:eta_c} Chiral and continuum extrapolation for the $\eta_c(1S)$ meson ($J^{\mathcal{PC}}=0^{-+}$). PDG value of the mass: 2.9836(6) GeV. Our lattice QCD result: 2.983(5) GeV (only statistical error, magenta), 2.983(6) GeV (total error, orange).}
\end{center}
\end{figure}
\subsection{Charmonium}
As in Ref.\ \cite{Kalinowski:2015bwa}, we neglect disconnected diagrams in the computation of charmonium states.
Although this introduces a systematic error, it is expected to be tiny -- much smaller than statistical errors and systematic errors from other sources.
The effects of disconnected diagrams were estimated in quenched lattice computations \cite{Levkova:2010ft} and also perturbatively \cite{Davies:2010ip,Gregory:2010gm,Donald:2012ga}, both estimates yielding shifts in the range of 1 MeV to 4 MeV.

\subsubsection{$A_1$ representation (spin $J=0$)}
The ground state charmonium is the $\eta_c(1S)$ meson ($J^{\mathcal{PC}}=0^{-+}$), see Figure\ \ref{fig:eta_c}. It can be extracted with excellent precision and the plateau quality is very good, such that the uncertainty from the choice of the fitting interval is negligible compared to the statistical error (see the upper left plot of Figure\ \ref{fig:syst}).
The slope of the chiral extrapolation has been determined rather precisely (with approximately 10\% precision) and again we find that the $(+,-)$ and $(+,+)$ discretizations have within errors the same pion mass dependence: $\alpha^{(+,-)}=-0.259(23)$ and $\alpha^{(+,+)}=-0.270(25)$.
In the end, we obtain excellent agreement of our continuum extrapolated lattice result with experiment.
It is interesting to note that we observe much smaller discretization effects in the $(+,+)$ setup than in the $(+,-)$ setup ($c^{(+,-)}=-13.1(9)$, $c^{(+,+)}=-5.0(9)$).
Hence, the often quoted rule that the $(+,-)$ setup leads to smaller cut-off effects (as is indeed the case e.g.\ for the mesons used in the present work for the tuning of strange and charm quark masses, i.e.\ the $\pi$, $K$ and $D$ mesons) is not universal, as will be discussed in more detail in Section\ \ref{sec:ansatz}.

\begin{figure}[t!]
\begin{center}
\includegraphics[width=0.52\textwidth,angle=-90]{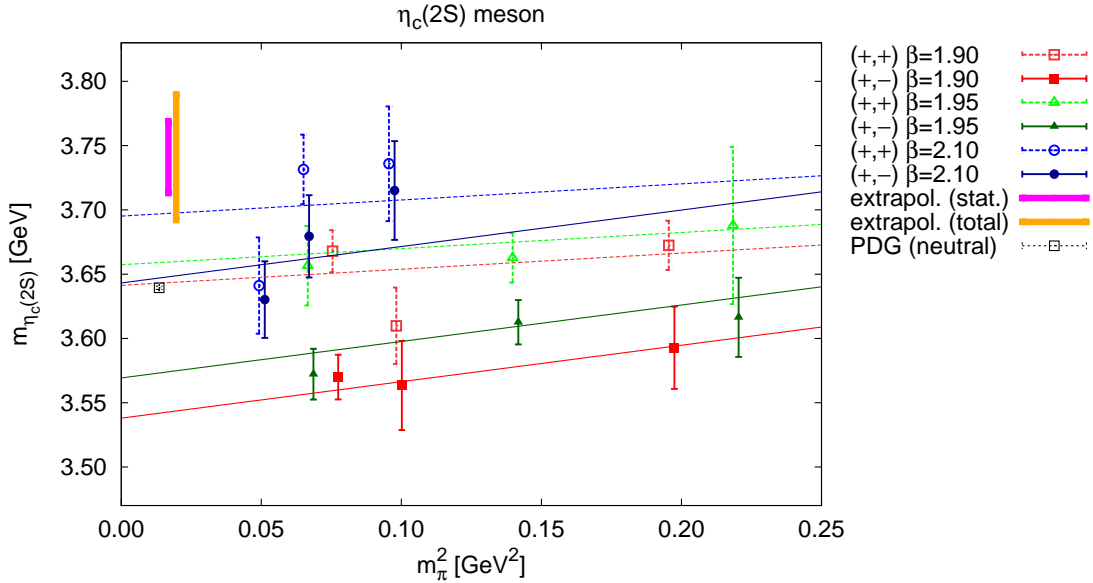}
\caption{\label{fig:eta_c2} Chiral and continuum extrapolation for the $\eta_c(2S)$ meson ($J^{\mathcal{PC}}=0^{++}$). PDG value of the mass: 3.6392(12) GeV. Our lattice QCD result: 3.741(28) GeV (only statistical error, magenta), 3.741(49) GeV (total error, orange).}
\end{center}
\end{figure}

The first excitation in the $J^{\mathcal{PC}}=0^{-+}$ channel is the $\eta_c(2S)$ meson (Figure\ \ref{fig:eta_c2}).
After combining our statistical and systematic uncertainties, we obtain a result which is slightly above the PDG value, by around $2\,\sigma$.
Our systematic procedure again reveals that the meson mass is decreasing with increasing shift parameter $n$, see the lower left plot of Figure\ \ref{fig:syst}.
To check the robustness of this result, one would need to go to even higher shifts (i.e.\ fitting ranges starting and ending at later Euclidean times) and for this, the data quality is not sufficient.
As before in cases of similar plateau quality, the extrapolation from the D ensembles alone, where the fitting range uncertainties are expected to be smaller (with small discretization effects in addition), leads to a value fully consistent with the PDG value.

\begin{figure}[t!]
\begin{center}
\includegraphics[width=0.52\textwidth,angle=-90]{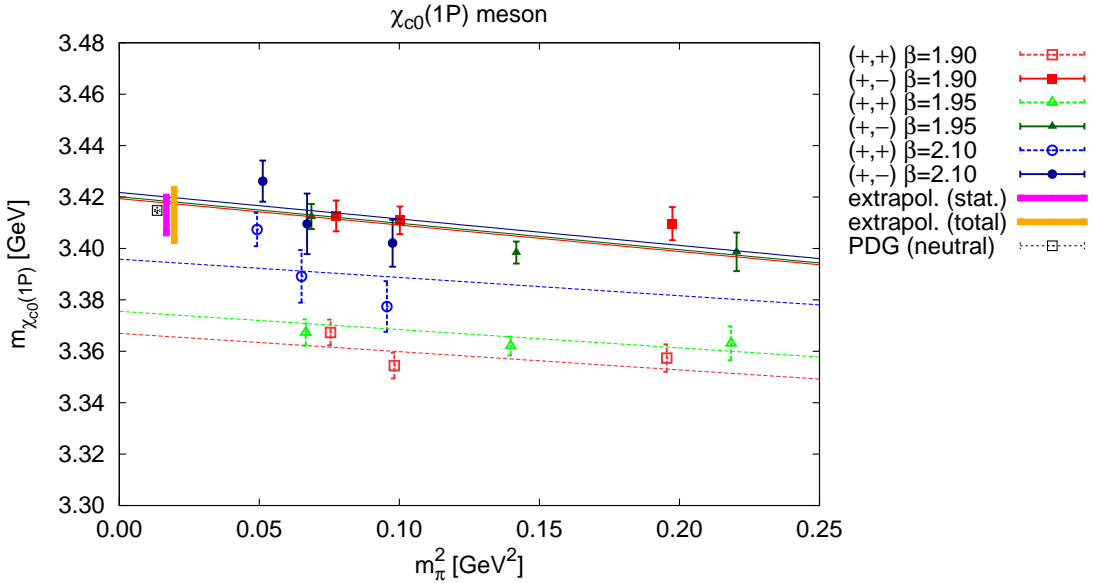}
\caption{\label{fig:chi_c0} Chiral and continuum extrapolation for the $\chi_{c0}(1P)$ meson ($J^{\mathcal{PC}}=0^{++}$). PDG value of the mass: 3.41475(31) GeV. Our lattice QCD result: 3.413(7) GeV (only statistical error, magenta), 3.413(10) GeV (total error, orange).}
\end{center}
\end{figure}

\begin{figure}[t!]
\begin{center}
\includegraphics[width=0.52\textwidth,angle=-90]{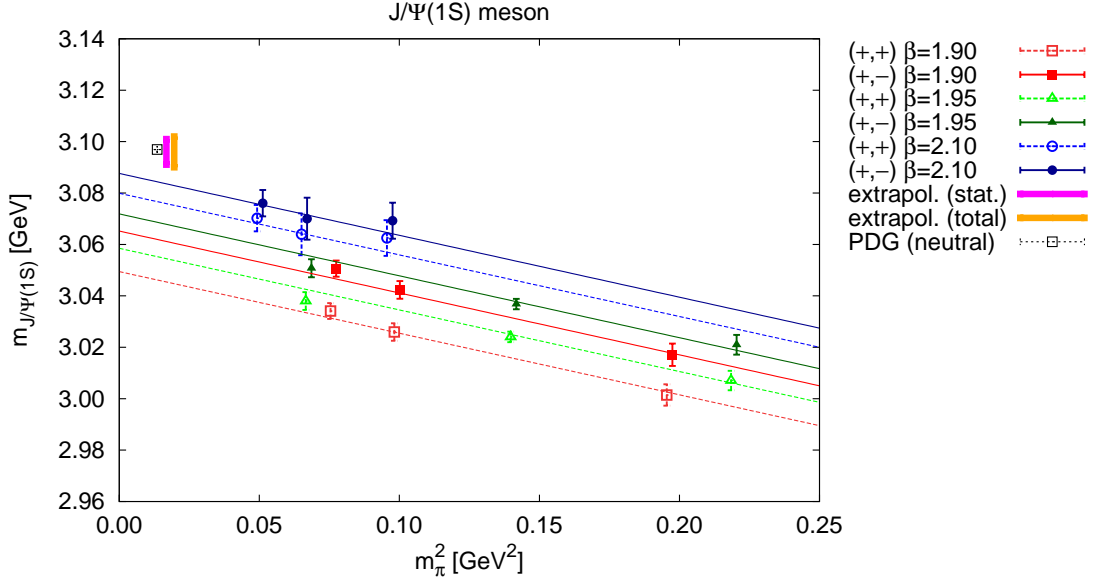}
\caption{\label{fig:J-psi} Chiral and continuum extrapolation for the $J/\psi(1S)$ meson ($J^{\mathcal{PC}}=1^{--}$). PDG value of the mass: 3.096916(11) GeV. Our lattice QCD result: 3.096(5) GeV (only statistical error, magenta), 3.096(6) GeV (total error, orange).}
\end{center}
\end{figure}

We also observe a crude signal of the next excitation, the $\eta_c(3S)$ meson, which is not included in PDG, but has been previously observed in lattice QCD computations \cite{Liu:2012ze,Mohler:2012na} at unphysically heavy pion masses.
The signal-to-noise ratio of our data and hence the plateau quality in our computation are, however, not good enough for solid quantitative statements.

The chiral and continuum extrapolation for the positive parity ground state meson (which is due to twisted mass parity breaking in the same twisted mass sector as the $\eta_c$ mesons), the $\chi_{c0}(1P)$ meson ($J^{\mathcal{PC}}=0^{++}$), is shown in Figure\ \ref{fig:chi_c0}. We observe very small discretization effects in the $(+,-)$ data and much larger effects for $(+,+)$, just opposite as for the $\eta_c(1S)$ meson. In both cases, the pion mass dependence is very mild, with the slope almost compatible with zero. The agreement with the PDG value is excellent.

\begin{figure}[t!]
\begin{center}
\includegraphics[width=0.52\textwidth,angle=-90]{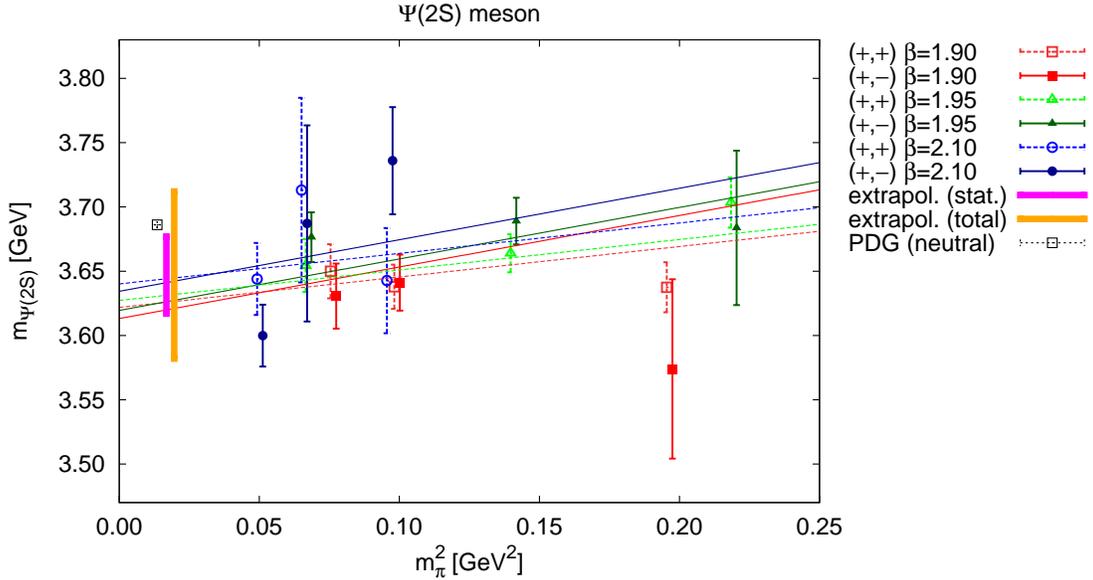}
\caption{\label{fig:psi} Chiral and continuum extrapolation for the $\psi(2S)$ meson ($J^{\mathcal{PC}}=1^{--}$). PDG value of the mass: 3.686109(14) GeV. Our lattice QCD result: 3.647(30) GeV (only statistical error, magenta), 3.647(65) GeV (total error, orange).}
\end{center}
\end{figure}

\begin{figure}[h!]
\begin{center}
\includegraphics[width=0.52\textwidth,angle=-90]{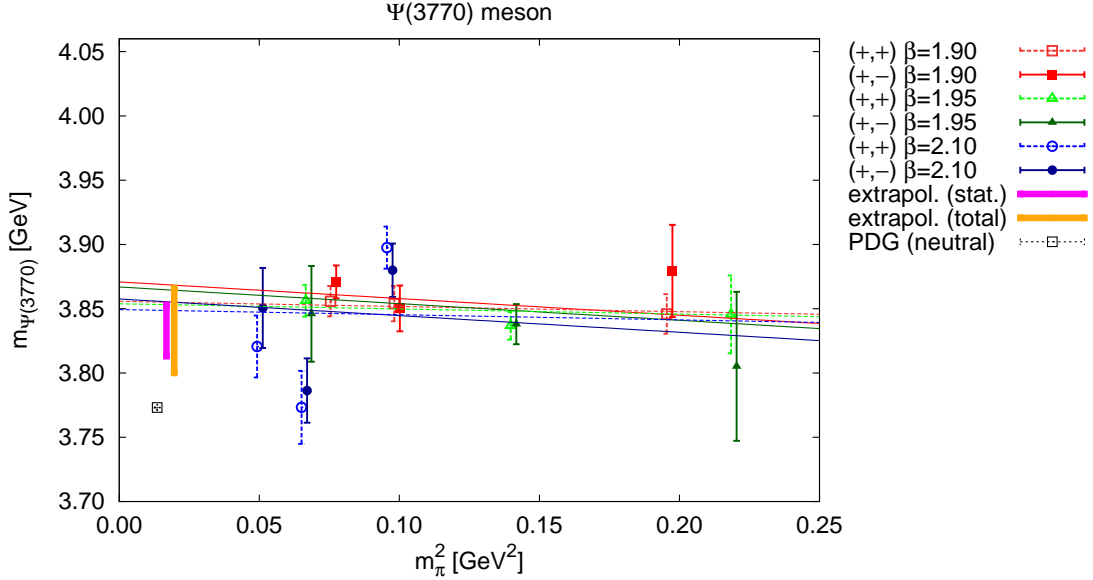}
\caption{\label{fig:psi3770} Chiral and continuum extrapolation for the $\psi(3770)$ meson ($J^{\mathcal{PC}}=1^{--}$). PDG value of the mass: 3.77315(33) GeV. Our lattice QCD result: 3.833(20) GeV (only statistical error, magenta), 3.833(33) GeV (total error, orange).}
\end{center}
\end{figure}

\begin{figure}[h!]
\begin{center}
\includegraphics[width=0.52\textwidth,angle=-90]{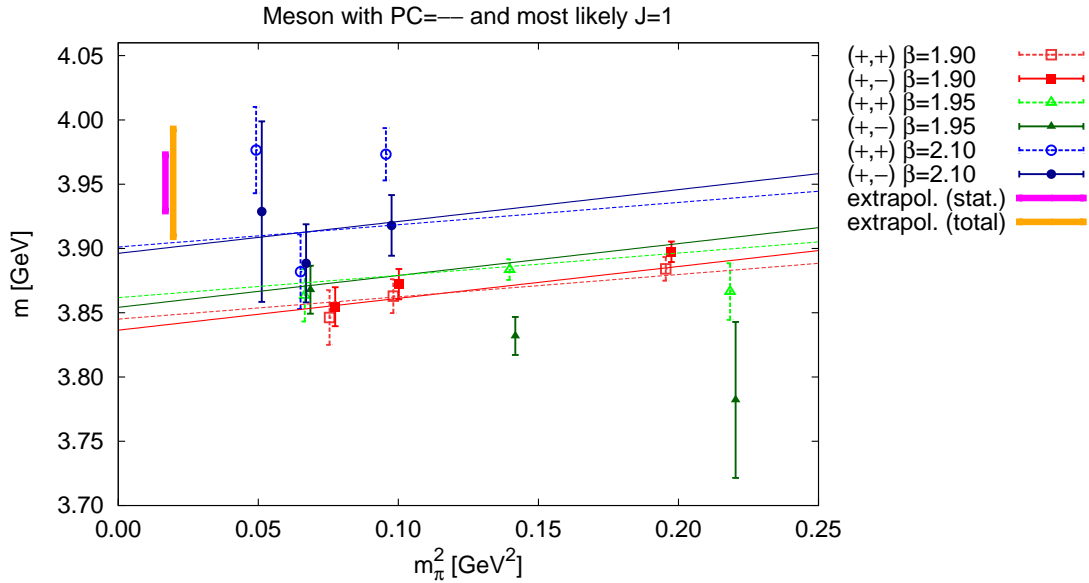}
\caption{\label{fig:psiT1} Chiral and continuum extrapolation for the third excited state with quantum numbers $\mathcal{PC}=--$ and most likely $J=1$. An obvious experimental counterpart is not available. Our lattice QCD result: 3.951(22) GeV (only statistical error, magenta), 3.951(42) GeV (total error, orange).}
\end{center}
\end{figure}

\begin{figure}[h!]
\begin{center}
\includegraphics[width=0.52\textwidth,angle=-90]{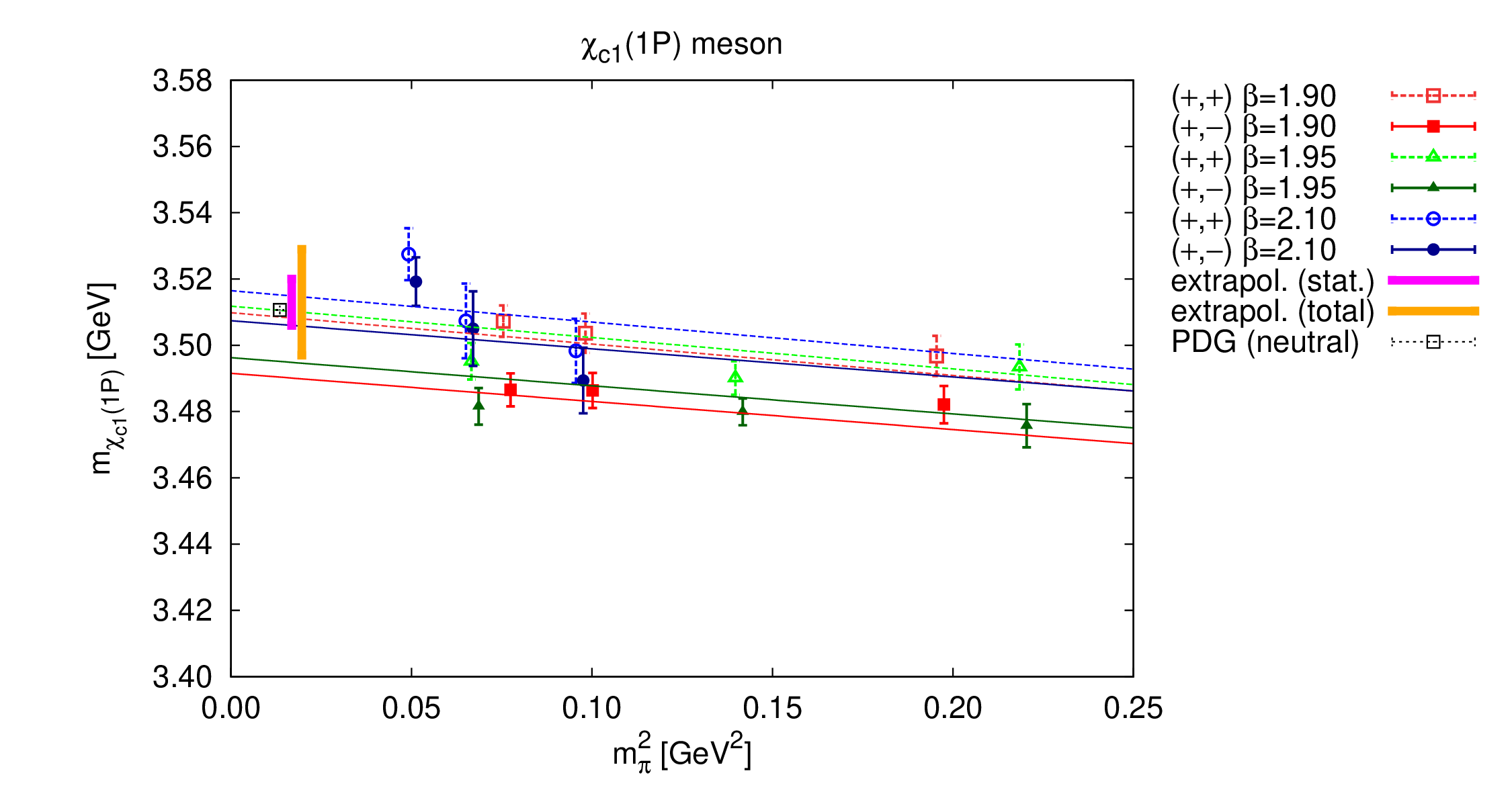}
\caption{\label{fig:chi_c1} Chiral and continuum extrapolation for the $\chi_{c1}(1P)$ meson ($J^{\mathcal{PC}}=1^{++}$). PDG value of the mass: 3.51066(7) GeV. Our lattice QCD result: 3.513(7) GeV (only statistical error, magenta), 3.513(16) GeV (total error, orange).}
\end{center}
\end{figure}

\begin{figure}[t!]
\begin{center}
\includegraphics[width=0.52\textwidth,angle=-90]{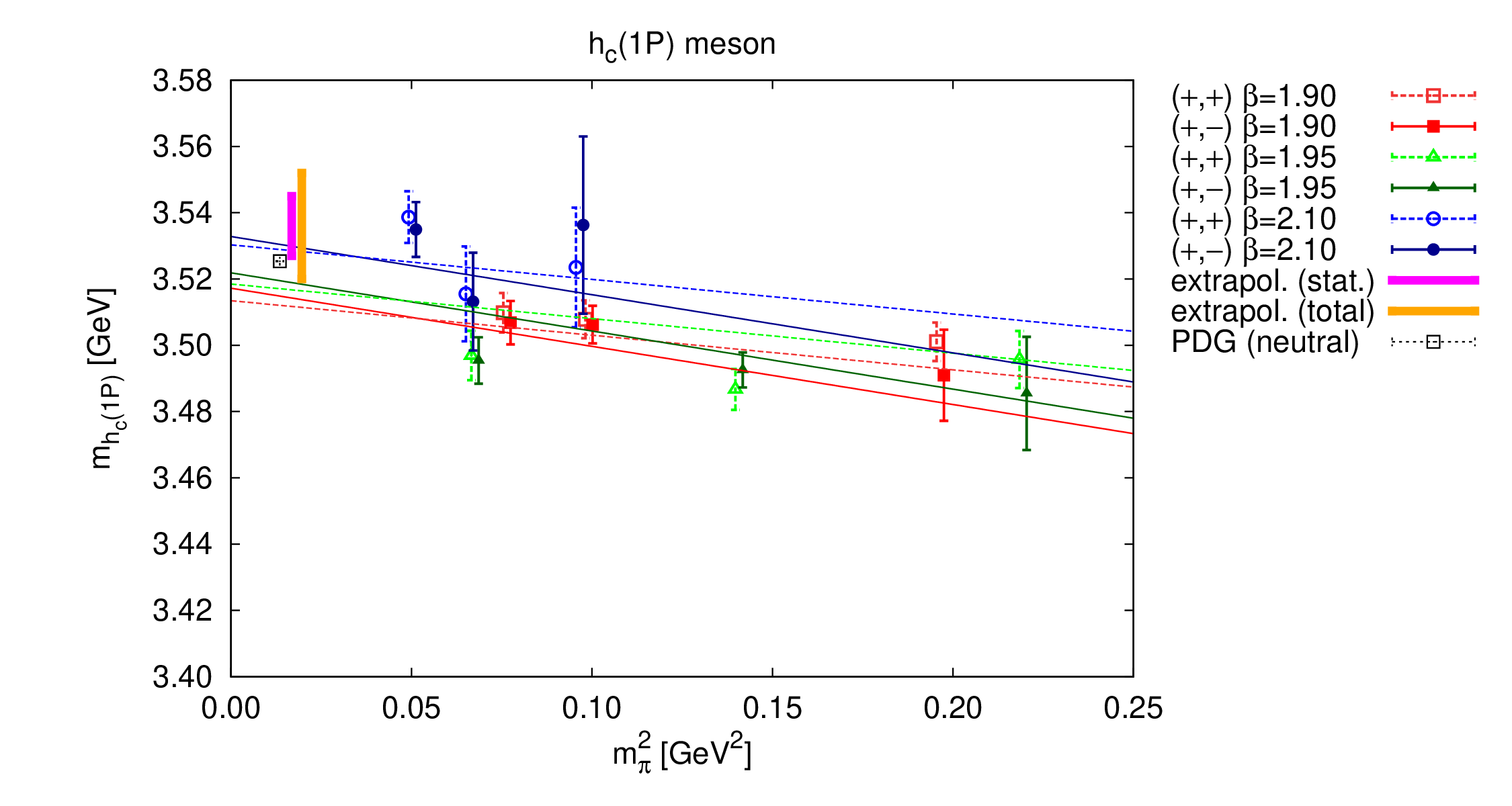}
\caption{\label{fig:h_c} Chiral and continuum extrapolation for the $h_c(1P)$ meson ($J^{\mathcal{PC}}=1^{+-}$). PDG value of the mass: 3.52538(11) GeV. Our lattice QCD result: 3.536(9) GeV (only statistical error, magenta), 3.536(16) GeV (total error, orange).}
\end{center}
\end{figure}

\begin{figure}[t!]
\begin{center}
\includegraphics[width=0.52\textwidth,angle=-90]{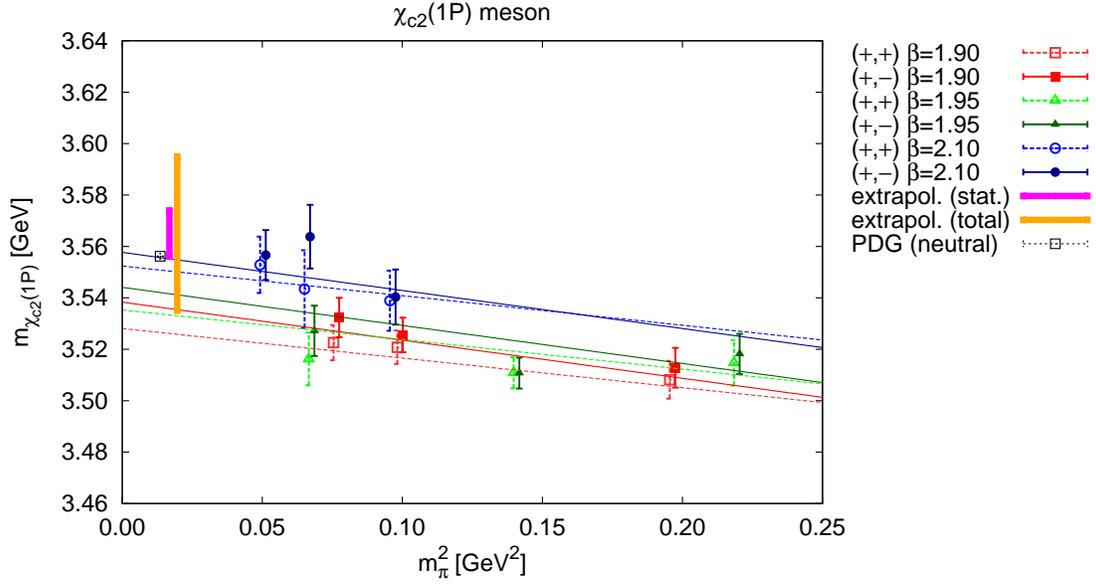}
\caption{\label{fig:chi_c2} Chiral and continuum extrapolation for the $\chi_{c2}(1P)$ meson ($J^{\mathcal{PC}}=2^{++}$). PDG value of the mass: 3.55620(9) GeV. Our lattice QCD result ($E$ representation): 3.565(9) GeV (only statistical error, magenta) 3.565(30) (total error, orange).}
\end{center}
\end{figure}

\begin{figure}[t!]
\begin{center}
\includegraphics[width=0.52\textwidth,angle=-90]{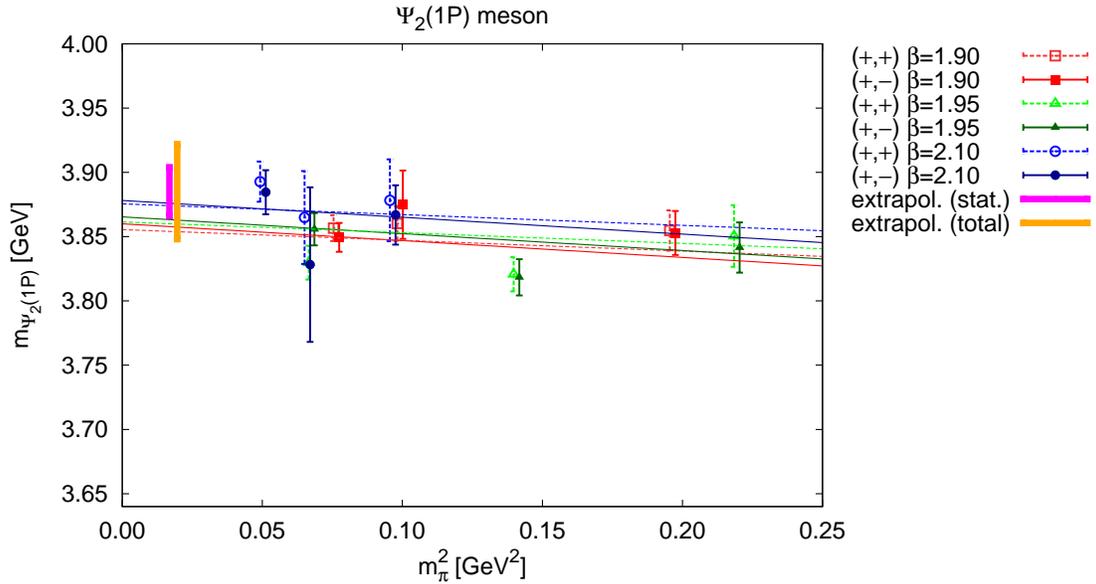}
\caption{\label{fig:psi1P} Chiral and continuum extrapolation for the $\psi_2(1P)$ meson ($J^{\mathcal{PC}}=2^{--}$). No PDG value of the mass.
Our lattice QCD result ($E$ representation): 3.885(19) GeV (only statistical error, magenta), 3.885(37) (total error, orange).}
\end{center}
\end{figure}

\begin{figure}[t!]
\begin{center}
\includegraphics[width=0.52\textwidth,angle=-90]{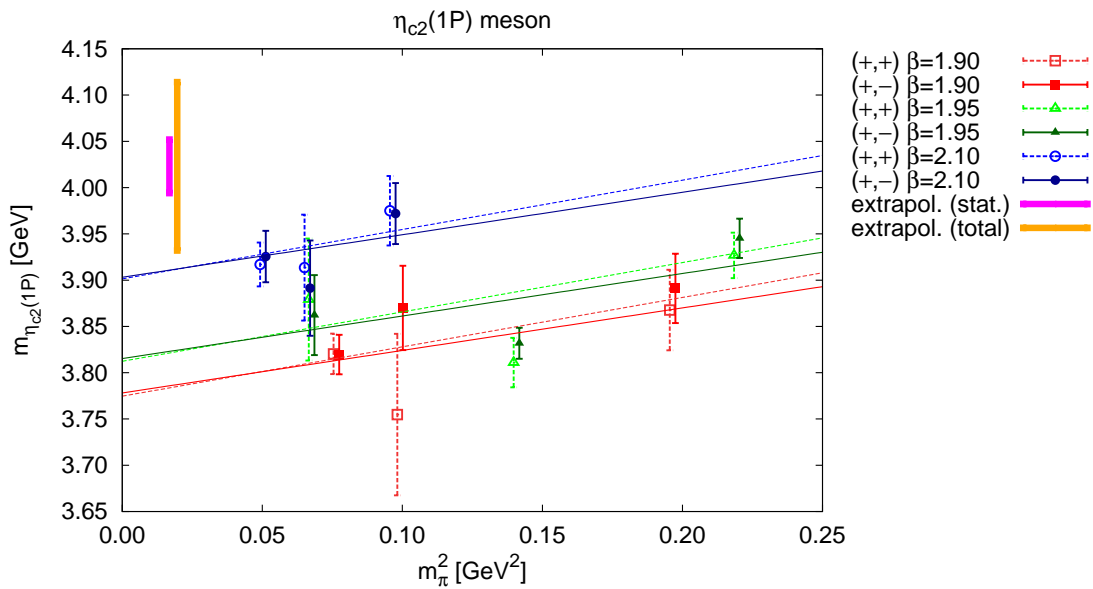}
\caption{\label{fig:eta_c2E} Chiral and continuum extrapolation for the $\eta_{c2}(1P)$ meson ($J^{\mathcal{PC}}=2^{-+}$). No PDG value of the mass. Our lattice QCD result ($E$ representation): 4.023(29) GeV (only statistical error, magenta), 4.023(91) (total error, orange).}
\end{center}
\end{figure}
\begin{figure}[t!]
\begin{center}
\includegraphics[width=0.52\textwidth,angle=-90]{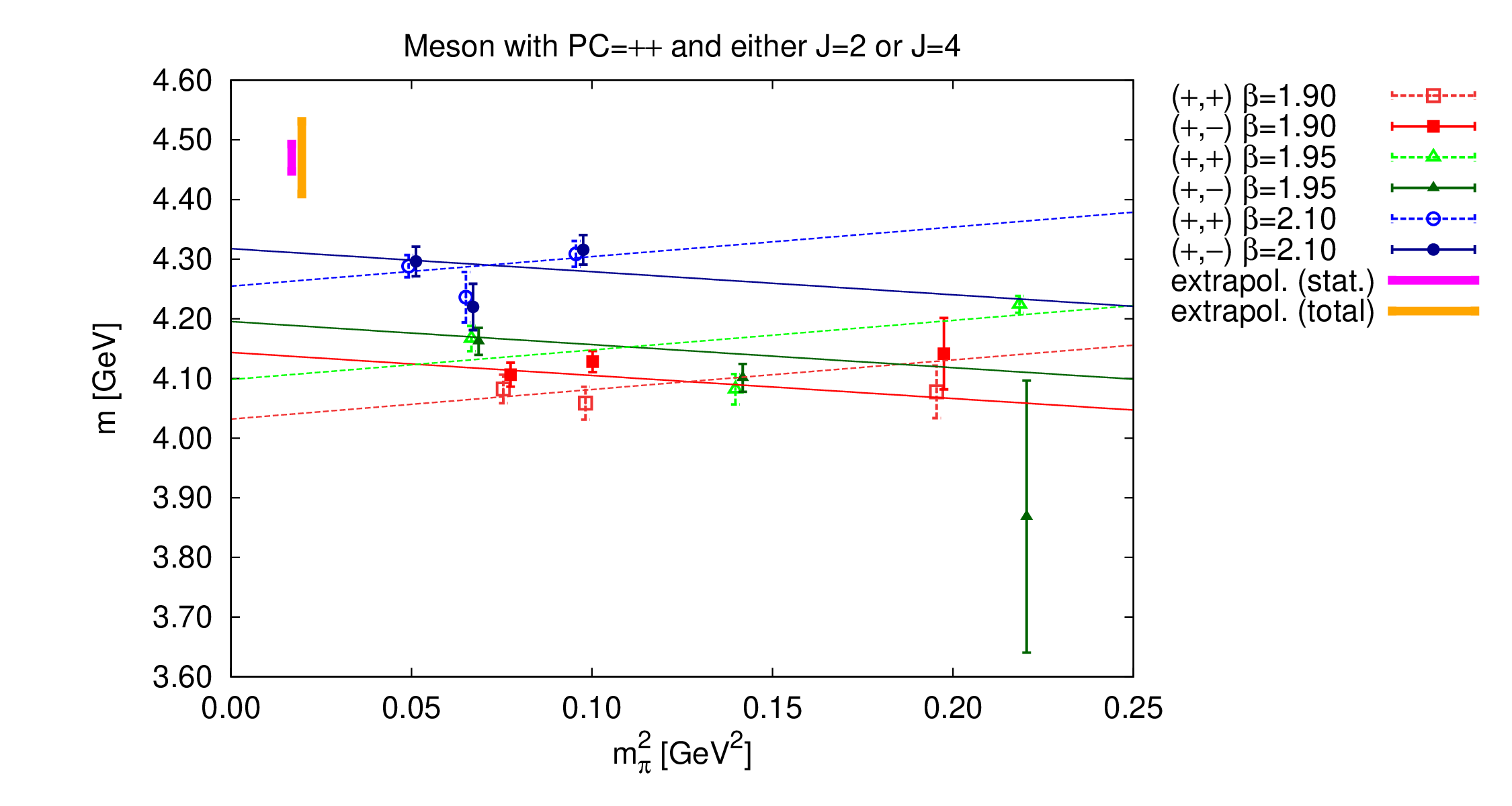}
\caption{\label{fig:chi_c2a} Chiral and continuum extrapolation for the first excited state in the $E$ representation with quantum numbers ${\mathcal{PC}}={++}$  ($J^{\mathcal{PC}}=2^{++}$ or $J^{\mathcal{PC}}=4^{++}$). The analogous state exists also in the $T_2$ representation. No PDG value of the mass. Our lattice QCD result: 4.470(23) GeV (only statistical error, magenta), 4.470(61) GeV (total error, orange).}
\end{center}
\end{figure}

\subsubsection{$T_1$ representation (spin $J=1$)}
\label{sec:ccT1}

We are able to extract four low lying $J^{\mathcal{PC}}=1^{--}$ states in the $T_1$ representation.
The mass of the ground state, the $J/\psi(1S)$ meson, in the continuum limit and at physical quark masses (Figure\ \ref{fig:J-psi}), has been obtained with a relative error of less than two per mille.
It allows again a precise comparison of the slopes of the chiral extrapolations: $\alpha^{(+,-)}=-0.241(26)$ and $\alpha^{(+,+)}=-0.240(24)$, i.e.\ there is no difference between the two employed discretizations of valence quarks.
The magnitude of cut-off effects is also similar for $(+,-)$ and $(+,+)$, but slightly smaller for the former ($c^{(+,-)}=-5.5(8)$, $c^{(+,+)}=-7.5(8)$).
We also note very good effective mass plateau quality, resulting in negligible uncertainty from the choice of the fitting interval.

As discussed in Section~\ref{SEC811} we also determine differences of meson masses, including the difference between the $J/\psi(1S)$ and $\eta_c(1S)$. This difference can be computed rather precisely, and hence is frequently studied with lattice QCD. We find
\begin{displaymath}
m_{J/\psi(1S)} - m_{\eta_c(1S)} = 0.1194(33)(14)(38) \, \textrm{GeV} = 0.119(5) \, \textrm{GeV} ,
\end{displaymath}
where the uncertainties are statistical, associated with the choice of the fitting range and with scale setting (cf.\ Section~\ref{sec:latspac}), respectively. This is compatible with the PDG value $0.1133(6) \, \textrm{GeV}$. Note that we observe a certain reduction of the statistical uncertainty compared to those of $\eta_c(1S)$ and $J/\psi(1S)$, which are both $0.005 \, \textrm{GeV}$, because correlations between the two effective masses are eliminated. The other two uncertainties (fitting range and scale setting), however, are comparable to the ones found for the two individual meson masses -- hence the rather modest reduction of the total error, which is $0.006 \, \textrm{GeV}$ for both the $\eta_c(1S)$ and $J/\psi(1S)$.

The first excitation with $J^{\mathcal{PC}}=1^{--}$ corresponds to the $\psi(2S)$ meson (Figure\ \ref{fig:psi}) and its experimental mass is well reproduced in our computation.
For the second excitation, the $\psi(3770)$ meson (Figure \ref{fig:psi3770}), we obtain a mass slightly larger than the PDG value, but still compatible within errors (discrepancy of around $1.5\,\sigma$).

The next excited state with ${\mathcal{PC}}={--}$ could be either the next excitation with $J^{\mathcal{PC}}=1^{--}$ or a state with higher spin (most probably $J^{\mathcal{PC}}=3^{--}$).
The former interpretation is favoured, since we obtain a mass of 3.951(42) GeV (cf. Figure \ref{fig:psiT1}) and we do not observe a state with a similar mass in the $T_2$ representation (thus $J=3$ seems unlikely; however, it cannot be excluded, since the signal for it might be too weak in our computation) and neither any obvious counterpart in $A_1$ nor in $E$ (which should exclude spin $J=4$).
The effective mass plateau quality is reasonable for this state, as well as the shift parameter plateau.
Moreover, the extracted continuum value is compatible with the one from D ensembles alone.
This indicates that the estimated uncertainties are under control.
Since there is no experimental result available, we compare to another lattice calculation \cite{Liu:2012ze} (at a single lattice spacing of 0.12 fm, much coarser than our coarsest one) and to Dyson-Schwinger and Bethe-Salpeter approaches \cite{Fischer:2014cfa}, which both give masses in the range between 3.85 GeV and 3.9 GeV, i.e.\ are qualitatively consistent with our result.

We find two mesons of positive parity in the $T_1$ representation -- the $\chi_{c1}(1P)$ meson ($J^{\mathcal{PC}}=1^{++}$, Figure\ \ref{fig:chi_c1}) and the $h_c(1P)$ meson ($J^{\mathcal{PC}}=1^{+-}$, Figure\ \ref{fig:h_c}).
The data quality for both these states is very good and we obtain excellent agreement with experimental values.

\subsubsection{$E$ and $T_2$ representations (spin $J=2$)}

The lowest lying state in both the $E$ and $T_2$ representations has a similar continuum value at physical quark masses and hence it can be attributed spin $J=2$.
Experimentally, it corresponds to the $\chi_{c2}(1P)$ meson ($J^{\mathcal{PC}}=2^{++}$).
We obtain agreement with the experiment and the total error is dominated by the fitting range uncertainty, which is much larger than the statistical error.
We show our fits for the $E$ representation in Figure\ \ref{fig:chi_c2} (the plot for the $T_2$ representation is very similar).

The next extracted state in the $E$ representation (Figure\ \ref{fig:psi1P}) has quantum numbers ${\mathcal{PC}}={--}$. 
Most probably, it corresponds to $J=2$, since a state with similar mass is also found in the $T_2$ representation.
Assuming $J=2$, it corresponds to the experimentally not established $\psi_2(1P)$ meson.
We again compare to Refs.\ \cite{Liu:2012ze} and \cite{Fischer:2014cfa}.
Our continuum and physical pion mass result is compatible with the lattice QCD result of Ref.\ \cite{Liu:2012ze} (only a single lattice spacing $a\approx0.12$ fm) and around 200 MeV higher than the Bethe-Salpeter result of Ref.\ \cite{Fischer:2014cfa}.

The second excitation in the $E$ representation corresponds to ${\mathcal{PC}}={-+}$. It agrees within the total error with the lowest $T_2$ representation state with ${\mathcal{PC}}={-+}$ -- hence, it can be attributed $J=2$.
Such a state is again experimentally not clearly identified, i.e.\ it is not included in the PDG review.
The naming scheme would suggest the name $\eta_{c2}(1P)$.
Our continuum and chiral extrapolation for the $E$ representation state is shown in Figure\ \ref{fig:eta_c2E}.
The total error is dominated by the fitting uncertainty and we do not observe a clear plateau in the shift parameter dependence of the meson mass.
Hence, our result should be treated with caution and needs to be confirmed in the future.
Our continuum and physical pion mass result is around 100 MeV higher than the lattice QCD and Bether-Salpeter results \cite{Liu:2012ze,Fischer:2014cfa}, but we note that our result using only the coarsest lattice spacing (A ensembles) is compatible with Ref.\ \cite{Liu:2012ze} at a single and even coarser lattice spacing.

Finally, we extract one more state in the $E$ and $T_2$ representations, with quantum numbers ${\mathcal{PC}}={++}$ (see Figure\ \ref{fig:chi_c2a}).
The mass is close to 4.5 GeV, with a statistical uncertainty of around 30 MeV and a fitting range uncertainty between 50 MeV and 60 MeV.
Since it appears in both representations, it could have $J=2$ but also $J=4$ cannot be excluded, taking into account the large value of the mass.
At present, we are not able to clearly associate it to any experimentally observed state.
Comparing to Ref.\ \cite{Liu:2012ze}, we find that our A ensembles results (around 4.05...4.15 GeV) are compatible with the result of this paper ($a\approx0.12$ fm), both for $J^{\mathcal{PC}}=2^{++}$ and $J^{\mathcal{PC}}=4^{++}$ (around 4 GeV and 4.1 GeV, respectively).
However, we predict that possible future experiments will find a higher mass, because our continuum result is around 4.5 GeV and the one from the D ensembles is between 4.2 GeV and 4.3 GeV.
Because of the rather poor signal quality and large cut-off effects, we do not quote a final value for this state.

\subsection{Analysis of further systematic effects}
The systematic errors are dominated by the fitting range uncertainties, which have already been discussed extensively.
We now investigate and quantify other sources of systematic errors.

\subsubsection{Finite volume effects}
\label{sec:FVE}
We expect that the finite volume effects (FVE) are small in our computation -- all our ensembles have rather large $m_\pi L$ (with smallest $m_\pi L$ of 3.4 and only three ensembles with $m_\pi L<4$) and moreover, FVE are probably strongly suppressed because our meson creation operators do not directly generate additional pions but only excite much heavier mesons (none of our creation operators contains a light quark and a light antiquark).
Nonetheless, we performed a dedicated analysis of FVE comparing ensembles A40.24 and A40.32, i.e.\ two volumes with the same lattice spacing, with spatial lattice extents of around 2.1 fm ($m_\pi L\approx3.5$) and 2.8 fm ($m_\pi L\approx4.5$), respectively.

\begin{figure}[t!]
\begin{center}
\includegraphics[width=0.345\textwidth,angle=-90]{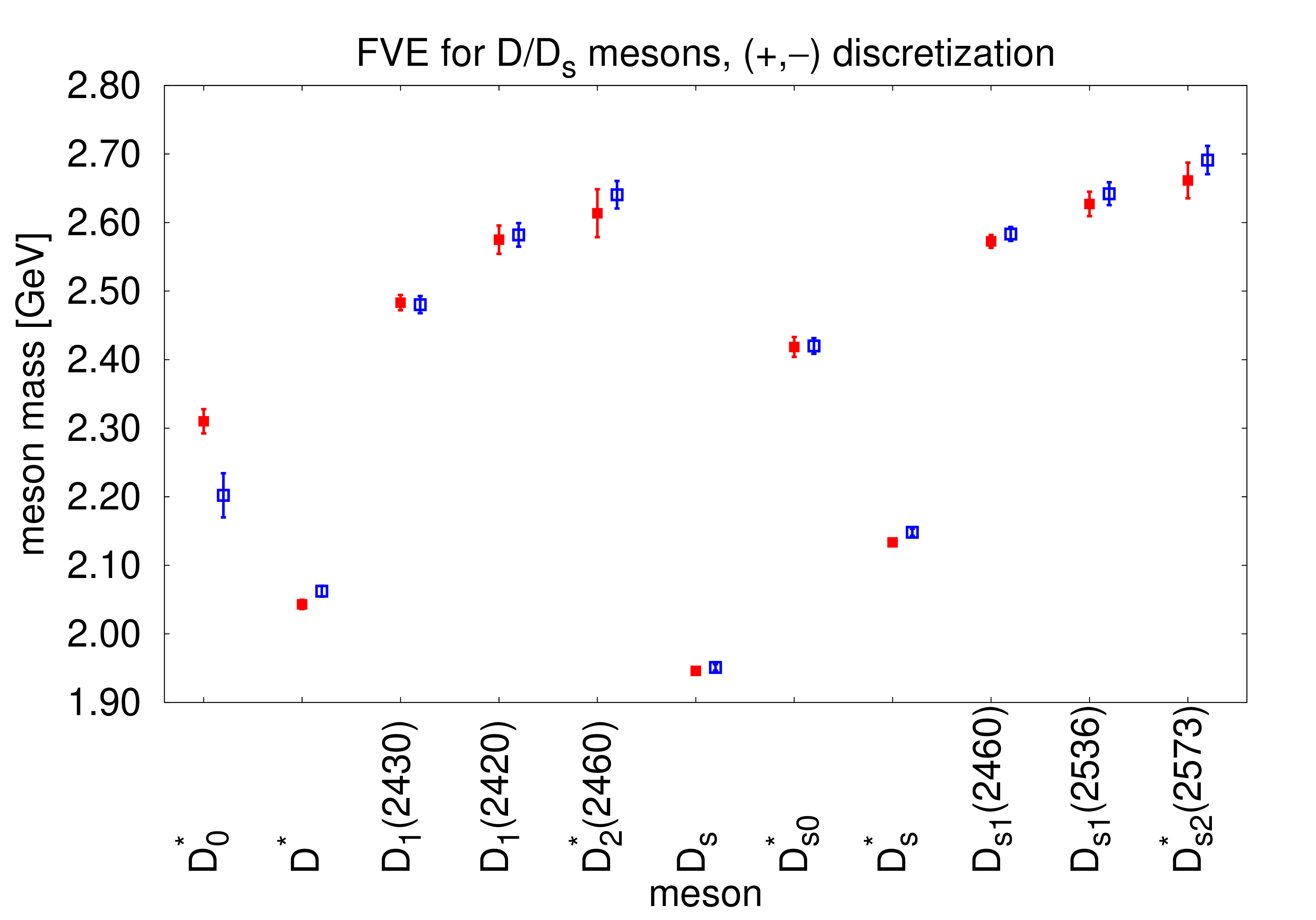}
\includegraphics[width=0.345\textwidth,angle=-90]{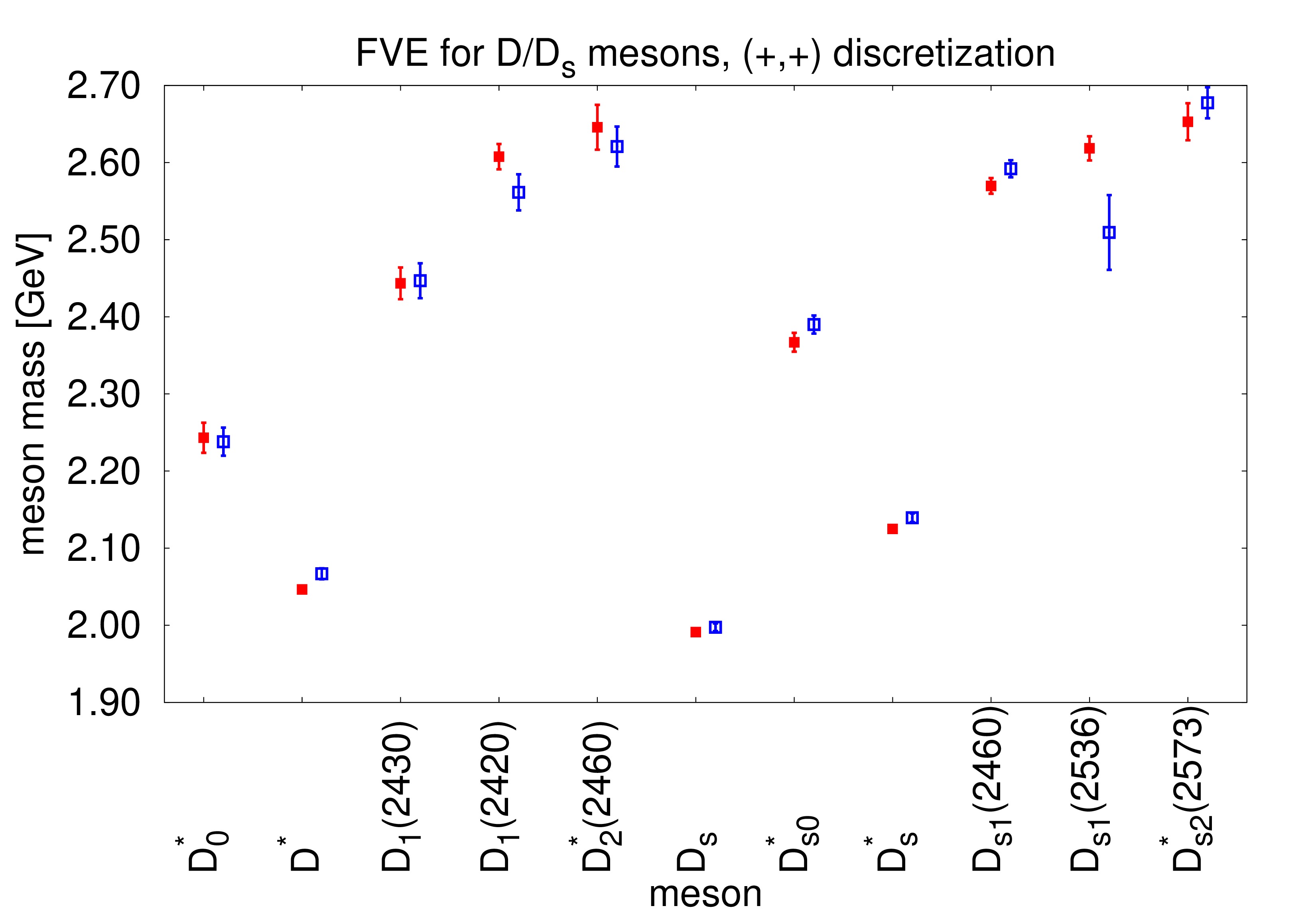}
\caption{\label{fig:FVE} Comparison of $D$ and $D_s$ meson masses from ensembles A40.32 (filled red squares) and A40.24 (open blue squares) for the $(+,-)$ setup (left) and the $(+,+)$ setup (right). The agreement between the two volumes (spatial lattice extents of 2.1 fm and 2.8 fm, respectively) confirms that finite volume effects are negligible. The errors are only statistical, but the same temporal fitting intervals are chosen for both ensembles.}
\end{center}
\end{figure}

\begin{figure}[t!]
\begin{center}
\includegraphics[width=0.345\textwidth,angle=-90]{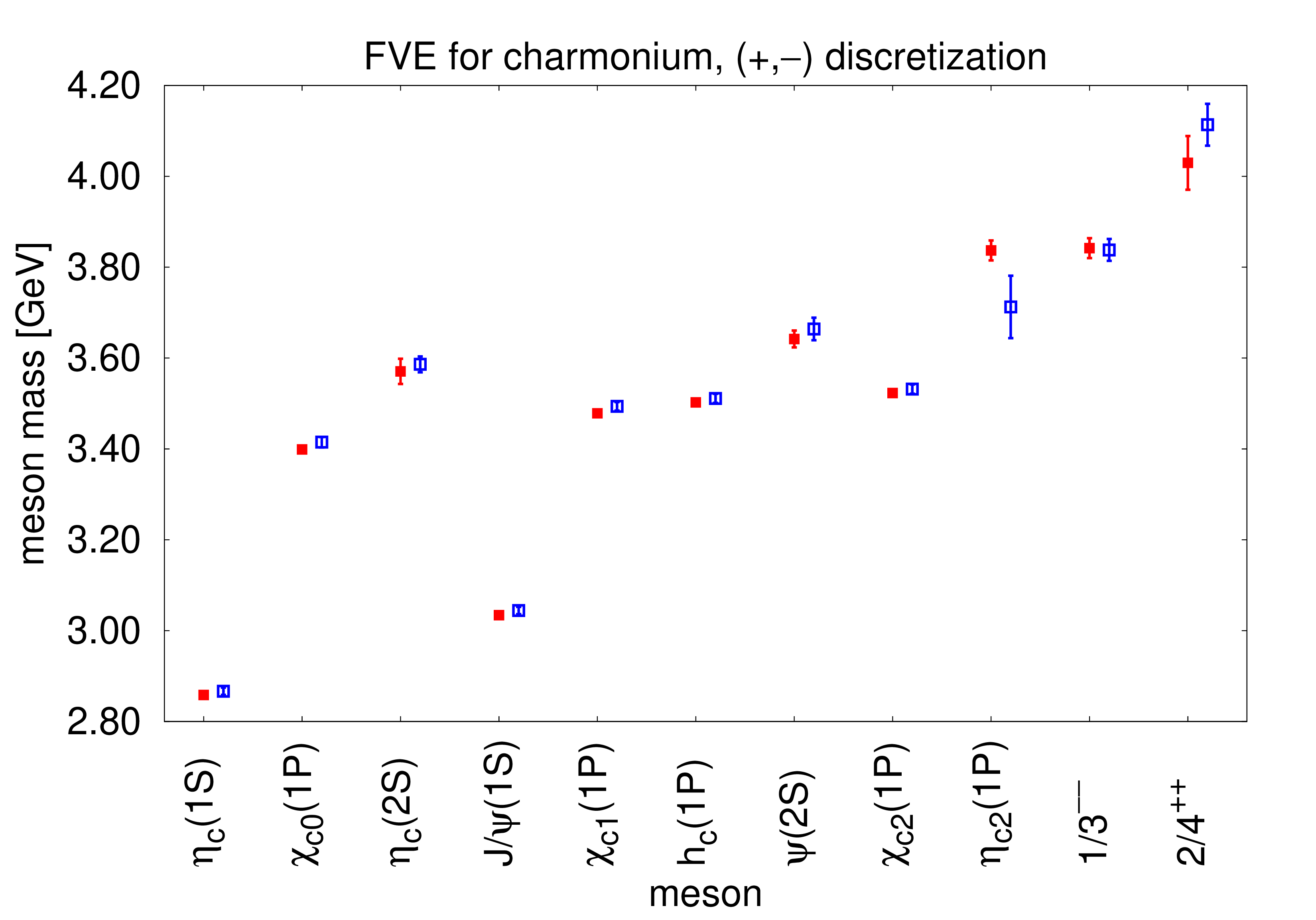}
\includegraphics[width=0.345\textwidth,angle=-90]{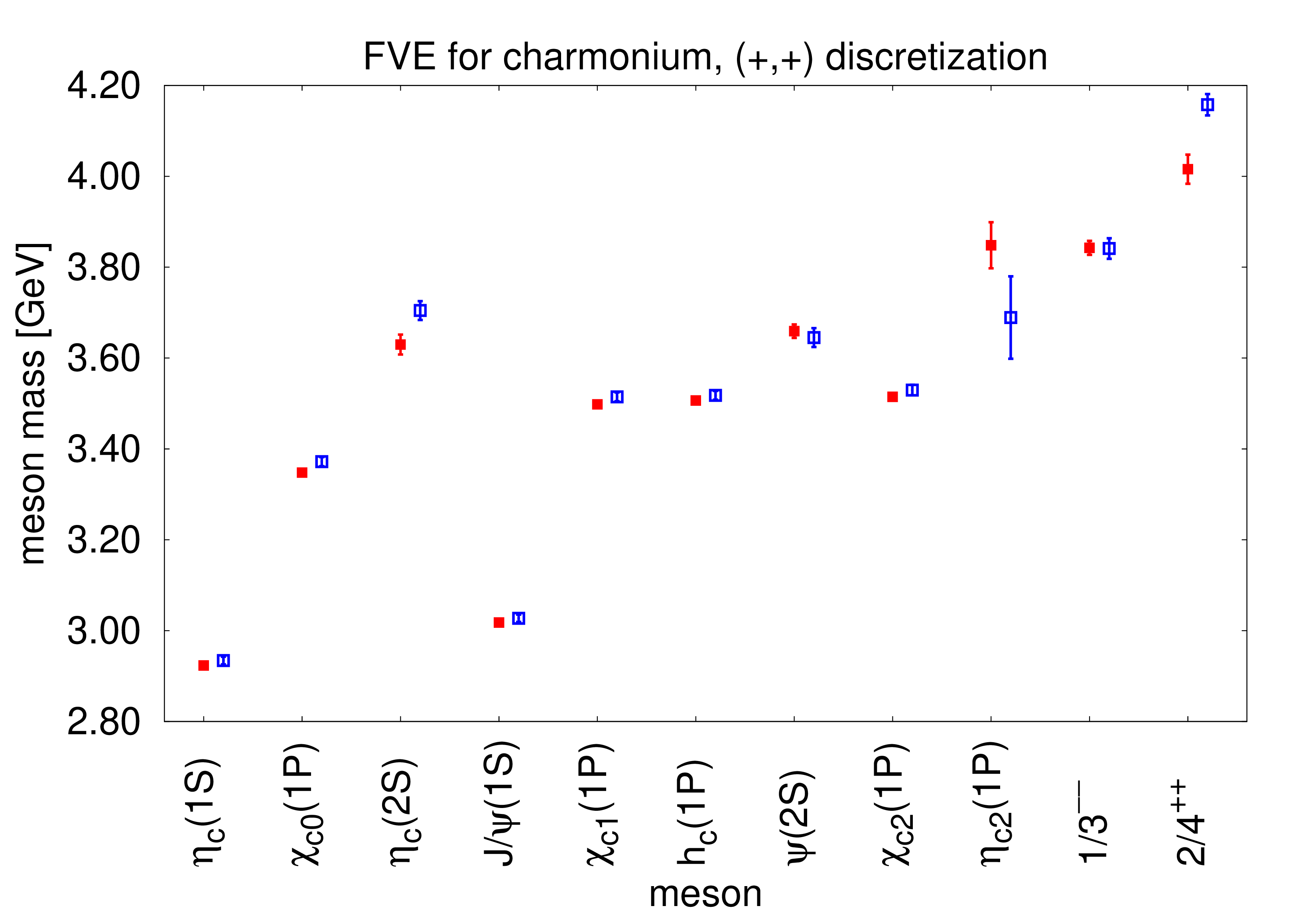}
\caption{\label{fig:FVE-cc} Comparison of charmonium masses from ensembles A40.32 (filled red squares) and A40.24 (open blue squares) for the $(+,-)$ setup (left) and the $(+,+)$ setup (right). The agreement between the two volumes (spatial lattice extents of 2.1 fm and 2.8 fm, respectively) confirms that finite volume effects are negligible. The errors are only statistical, but the same temporal fitting intervals are chosen for both ensembles.}
\end{center}
\end{figure}

In Figures\ \ref{fig:FVE} and \ref{fig:FVE-cc}, we compare the $D$ and $D_s$ meson and charmonium masses computed for these ensembles.
We observe fully compatible results for both valence quark discretizations.
In most cases, the differences between the two volumes are below $1\,\sigma$ and the number of cases where this difference is larger than $1\,\sigma$ is close to the expected 32\%. 
This is confirmed by the fact that in all cases where the difference between the two volumes is quite large e.g.\ for the $D_0^*$ in the $(+,-)$ discretization, it is almost perfectly zero in the other discretization, i.e.\ a difference significantly smaller than $1 \, \sigma$. Similarly for the $D_{s1}(2536)$ or $\eta_c(2S)$, for which the difference between the volumes is almost $2\,\sigma$ for $(+,+)$ and well within $1\,\sigma$ for $(+,-)$.
We therefore conclude that, according to our expectations, FVE are negligible when computing masses of mesons containing charm quarks for our gauge field ensembles and any difference between values obtained for the same physical parameters and only different volumes is not a finite size effect, but rather a statistical fluctuation.

\subsubsection{Choice of the fitting ansatz for the combined chiral and continuum extrapolation}
\label{sec:ansatz}
\begin{table}[t!]
\centering
\begin{tabular}{lccccc}
\hline
\multirow{2}*{meson} & \multirow{2}*{$J^{\mathcal{P(C)}}$} & Strategy 2 & Strategy 3 & \multicolumn{2}{c}{Strategy 1}\\
&  & $m$ & $m$ & $m^{(+,-)}$ & $m^{(+,+)}$ \\
\hline
$D_0^*$ & $0^+$ & 2.325(10) & 2.324(10) & 2.318(16) & 2.331(15)\\
$D^*$ & $1^-$ & 2.027(7) & 2.027(7) & 2.031(11) & 2.024(9) \\
$D_1(2430)$ & $1^+$ & 2.464(15) & 2.463(15) & 2.454(20) & 2.478(23)\\
$D_1(2420)$ & $1^+$ & 2.631(22) & 2.631(22) & 2.591(29) & 2.682(35)\\
$D_2^*(2460)$ ($E$) & $2^+$ & 2.743(30) & 2.731(30) & 2.728(40) & 2.763(48)\\
$D_2^*(2460)$ ($T_2$) & $2^+$ & 2.707(27) & 2.702(26) & 2.726(34) & 2.674(44)\\
\hline
$D_s$ & $0^-$ & 1.9679(27) & 1.9679(28) & 1.9721(39) & 1.9636(39) \\
$D_{s0}^*$ & $0^+$ & 2.390(11) & 2.390(11) & 2.384(16) & 2.396(15) \\
$D_s^*$ & $1^-$ & 2.1226(38) & 2.1226(38) & 2.124(5) & 2.122(5) \\
$D_{s1}(2460)$ & $1^+$ & 2.556(10) & 2.556(10) & 2.552(15) & 2.559(14)\\
$D_{s1}(2536)$ & $1^+$ & 2.617(22) & 2.618(22) & 2.592(31) & 2.641(29)\\
$D_{s2}^*(2573)$ ($E$) & $2^+$ & 2.734(25) & 2.734(25) & 2.751(36) & 2.718(35)\\
$D_{s2}^*(2573)$ ($T_2$) & $2^+$ & 2.690(27) & 2.690(27) & 2.703(37) & 2.676(38)\\
\hline
$\eta_c(1S)$ & $0^{-+}$ & 2.9828(54) & 2.9828(57) & 2.990(8) & 2.976(8)\\
$\eta_c(2S)$ & $0^{-+}$ & 3.741(28) & 3.742(28) & 3.743(40) & 3.739(42)\\
$\chi_{c0}(1P)$ & $0^{++}$ & 3.413(7) & 3.413(7) & 3.406(11) & 3.419(10) \\
$J/\psi(1S)$ & $1^{--}$ & 3.0961(51) & 3.0961(51) & 3.094(7) & 3.098(7)\\
$\psi(2S)$ & $1^{--}$ & 3.647(30) & 3.648(29) & 3.629(42) & 3.666(43)\\
$\psi(3770)$ & $1^{--}$ & 3.833(20) & 3.832(19) & 3.812(31) & 3.846(26)\\
? ($T_1$) & $1^{--}/3^{--}$ & 3.951(22) & 3.951(21) & 3.883(32) & 3.999(29)\\
$\chi_{c1}(1P)$ & $1^{++}$ & 3.513(7) & 3.513(8) & 3.515(11) & 3.510(11)\\
$h_c(1P)$ & $1^{+-}$ & 3.536(9) & 3.536(9) & 3.538(13) & 3.535(12)\\
$\chi_{c2}(1P)$ ($E$) & $2^{++}$ & 3.565(9) & 3.565(9) & 3.569(13) & 3.560(14)\\
$\chi_{c2}(1P)$ ($T_2$) & $2^{++}$ & 3.562(9) & 3.562(9) & 3.563(13) & 3.560(13)\\
$\psi_2(1P)$ ($E$) & $2^{--}$ & 3.885(19) & 3.885(18) & 3.883(27) & 3.887(25)\\
$\psi_2(1P)$ ($T_2$) & $2^{--}$ & 3.935(21) & 3.935(21) & 3.973(29) & 3.885(30)\\
$\eta_{c2}(1P)$ ($E$) & $2^{-+}$ & 4.023(29) & 4.023(30) & 4.017(44) & 4.029(41) \\
$\eta_{c2}(1P)$ ($T_2$) & $2^{-+}$ & 4.034(29) & 4.031(28) & 4.032(43) & 4.037(41) \\
? ($E$) & $2^{++}/4^{++}$  & 4.470(23) & 4.458(23) & 4.455(39) & 4.480(31)\\
? ($T_2$) & $2^{++}/4^{++}$  & 4.529(30) & 4.531(31) & 4.531(43) & 4.528(45)\\
\hline
\end{tabular}
\caption{\label{tab:summary}Neutral $D$ meson, $D_s$ meson and charmonium masses using different strategies for the combined chiral and continuum extrapolation (see Section\ \ref{sec:fits}).}
\end{table}

Another source of systematic uncertainty is related to the combined chiral and continuum extrapolation.
We try to quantify this uncertainty by considering and comparing three different types of fitting ans\"atze, with six, five or four fitting parameters, as discussed in Section\ \ref{sec:fits} (Strategy 1, Strategy 2 and Strategy 3, respectively).
In Table\ \ref{tab:summary}, we list the chirally and continuum extrapolated values of the meson masses for all considered states (for charm-light mesons, we show results corresponding to neutral $D$ mesons; the results for the charged counterparts are collected in Table\ \ref{tab:all} only for Strategy 2, but differences due to the variation of fitting strategies are nearly identical).
Fitting Strategy 2 and Strategy 3 assume a common continuum limit for the $(+,-)$ and $(+,+)$ valence quark discretizations and differ only in the additional assumption of a common (Strategy 3) or possibly different (Strategy 2) slope of the chiral extrapolation.
As can be seen in the table, the resulting meson masses are almost identical -- hence, the assumption about a discretization independent slope of the chiral extrapolation seems to be justified.

On the other hand, Strategy 2 and Strategy 3 differ with respect to Strategy 1 in terms of the assumption about the common (Strategy 2 and Strategy 3) or possibly different (Strategy 1) final values for the meson masses obtained from both discretizations. Although common values are guaranteed by universality, both discretizations differ by cut-off effects and it is an important cross-check to confirm that universality is indeed satisfied.
We find that the continuum extrapolated values from both discretizations agree with each other for almost all of our states within one standard deviation, with only three exceptions of higher excited states.
Hence, these observations confirm universality.

In addition to testing universality, we also check whether the slopes of the chiral extrapolation agree within errors for both discretizations.
Table\ \ref{tab:summary2} contains the values of these slopes for all three fitting strategies (again, for $D$ mesons, we show results only for the neutral case; values for charged mesons are almost identical).
We note that the statistical errors are rather similar for both discretizations and hence the common $\alpha\equiv\alpha^{(+,-)}=\alpha^{(+,+)}$ (of Strategy 3) is always close to the average of $\alpha^{(+,-)}$ and $\alpha^{(+,+)}$ (from Strategy 2 or Strategy 1 -- both yield very similar values).
Moreover, $\alpha^{(+,-)}$ and $\alpha^{(+,+)}$ are compatible with each other within one standard deviation for almost all cases.
This implies that, as expected, the dependence of the slopes of the chiral extrapolation on the discretization can be neglected as a higher-order effect in our computation, i.e.\ assuming or not assuming the equality of $\alpha^{(+,-)}$ and $\alpha^{(+,+)}$ has a negligible effect compared to statistical and other types of systematic uncertainties.

\addtolength{\tabcolsep}{-3pt} 
\begin{table}[t!]
\centering
\begin{tabular}{lccccccc}
\hline
\multirow{2}*{meson} & \multicolumn{4}{c}{Strategy 2} & Strategy 3 & \multicolumn{2}{c}{Strategy 1}\\
 & $\alpha^{(+,-)}$ & $\alpha^{(+,+)}$ & $c^{(+,-)}$ & $c^{(+,+)}$ & $\alpha$ & $\alpha^{(+,-)}$ & $\alpha^{(+,+)}$ \\
\hline
$D_0^*$ & 0.97(11) & 0.77(8) & -14.3(2.7) & -18.2(2.4) & 0.83(6) & 0.96(11) & 0.77(8)\\
$D^*$ & 0.065(37) & 0.060(29) & 2.7(1.1) & 1.8(1.1) & 0.062(25) & 0.063(38) & 0.062(29) \\
$D_1(2430)$ & 0.87(10) & 0.80(9) & -11.5(3.1) & -9.5(2.9) & 0.83(7) & 0.86(9) & 0.81(9) \\
$D_1(2420)$ & 0.09(10) & 0.13(15) & -6.6(3.9) & -8.7(4.3) & 0.10(9) & 0.09(10) & 0.12(15) \\
$D_2^*(2460)$ ($E$) & 0.81(21) & 0.08(24) & -26(7) & -16(7) & 0.49(15) & 0.78(21) & 0.08(23)\\
$D_2^*(2460)$ ($T_2$) & 0.02(14) & 0.26(14) & -5.5(3.9) & -10(5) & 0.14(10) & 0.01(14) & 0.24(14)\\
\hline
$D_s$ & -0.166(12) & -0.168(13) & -1.0(4) & -4.7(4) & -0.167(9) & -0.168(12) & -0.165(13)\\
$D_{s0}^*$ & 0.34(9) & 0.27(8) & 0.1(2.1) & -5.8(2.2) & 0.30(6) & 0.34(9) & 0.27(9)\\
$D_s^*$ & -0.098(22) & -0.106(18) & 2.7(6) & 1.6(6) & -0.103(15) & -0.099(23) & -0.106(19)\\
$D_{s1}(2460)$ & 0.32(6) & 0.27(7) & -3.4(1.9) & -1.0(2.0) & 0.29(5) & 0.32(7) & 0.27(7)\\
$D_{s1}(2536)$ & 0.28(15) & 0.07(17) & -2.4(4.0) & -0.1(4.4) & 0.18(11) & 0.29(15) & 0.08(16)\\
$D_{s2}^*(2573)$ ($E$) & 0.24(12) & 0.24(17) & -12(4) & -13(5) & 0.24(10) & 0.24(12) & 0.23(17)\\
$D_{s2}^*(2573)$ ($T_2$) & 0.30(13) & 0.22(14) & -6(4) & -5(5) & 0.27(10) & 0.30(13) & 0.22(15)\\
\hline
$\eta_c(1S)$ & -0.259(23) & -0.270(24) & -13.1(8) & -5.0(8) & -0.264(18) & -0.263(25) & -0.266(25)\\
$\eta_c(2S)$ & 0.29(16) & 0.12(18) & -26(5) & -13(5) & 0.21(12) & 0.00(16) & -0.23(17)\\
$\chi_{c0}(1P)$ & -0.10(4) & -0.070(38) & -0.5(1.2) & -7.1(1.1) & -0.089(27) & -0.10(4) & -0.071(37)\\
$J/\psi(1S)$ & -0.241(26) & -0.240(24) & -5.5(8) & -7.5(8) & -0.240(17) & -0.240(24) & -0.240(23)\\
$\psi(2S)$ & 0.42(24) & 0.24(13) & -4.5(5.4) & -3.6(4.8) & 0.28(11) & 0.43(24) & 0.24(13)\\
$\psi(3770)$ & -0.13(19) & -0.04(11) & 3.7(3.4) & 1.9(3.2) & -0.06(10) & -0.09(20) & -0.05(11)\\
? ($T_1$) & 0.25(11) & 0.17(11) & -15.0(3.6) & -14.0(3.8) & 0.22(8) & 0.24(11) & 0.18(12)\\
$\chi_{c1}(1P)$ & -0.085(38) & -0.09(4) & -3.9(1.2) & -1.6(1.2) & -0.129(29) & -0.09(4) & -0.09(4)\\
$h_c(1P)$ & -0.18(7) & -0.10(5) & -3.8(1.6) & -4.2(1.5) & -0.094(27) & -0.18(7) & -0.10(5)\\
$\chi_{c2}(1P)$ ($E$) & -0.15(5) & -0.12(5) & -4.7(1.6) & -5.9(1.6) & -0.132(37) & -0.15(5) & -0.11(5)\\
$\chi_{c2}(1P)$ ($T_2$) & -0.12(5) & -0.21(6) & -4.6(1.5) & -4.0(1.6) & -0.16(4) & -0.12(5) & -0.21(6)\\
$\psi_2(1P)$ ($E$) & -0.13(11) & -0.08(11) & -4.5(3.1) & -4.9(3.2) & -0.11(8) & -0.13(11) & -0.08(11)\\
$\psi_2(1P)$ ($T_2$) & -0.02(12) & 0.24(9) & -10.5(3.2) & -14.0(3.0) & 0.14(7) & -0.05(11) & 0.28(9)\\
$\eta_{c2}(1P)$ ($E$) & 0.53(16) & 0.46(19) & -32(5) & -31(6) & 0.50(13) & 0.54(17) & 0.46(18)\\
$\eta_{c2}(1P)$ ($T_2$) & -0.22(18) & 0.19(20) & -26(4) & -29(5) & -0.04(14) & -0.21(18) & 0.19(21)\\
? ($E$) & -0.38(29) & 0.47(12) & -43(5) & -55(5) & 0.34(12) & -0.34(29) & 0.47(12)\\
? ($T_2$) & -0.17(26) & -0.67(27) & -57(6) & -47(6) & -0.43(18) & -0.17(27) & -0.69(26)\\
\hline
\end{tabular}
\caption{\label{tab:summary2}Chiral extrapolation fitting parameters $\alpha^{(+,-)}$, $\alpha^{(+,+)}$, $\alpha\equiv\alpha^{(+,-)}=\alpha^{(+,+)}$ using different strategies for the combined chiral and continuum extrapolation (see Section\ \ref{sec:fits}).  For Strategy 2, we also list fitting parameters describing discretization effects, $c^{(+,-)}$ and $c^{(+,+)}$.}
\end{table}
\addtolength{\tabcolsep}{3pt}

In Table\ \ref{tab:summary2}, we also compare cut-off effects for all the meson masses for the two considered discretizations.
The expectation from previous ETMC investigations of light pseudoscalar mesons is that the $(+,-)$ setup has smaller discretization effects than the $(+,+)$ setup.
However, as we already mentioned in the case of $\eta_c(1S)$, this is not always true.
Only for five of the considered states, $|c^{(+,-)}|<|c^{(+,+)}|$ (within one standard deviation), while in two cases, $|c^{(+,-)}|>|c^{(+,+)}|$.
In the remaining around 75\% of the cases, $c^{(+,-)}$ and $c^{(+,+)}$ are compatible within statistical errors.
Thus, the conclusion is that cut-off effects tend to be quite similar in both setups, with only a slight tendency for smaller discretization effects in the $(+,-)$ setup.
However, we confirm previous ETMC statements that for the states relevant for the tuning of the strange and charm quark mass (the pion, the kaon and the $D$ meson), cut-off effects are much smaller in the $(+,-)$ setup.

For the final results presented in Section\ \ref{sec:summary}, our preferred fitting strategy is Strategy 2, a compromise that assumes universality of the continuum limit, but leaves independent slopes of the chiral extrapolation.
However, as detailed above, the uncertainty associated with the choice of the fitting ansatz has negligible impact on our final results.

\begin{table}[t!]
\centering
\begin{tabular}{lccc}
\hline
set & $a(\beta=1.90)$ & $a(\beta=1.95)$ & $a(\beta=2.10)$ \\
\hline
I & 0.0885(36) & 0.0815(30) & 0.0619(18)\\
II & 0.0886(27) & 0.0815(21) & 0.0619(11)\\
III & 0.0899(31) & 0.0827(25) & 0.0628(13)\\
IV & 0.0868(33) & 0.0799(27) & 0.0607(14)\\
V & 0.0892(34) & 0.0820(28) & 0.0623(15)\\
VI & 0.0887(27) & 0.0816(21) & 0.0620(11)\\
VII & 0.0898(31) & 0.0826(25) & 0.0627(13)\\
VIII & 0.0865(34) & 0.0796(28) & 0.0604(15)\\
IX & 0.0888(35) & 0.0817(29) & 0.0620(15)\\
\hline
\end{tabular}
\caption{\label{tab:latspac}Sets of lattice spacing values in fm used in our computations, together with their uncertainties given in parentheses, taken from Ref.~\cite{Carrasco:2014cwa}, Tabs.~8 and 9. Our preferred values are those from set I. The lattice spacing uncertainty is estimated from the variation of meson mass results when performing nine independent identical computations using sets I--IX. }
\end{table}

\subsubsection{Lattice spacing uncertainty}
\label{sec:latspac}
The lattice spacings that we have used in this paper were determined in Ref.~\cite{Carrasco:2014cwa}.
The authors of this paper used a rather involved procedure combining the lattice data from ensembles at different gauge couplings and quark masses, in the framework of chiral perturbation theory, with other inputs such as the values of the renormalization constant $Z_P$ and $r_0/a$.
The final values of the lattice spacings come with uncertainties, for both the relative scale setting (the values of $r_0/a$) and the absolute one (the value of $r_0$ in physical units).
However, we do not have access to the correlations between the different inputs to the procedure and hence, we cannot clearly separate absolute and relative lattice spacing uncertainties in our analysis.

Nevertheless, to have an estimate of the lattice spacing uncertainties of meson masses determined in this paper, we repeated our analysis on nine sets of lattice spacing values, given in Table~\ref{tab:latspac}. These sets of lattice spacings were taken from Table~8 and Table~9 of Ref.~\cite{Carrasco:2014cwa} and reflect different strategies that were followed to estimate the lattice spacing values in physical units.
The values of meson masses that we finally quote correspond to set I (the final values of lattice spacings from Eq.~(31) of Ref.~\cite{Carrasco:2014cwa}) and the scale setting uncertainty is taken as the second largest difference between set I and other sets (this procedure is quite similar to our procedure to estimate the fitting range uncertainty described in Section~\ref{sec:analysis}; it is even more conservative, since we discard only a single set, i.e.\ the remaining sets cover significantly more than a $1 \, \sigma$ region assuming a Gaussian distribution). Note that the different sets of lattice spacings influence our combined chiral and continuum extrapolations not only directly, but also indirectly, via the strange and charm quark mass tuning, as well as by influencing the values of the pion masses.

We list the obtained lattice spacing uncertainties in Table~\ref{tab:all}.
In most cases, they are smaller than the corresponding statistical and/or fitting range uncertainties with only few exceptions (e.g.\ $D_0^*(2400)$, $D_s$, $\eta_c(1S)$ [same order of magnitude] and $D_{s1}(2460)$, $\psi(2S)$ [even larger]). In any case, the obtained lattice spacing uncertainties are rather small, typically a few per mille, which is much smaller than the errors associated with the lattice spacings themselves. One of the main reasons for this is that we perform a tuning of the strange and charm quark mass at the beginning of each computation, which depends on the given set of lattice spacings, such that the resulting $D$ and $D_s$ meson masses agree with their experimental counterparts (cf.\ Section~\ref{SEC595}). Similarly, chiral and continuum extrapolations also depend on the given set of lattice spacings. Such a procedure is, of course, much more robust and powerful than e.g.\ performing a quark mass tuning only once for a given value of the lattice spacing. Note that we performed all computations for two values of the strange and for two values of the charm quark mass to be able to carry out such an evolved analysis (cf.\ also the discussion in our previous technical paper \cite{Kalinowski:2015bwa}).
As a bottom line, the 2-3\% differences between sets of lattice spacing values in Table~\ref{tab:latspac} translate to 2-3\% lattice spacing uncertainties on the meson mass splittings. Note that this is only a rough expectation and the true uncertainty depends on the variation of results from different sets of values in Table~\ref{tab:latspac}, e.g.\ in the ideal case of no difference in the respective mass splitting (indicating small cut-off effects in the sense of close to zero values of the coefficients $c^{(+,\pm)}$ in our combined chiral and continuum extrapolations), the resulting lattice spacing uncertainty could be zero (which explains that in certain cases the observed uncertainty in the mass splitting is smaller than the expected 2-3\%).

As a further cross-check of the reliability of our lattice spacing uncertainties, we also computed all meson masses considered in this paper using an alternative scale setting of Ref.~\cite{Alexandrou:2014sha}, based on the baryonic sector.
The following values of lattice spacings were obtained: $a=0.0936(13)$\ fm ($\beta=1.90$), $a=0.0823(10)$\ fm ($\beta=1.95$) and $a=0.0646(7)$\ fm ($\beta=2.10$).
Using these values in our procedure, we found results in perfect agreement with the ones obtained from set I of Table~\ref{tab:latspac} (within the quoted lattice spacing uncertainty).
This is reassuring and a strong indication that the lattice spacing uncertainties that we quote are realistic.

\subsubsection{Isospin breaking effects}
\label{sec:isospin}
Another type of systematic uncertainty is the breaking of isospin symmetry by electromagnetic effects (different electric charges of up and down quarks) and different masses of up and down quarks.
We are not able to address this issue directly and rigorously, since it would require working in a 1+1+1+1 setup with different light quark masses and an inclusion of electromagnetism, which goes beyond the scope of the present work (for pioneering work regarding such computations, cf. Refs.\ \cite{deDivitiis:2013xla,Borsanyi:2013lga}).

One simple possibility to estimate the magnitude of these effects is to take the experimentally known difference of the masses of the charged and neutral $D$ mesons, which is $\mathcal{O}(5)$ MeV.
Alternatively, we try to largely eliminate these effects by performing tuning of the strange and charm quark masses with either the neutral or the charged $\pi$, $K$ and $D$ meson masses. 
As discussed in Section\ \ref{sec:tuning} and as indicated by the perfect agreement of our lattice results for e.g.\ $D_s$, $\eta_c(1S)$ and $J/\psi(1S)$, we estimate the remaining systematic errors to be well beyond 5 MeV, which is negligible compared to the statistical errors and the fitting range uncertainties.
As an example, we note that in the case of the $D^*$ meson (see Table\ \ref{tab:all}), our crude method of estimating the splitting between the charged and neutral meson masses by comparing results from different input $\pi$, $K$ and $D$ meson masses for the strange and charm quark masses tuning, gives 4 MeV, as compared to the approximately 3.3 MeV experimental splitting.

\section{\label{sec:summary}Conclusions}

\begin{figure}[htb]
\begin{center}
\includegraphics[width=0.52\textwidth,angle=270]{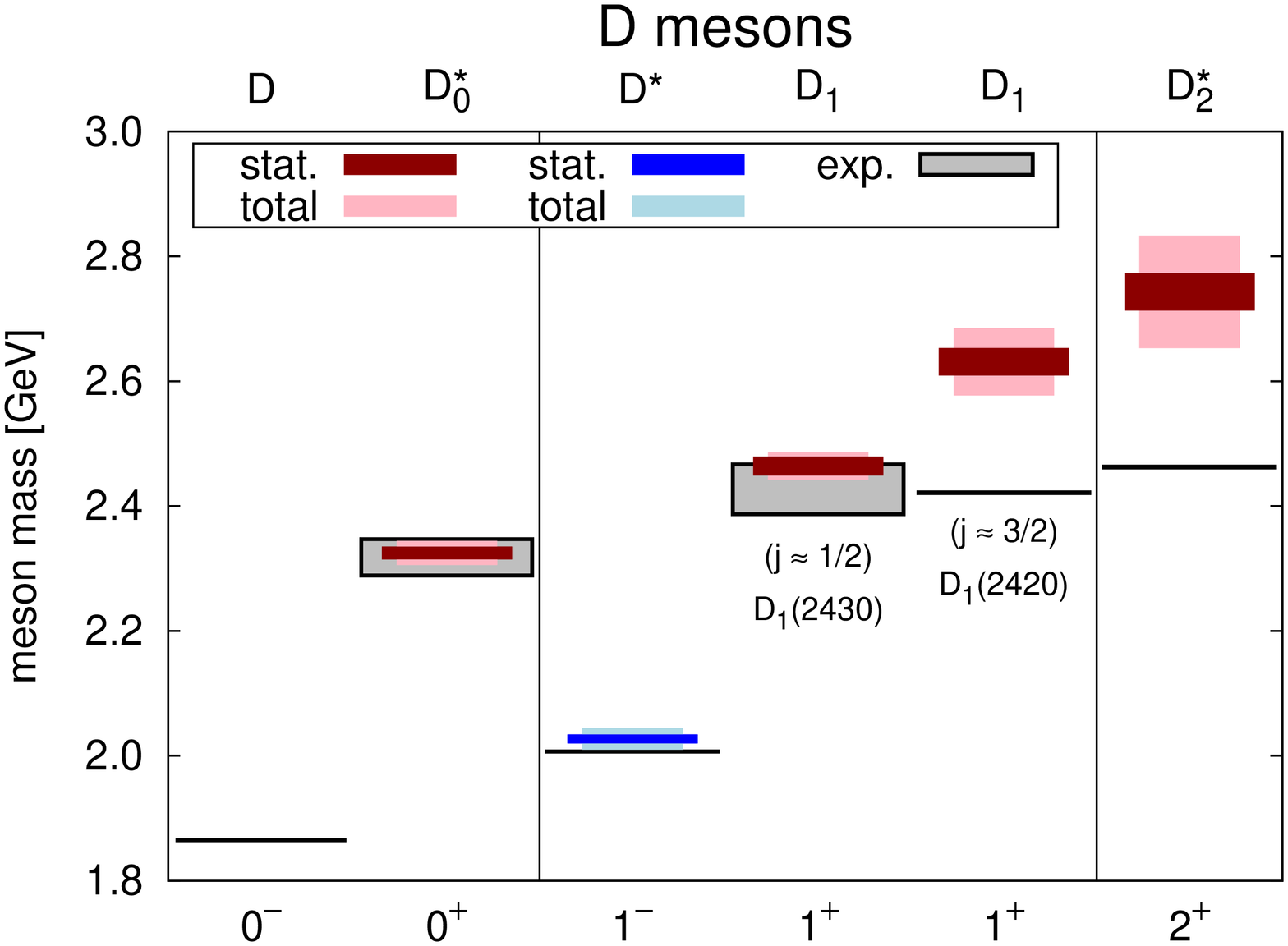}
\caption{\label{fig:specD}Chirally and continuum extrapolated lattice QCD results for $D$ mesons (dark colored boxes represent statistical, light colored boxes total errors; dark red/pink boxes correspond to $\mathcal{P}=+$, blue/light blue boxes to $\mathcal{P}=-$) compared to experimental results (black/gray boxes).}
\end{center}
\end{figure}

Our final results are collected in Table~\ref{tab:all} and in Figures~\ref{fig:specD}, \ref{fig:specDs} and \ref{fig:speccc}. In Table~\ref{tab:all}, we list meson names, $J^{\mathcal{P (C)}}$ quantum numbers, experimental results according to the PDG \cite{PDG} and our lattice QCD results, for which we show mean values, statistical errors, uncertainties associated with the choice of the fitting range and with the lattice spacing and total errors (statistical errors and fitting range uncertainties combined in quadrature). Since finite volume effects as well as isospin breaking effects are negligible as detailed in Sections~\ref{sec:FVE} and \ref{sec:isospin}, these total errors represent an estimate of the combined uncertainties of all possible sources. Consequently, the lattice QCD results (mean values and total errors) provided in Table~\ref{tab:all} can be compared in a direct and meaningful way to experiment or other theoretical work providing a full error budget. For states with total angular momentum $J = 2$, always two numerical results are available, one from the $E$ representation, the other from the $T_2$ representation. In all cases, these results are compatible within errors. As final results we use the ones from the $E$ representation, because for those, contamination by $J = 3$ states is excluded. Similarly, in Figures~\ref{fig:specD}, \ref{fig:specDs} and \ref{fig:speccc}, we summarize and compare our lattice QCD results (dark colored boxes represent statistical, light colored boxes total errors) to experimental results (gray boxes with black edges).

\addtolength{\tabcolsep}{-3pt} 
\begin{table}[p]
\centering
\begin{tabular}{lc|c|cccccc}
\hline
\multirow{2}*{name} & \multirow{2}*{$J^{\mathcal{P(C)}}$} & experiment & lattice & stat.\ & fitting & lat.spac. & total & \\
&  & (PDG) & QCD & error & error & error & error & \\
\hline
$D^0 \ $ ($(+,+)$ discr.) & $0^-$ & 1.86484(5)\phantom{0} & 1.8651 & 0.0033 & --  & 0.0003 & 0.0033 & (A) \\
$D^\pm \ $ ($(+,+)$ discr.) & $0^-$ & 1.86961(9)\phantom{0} & 1.8699 & 0.0033 & -- & 0.0003 & 0.0033 & (A) \\
\hline
$D_0^*(2400)^0$ & $0^+$ & 2.318(29)\phantom{00} & \phantom{0}2.325\phantom{0} & \phantom{0}0.010\phantom{0} & \phantom{0}0.012\phantom{0} & 0.010 & \phantom{0}0.019\phantom{0} & \\
$D_0^*(2400)^\pm$ & $0^+$ & 2.403(40)\phantom{00} & 2.330 & 0.010 & 0.012 & 0.009 &  0.018 & \\
$D^*(2007)^0$ & $1^-$ & 2.00697(8)\phantom{0} & 2.027 & 0.007 & 0.015 & 0.005 & 0.017 & \\
$D^*(2010)^\pm$ & $1^-$ & 2.01027(5)\phantom{0} & 2.031 & 0.007 & 0.015 & 0.005 & 0.017 & \\
$D_1(2430)^0 \ $ ($j \approx 1/2$) & $1^+$ & -- & 2.464 & 0.015 & 0.007 & 0.014 & 0.022 & \\
$D_1(2430)^\pm \ $ ($j \approx 1/2$) & $1^+$ & 2.427(40)\phantom{00} & 2.468 & 0.015 & 0.007 & 0.014 & 0.022 & \\
$D_1(2420)^0 \ $ ($j \approx 3/2$) & $1^+$ & 2.4214(6)\phantom{00} & 2.631 & 0.022 & 0.047 & 0.015 & 0.054 & (B) \\
$D_1(2420)^\pm \ $ ($j \approx 3/2$) & $1^+$ &  2.4232(24)\phantom{0} & 2.636 & 0.022 & 0.047 & 0.015 & 0.054 & (B) \\
$D_2^*(2460)^0 \ $ ($E$ rep.) & $2^+$ & 2.4626(6)\phantom{00} & 2.743 & 0.030 & 0.083 & 0.019 & 0.090 & (B) \\
$D_2^*(2460)^\pm \ $ ($E$ rep.) & $2^+$ &  2.4643(16)\phantom{0} & 2.747 & 0.030 & 0.083 & 0.020 & 0.090 & (B) \\
\hline
$D_s$ & $0^-$  & 1.96830(10) & 1.9679 & 0.0027 & 0.0020 &  0.0019 & 0.0040 & \\
$D_{s0}^*(2317)$ & $0^+$  & 2.3177(6)\phantom{00} & 2.390 & 0.011 & 0.021 & 0.008 & 0.025 & (C) \\
$D_s^*$ & $1^-$  & 2.1121(4)\phantom{00} & 2.123 & 0.0038 & 0.010 &  0.004 & 0.011 & \\
$D_{s1}(2460) \ $ ($j \approx 1/2$) & $1^+$  & 2.4595(6)\phantom{00} & 2.556 & 0.010 & 0.006 & 0.012 &  0.017 & (C) \\
$D_{s1}(2536) \ $ ($j \approx 3/2$) & $1^+$  & 2.53511(6)\phantom{0} & 2.617 & 0.022 & 0.085 &  0.016 & 0.089 & \\
$D_{s2}^*(2573)$ ($E$ rep.) & $2^+$  & 2.5719(8)\phantom{00} & 2.734 & 0.025 & 0.079  & 0.012 & 0.084 & \\
\hline
$\eta_c(1S)$ & $0^{-+}$ & 2.9836(6)\phantom{00} & 2.983 & 0.005 & 0.0008  & 0.004 & 0.006 \\
$\eta_c(2S)$ & $0^{-+}$ & 3.6392(12)\phantom{0} & 3.741 & 0.028 & 0.039  & 0.011 & 0.049 \\
$\chi_{c0}(1P)$ & $0^{++}$ & 3.41475(31) & 3.413 & 0.007 & 0.007  & 0.001 & 0.010\\
$J/\psi(1S)$ & $1^{--}$ & 3.09692(1)\phantom{0} & 3.096 & 0.005 & 0.0012  & 0.004 & 0.006\\
$\psi(2S)$ & $1^{--}$ & 3.68611(1)\phantom{0} & 3.647 & 0.030 & 0.040  & 0.042 & 0.065\\
$\psi(3770)$ & $1^{--}$ & 3.77315(33) & 3.833 & 0.020 & 0.026  & 0.001 & 0.033\\
? ($T_1$ rep.) & $1^{--}$ & -- & \multirow{2}*{3.951} & \multirow{2}*{0.022} & \multirow{2}*{0.028} & \multirow{2}*{0.022} & \multirow{2}*{0.042} & \multirow{2}*{(D,E)} \\
? ($T_1$ rep.) & $3^{--}$ & --                    & & & & & \\
$\chi_{c1}(1P)$ & $1^{++}$ & 3.51066(7)\phantom{0} & 3.513 & 0.007 & 0.014  & 0.004 & 0.016 \\
$h_c(1P)$ & $1^{+-}$ & 3.52538(11) & 3.536 & 0.009 & 0.012  & 0.006 & 0.016 \\
$\chi_{c2}(1P)$ ($E$ rep.) & $2^{++}$ & 3.55620(9)\phantom{0} & 3.565 & 0.009 & 0.028 & 0.006 & 0.030 \\
$\psi_2(1P)$ ($E$ rep.) & $2^{--}$  & -- & 3.885 & 0.019 & 0.028 & 0.014 & 0.037 & (D) \\
$\eta_{c2}(1P)$ ($E$ rep.) & $2^{-+}$  & -- & 4.023 & 0.029 & 0.085 & 0.017 & 0.091 & (D) \\ 
\hline
\end{tabular}
\caption{\label{tab:all}Summary of our lattice QCD results for $D$ meson, $D_s$ meson and charmonium masses in GeV (central value, statistical error, fitting range uncertainty, propagated lattice spacing uncertainty, total error). For comparison, we also list experimental results according to the PDG \cite{PDG}. (A)~Since the $(+,-)$ $D$ meson mass is used for valence charm quark mass tuning, not a prediction, just a check of universality of our two discretizations $(+,-)$ and $(+,+)$. (B)~Discrepancy of $\approx 3 \ldots 4 \, \sigma$ to experiment, presumably due to contamination by excited states. (C)~Discrepancy of $\approx 3 \ldots 8 \, \sigma$ to experiment, presumably an indication that both $D_{s0}^*(2317)$ and $D_{s1}(2460)$ have a non-$q \bar{q}$-like structure. (D)~Our theoretical prediction, no established experimental counterpart available. (E)~No clear assignment of total angular momentum possible, $J = 1$ favored, but $J = 3$ not completely ruled out. A candidate for an experimental result for the third $J^{\mathcal{P C}} = 1^{- -}$ excitation is $\psi(4040)$ with mass $4.039(1) \, \textrm{GeV}$. }
\end{table}
\addtolength{\tabcolsep}{3pt}

\addtolength{\tabcolsep}{-3pt} 
\begin{table}[p]
\centering
\begin{tabular}{lc|c|cccccc}
\hline
\multirow{2}*{name} & \multirow{2}*{$J^{\mathcal{P(C)}}$} & experiment & lattice & stat.\ & fitting & lat.spac. & total & \\
&  & (PDG) & QCD & error & error & error & error & \\
\hline
$D_0^*(2400)^0$                     $-D^0$ & $0^+$ & 0.453(29)\phantom{00}&0.461&0.011&0.010&0.010&0.018\\
$D_0^*(2400)^\pm$                   $-D^\pm$ & $0^+$ & 0.533(40)\phantom{00}&0.455&0.011&0.009&0.010&0.017\\
$D^*(2007)^0$                       $-D^0$   & $1^-$ & 0.14213(9)\phantom{0}&0.166&0.007&0.015&0.006&0.018\\
$D^*(2010)^\pm$                     $-D^\pm$ & $1^-$ & 0.14066(7)\phantom{0}&0.166&0.007&0.015&0.006&0.018\\
$D_1(2430)^0 \ $ ($j \approx 1/2$)  $-D^0$   & $1^+$ & --                &0.593&0.015&0.019&0.015&0.029\\
$D_1(2430)^\pm \ $ ($j \approx 1/2$)$-D^\pm$ & $1^+$ & 0.557(40)\phantom{00}&0.593&0.015&0.019&0.015&0.029\\
$D_1(2420)^0 \ $ ($j \approx 3/2$)  $-D^0$   & $1^+$ & 0.5566(6)\phantom{00}&0.764&0.022&0.024&0.015&0.036\\
$D_1(2420)^\pm \ $ ($j \approx 3/2$)$-D^\pm$ & $1^+$ & 0.5536(24)\phantom{0}&0.764&0.022&0.024&0.015&0.036\\
$D_2^*(2460)^0 \ $ ($E$ rep.)       $-D^0$   & $2^+$ & 0.5972(6)\phantom{00}&0.894&0.030&0.053&0.020&0.064\\
$D_2^*(2460)^\pm \ $ ($E$ rep.)     $-D^\pm$& $2^+$ & 0.5947(16)\phantom{0}&0.894&0.030&0.052&0.020&0.063\\
\hline
$D_{s0}^*(2317)$                    $-D_s$& $0^+$  & 0.3494(6)\phantom{00}&0.418&0.011&0.023&0.010&0.027\\
$D_s^*$                             $-D_s$& $1^-$  & 0.1440(4)\phantom{00}&0.158&0.0034&0.016&0.006&0.018\\
$D_{s1}(2460) \ $ ($j \approx 1/2$) $-D_s$& $1^+$  & 0.4912(6)\phantom{00}&0.588&0.011&0.012&0.013&0.021\\
$D_{s1}(2536) \ $ ($j \approx 3/2$) $-D_s$& $1^+$  & 0.56682(16)&0.643&0.022&0.086&0.018&0.091\\
$D_{s2}^*(2573)$ ($E$ rep.)         $-D_s$& $2^+$  & 0.6036(8)\phantom{00}&0.764&0.025&0.084&0.020&0.090\\
\hline
$\eta_c(2S)$              $-\eta_c(1S)$& $0^{-+}$ & 0.6556(13)\phantom{0}&0.765&0.028&0.044&0.020&0.055\\
$\chi_{c0}(1P)$            $-\eta_c(1S)$& $0^{++}$ & 0.4312(7)\phantom{00}&0.433&0.005&0.006&0.009&0.012\\
$J/\psi(1S)$               $-\eta_c(1S)$& $1^{--}$ & 0.1133(6)\phantom{00}&0.1194&0.0033&0.0014&0.0038&0.005\\
$\psi(2S)$                 $-\eta_c(1S)$& $1^{--}$ & 0.7025(6)\phantom{00}&0.663&0.030&0.040&0.034&0.060\\
$\psi(3770)$               $-\eta_c(1S)$& $1^{--}$ & 0.7896(7)\phantom{00}&0.855&0.018&0.031&0.008&0.036\\
? ($T_1$ rep.)             $-\eta_c(1S)$& $1^{--}$ & -- &\multirow{2}*{0.967}&\multirow{2}*{0.021}&\multirow{2}*{0.015}&\multirow{2}*{0.031}&\multirow{2}*{0.040}\\
? ($T_1$ rep.)             $-\eta_c(1S)$& $3^{--}$ & -- & &&&&\\
$\chi_{c1}(1P)$            $-\eta_c(1S)$& $1^{++}$ & 0.5271(6)\phantom{00}&0.530&0.006&0.014&0.014&0.021\\
$h_c(1P)$                  $-\eta_c(1S)$& $1^{+-}$ & 0.5418(6)\phantom{00}&0.554&0.009&0.011&0.015&0.021\\
$\chi_{c2}(1P)$ ($E$ rep.) $-\eta_c(1S)$& $2^{++}$ & 0.5726(6)\phantom{00}&0.584&0.008&0.028&0.015&0.033\\
$\psi_2(1P)$ ($E$ rep.)    $-\eta_c(1S)$& $2^{--}$ & --                &0.901&0.019&0.028&0.023&0.041\\
$\eta_{c2}(1P)$ ($E$ rep.) $-\eta_c(1S)$& $2^{-+}$ & --                &1.041&0.029&0.086&0.026&0.094\\
\hline
$                 D_s-D^\pm$& $0^-$  & 0.10346(7)\phantom{0}&0.1025&0.0028&0.0023&0.0018&0.004\\

\end{tabular}
\caption{\label{tab:all_diff}
Summary of our lattice QCD results for $D$ meson, $D_s$ meson and charmonium masses splittings in GeV (central value, statistical error, fitting range uncertainty, propagated lattice spacing uncertainty, total error). 
Each splitting is given with respect to the ground state $D$ meson, $D_s$ meson or charmonium.
In addition, we give the mass splitting between the ground state $D_s$ and $D$ mesons.
We also list experimental results according to the PDG \cite{PDG}. 
See also remarks in Tab.~\ref{tab:all} for specific states where we comment on possible sources of discrepancy with respect to the experimental result.
}
\end{table}
\addtolength{\tabcolsep}{3pt}

In Table~\ref{tab:all_diff}, we provide in the same way mass differences of mesons as discussed in Section~\ref{SEC811}.
In most cases, we find that the statistical uncertainty is slightly reduced, usually quite similar to the larger of the two statistical uncertainties associated with the individual meson masses, while the other errors are usually unaffected. This implies that the typical total error of a meson mass difference is similar to the larger of the two total errors of the individual meson masses. Obviously, this is smaller than naively assuming uncorrelated data, i.e.\ adding the total errors of the two individual meson masses in quadrature.


\subsection{$D$ mesons (charm-light mesons)}

The $D$ meson ($J^\mathcal{P} = 0^-$), i.e.\ the lightest charm-light meson, plays a special role in our computation. The $(+,-)$ lattice QCD results are used to tune the valence charm quark mass and, hence, cannot be considered as predictions. The corresponding $(+,+)$ results are in excellent agreement with experiment (cf.\ the first two lines in Table~\ref{tab:all}), which is a convincing test of universality, but again they are not lattice QCD predictions of the $D^0$ and $D^\pm$ meson masses. Moreover, we do not quote fitting interval uncertainties, since the effective mass plateaus for this state are extremely long and clear and the resulting uncertainties are, hence, negligible. The latter is one of the reasons for using this state for the valence charm quark mass tuning procedure discussed in Section~\ref{SEC595}.

For $D^\ast$ ($J^\mathcal{P} = 1^-$) as well as for $D_0^*(2400)$ and $D_1(2430)$ ($J^\mathcal{P} = 0^+$ and $1^+$; both $j \approx 1 /2$, where $j$ denotes the total angular momentum of the light degrees of freedom), we find agreement with experiment. Since the latter two states are quite unstable, a more rigorous computation of their masses (and also widths) would require one to treat them as resonances and use e.g.\ the L\"uscher's method in combination with suitable four-quark creation operators. Such a computation, however, is significantly more challenging and computer time consuming than the methods used in this work, which are based on creation operators of quark-antiquark type only. A recent exploratory resonance study of $D_0^*(2400)$ and $D_1(2430)$ using a single lattice spacing ($a \approx 0.124 \, \textrm{fm}$) and a single pion mass ($m_\pi \approx 266 \, \textrm{MeV}$) can be found in Ref.~\cite{Mohler:2012na}.

The resulting masses for the remaining two extracted states, $D_1(2420)$ and $D_2^*(2460)$ ($J^\mathcal{P} = 1^+$ and $2^+$; both $j \approx 3 /2$), are around $3 \ldots 4 \, \sigma$ above the experimental results\footnote{In this context, it is interesting to note that we have observed a very similar behavior for the corresponding $j \approx 3/2$ $B$ mesons \cite{Michael:2010aa}.}. We attribute this discrepancy to technical problems, in particular a not sufficiently optimized choice of the meson creation operators (cf.\ the $E$ and $T_2$ segments of Table~\ref{tab.operators}) and thus a contamination of effective mass plateaus by excited states. This is indicated by our procedure to determine the fitting range uncertainties, which exhibits a systematic decrease of the extracted meson masses when using fitting intervals starting at increasingly larger temporal separations. Before an unambiguous plateau is reached, the signal is lost in statistical noise. This interpretation is additionally supported when considering only gauge link ensembles with the finest lattice spacing (D ensembles). There, it is easier to identify effective mass plateaus, because a given temporal range in physical units corresponds to a larger number of discrete lattice separations. A chiral extrapolation from the D ensembles only results in masses compatible with experiments for both $D_1(2420)$ and $D_2^*(2460)$.


\subsection{$D_s$ mesons (charm-strange mesons)}

\begin{figure}[htb]
\begin{center}
\includegraphics[width=0.52\textwidth,angle=270]{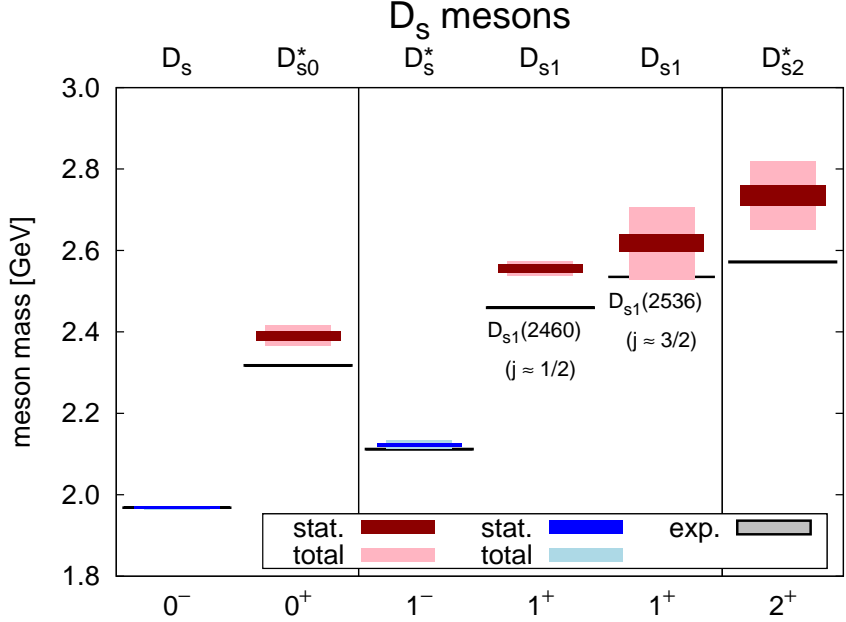}
\caption{\label{fig:specDs}Chirally and continuum extrapolated lattice QCD results for $D_s$ mesons (dark colored boxes represent statistical, light colored boxes total errors; dark red/pink boxes correspond to $\mathcal{P}=+$, blue/light blue boxes to $\mathcal{P}=-$) compared to experimental results (black/gray boxes).}
\end{center}
\end{figure}

The overall picture for charm-strange mesons is quite similar. We have obtained very robust results for the $D_s$ ($J^\mathcal{P} = 0^-$) and the $D_s^\ast$ ($J^\mathcal{P} = 1^-$) in perfect agreement with experiment.

For $D_{s0}^*$ and $D_{s1}(2460)$ ($J^\mathcal{P} = 0^+$ and $1^+$; both $j \approx 1 /2$), our resulting lattice QCD masses are significantly above their experimental counterparts with fitting range uncertainties which seem to be well controlled. Similar discrepancies to experimental results have been found in other lattice QCD studies based on quark-antiquark creation operators \cite{Mohler:2011ke} and in quark model calculations \cite{Ebert:2009ua}. A possible explanation could be that both the $D_{s0}^*$ and the $D_{s1}(2460)$ have a structure quite different from an ordinary quark-antiquark pair, e.g.\ a mesonic molecule or a diquark-antidiquark structure (cf.\ e.g.\ \cite{Dmitrasinovic:2005gc,Cleven:2014oka} where such scenarios are discussed theoretically). It would be interesting to include corresponding four-quark creation operators into our lattice QCD computation and to see whether the masses of these two states decrease and the associated eigenvector components indicate a sizable four-quark contribution. In a recent lattice QCD study using both two- and four-quark creation operators, strong evidence has been presented that four-quark creation operators are essential to properly resolve these states \cite{Lang:2014yfa}.

The resulting masses for the remaining two extracted states, $D_{s1}(2536)$ and $D_{s2}^*(2573)$ ($J^\mathcal{P} = 1^+$ and $2^+$; both $j \approx 3 /2$), are slightly above the experimental results, but still compatible within the total errors. Analogous to the corresponding $D$ mesons, we expect that an optimization of the meson creation errors could help to reduce the rather large fitting range uncertainty and thus lead to more precise predictions.


\subsection{Charmonium (charm-charm mesons)}

\begin{figure}[htb]
\begin{center}
\includegraphics[width=0.52\textwidth,angle=270]{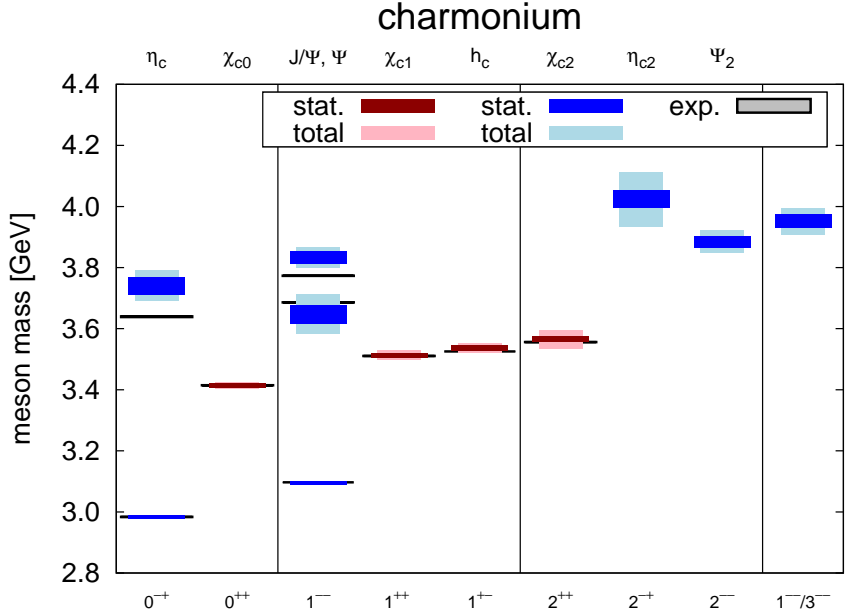}
\caption{\label{fig:speccc}Chirally and continuum extrapolated lattice QCD results for charmonium (dark colored boxes represent statistical, light colored boxes total errors; dark red/pink boxes correspond to $\mathcal{P}=+$, blue/light blue boxes to $\mathcal{P}=-$) compared to experimental results (black/gray boxes).}
\end{center}
\end{figure}

We were able to determine the masses of several states rather precisely and in excellent agreement with experimental results, including the $\eta_c(1S)$, $\chi_{c0}(1P)$, $J/\psi(1S)$, $\chi_{c1}(1P)$, $h_c(1P)$ and $\chi_{c2}(1P)$ (the ground states in the sectors $J^\mathcal{P C} = 0^{- +}, 0^{+ +}, 1^{- -}, 1^{+ +}, 1^{+ -}, 2^{+ +}$). For the masses of the excited states $\eta_c(2S)$ ($J^\mathcal{P C} = 0^{- +}$) and $\psi(2S)$, $\psi(3770)$ (both $J^\mathcal{P C} = 1^{- -}$), the total errors are larger, but results are consistent with experiment.

We were also able to predict three states, which have no established experimental counterparts yet: $\eta_{c2}(1P)$ ($J^{\mathcal{P C}} = 2^{- +}$), $\psi_2(1P)$ ($J^{\mathcal{P C}} = 2^{- -}$) and a state with either $J^{\mathcal{P C}} = 1^{- -}$ or $J^{\mathcal{P C}} = 3^{- -}$. For the latter, we prefer to interpret it as the third excitation in the $J^{\mathcal{P C}} = 1^{- -}$ sector, as discussed in Section~\ref{sec:ccT1}. These predictions are qualitatively similar to those from another recent lattice QCD study using only a single unphysically heavy $u/d$ quark mass ($m_\pi \approx 396 \, \textrm{MeV}$) and a single lattice spacing \cite{Liu:2012ze} and with a calculation based on Dyson-Schwinger/Bethe-Salpeter equations \cite{Fischer:2014cfa}. It should be noted that the fitting range uncertainties for these three predictions are comparably large. Therefore, the possibility of contamination by excited states cannot be fully excluded. We compare our results with corresponding chiral extrapolations from D ensemble data only, where effective mass plateaus can be identified most reliably. We find agreement in all three cases, which is reassuring and indicates that the total errors quoted in Table~\ref{tab:all} realistically reflect the uncertainties of these three predictions.


\subsection{Summary and outlook}

We have computed the low-lying $D$ meson, $D_s$ meson and charmonium spectra using Wilson twisted mass lattice QCD with 2+1+1 flavors of sea quarks and meson creation operators of quark-antiquark type. We were able to determine 5 $D$ meson masses, 6 $D_s$ meson masses and 12 charmonium masses, where three of the latter are theoretical predictions with no currently existing or clearly identified experimental counterparts. We have performed computations on nine different gauge link ensembles at three different lattice spacings in the range $a \approx 0.0619 \ldots 0.0885 \, \textrm{fm}$ and pion masses $m_\pi \approx 224 \ldots 468 \, \textrm{MeV}$. Moreover, all computations have been done with two different valence quark discretizations. Using the resulting 18 mass values for each meson state allowed us to perform combined continuum and chiral extrapolations. These extrapolations seem to be very solid and trustworthy as e.g.\ demonstrated by the corresponding plots for $D_s$ and $\eta_c(1S)$ (which have very small statistical errors; cf.\ Figure~\ref{fig:Ds} and Figure~\ref{fig:eta_c}). Similarly, we have performed computations also with two slightly different values of the valence strange quark mass as well as of the valence charm quark mass for each meson state. This allowed us not only to tune these quark masses precisely to their physical values, but also to crudely estimate electromagnetic effects and to reduce errors associated with the lattice spacing. The latter have been studied in detail by repeating the whole analysis for several different ETMC strategies to determine the lattice spacing and, thus, set the scale. Finally, by comparison with analogous computations on another ensemble with a rather small volume, we were able to demonstrate that finite volume effects are negligible. The masses collected in Tables~\ref{tab:all} and \ref{tab:all_diff} and in Figures~\ref{fig:specD}, \ref{fig:specDs} and \ref{fig:speccc} are listed and shown with a total error accounting for all these effects and, hence, can be compared to existing or future experimental data in a direct and meaningful way.

We were also able to clearly distinguish the two close-by $J^\mathcal{P} = 1^+$ $D$ meson states $D_1(2430)$ and $D_1(2420)$ and the two analogous $D_s$ meson states $D_{s1}(2460)$ and $D_{s1}(2536)$ according to their light total angular momentum $j \approx 1/2$ and $j \approx 3/2$. This is important not only when comparing with experimental results, but also when studying semileptonic decays $B \rightarrow D^{\ast \ast}$, where $D^{\ast \ast} = \{ D_0^\ast(2400) \, , \, D_1(2430) \, , \, D_1(2420) \, , \, D_2^\ast(2460) \}$. Such decays are of particular interest, because there is a long-standing conflict between theory and experiment (QCD sum rules, model calculations), the so-called 1/2 versus 3/2 puzzle \cite{Bigi:2007qp}. Recently, decays $B \rightarrow D_0^\ast$ and $B \rightarrow D_2^\ast$ have been studied on the same gauge link ensembles for the first time with non-static $b$ and $c$ quarks \cite{Atoui:2013sca,Atoui:2013ksa}. Using the same techniques and the results on $D_1(2430)$ and $D_1(2420)$ presented in this work will allow one to also include $B \rightarrow D_1$. This in turn should provide important insights regarding the 1/2 versus 3/2 puzzle from theoretical methods based on first principles.

Our future plans are mainly focused on studying specific $D$ mesons, $D_s$ mesons or charmonium states with a larger set of creation operators, including in particular four-quark operators of either mesonic molecule, diquark-antidiquark or two-meson type. Existing lattice QCD studies \cite{Gong:2011nr,Mohler:2012na,Liu:2012zya,Prelovsek:2013cra,
Prelovsek:2013xba,Mohler:2013rwa,Ikeda:2013vwa,Lang:2014yfa,Prelovsek:2014swa,Guerrieri:2014nxa,Padmanath:2015era,Lang:2015sba,Moir:2016srx} strongly indicate that such techniques are mandatory for a rigorous treatment of certain mesons, e.g.\ the tetraquark candidates $D_{s0}^\ast(2317)$ and $D_{s1}(2460)$. We are currently in the process of developing such techniques, in particular to efficiently compute corresponding correlation matrix elements \cite{Alexandrou:2012rm,Abdel-Rehim:2014zwa,Berlin:2015faa}.



\section*{Acknowledgments}

It is a pleasure to thank V.~O.~Galkin for many useful discussions.
Moreover, we acknowledge useful conversations with J.~Berlin, B.~Blossier,
F.~Giacosa, M.~F.~M.~Lutz, O.~P\`ene, D.~H.~Rischke and R.~Sommer.

M.K.\ and M.W.\ acknowledge support by the Emmy Noether Programme of the
DFG (German Research Foundation), grant WA 3000/1-1.

This work was supported in part by the Helmholtz International Center for
FAIR within the framework of the LOEWE program launched by the State of
Hesse.

K.C.\ was supported in part by the Deutsche Forschungsgemeinschaft (DFG), project nr. CI 236/1-1 (Sachbeihilfe).

Calculations on the LOEWE-CSC and on the on the FUCHS-CSC high-performance
computer of the Frankfurt University were conducted for this research. We
would like to thank HPC-Hessen, funded by the State Ministry of Higher
Education, Research and the Arts, for programming advice.



\begin{thebibliography}{99}

\bibitem{PDG}
  K.~A.~Olive {\it et al.} [Particle Data Group],
  ``2014 Review of particle physics,''
  Chin.\ Phys.\ C, {\bf 38}, 090001 (2014). 

\bibitem{Aubert:2003fg} 
  B.~Aubert {\it et al.} [BaBar Collaboration],
  ``Observation of a narrow meson decaying to $D_s^+ \pi^0$ at a mass of $2.32 \, \textrm{GeV}/c^2$,''
  Phys.\ Rev.\ Lett.\ {\bf 90}, 242001 (2003)
  [hep-ex/0304021].

\bibitem{Besson:2003cp} 
  D.~Besson {\it et al.} [CLEO Collaboration],
  ``Observation of a narrow resonance of mass $2.46 \, \textrm{GeV}/c^2$ decaying to $D_s^{\ast +} \pi^0$ and confirmation of the $D_{sJ}^\ast(2317)$ state,''
  Phys.\ Rev.\ D {\bf 68}, 032002 (2003)
  [hep-ex/0305100].

\bibitem{Choi:2003ue} 
  S.~K.~Choi {\it et al.} [Belle Collaboration],
  ``Observation of a narrow charmonium-like state in exclusive $B^\pm \rightarrow K^\pm \pi^+ \pi^- J/\Psi$ decays,''
  Phys.\ Rev.\ Lett.\ {\bf 91}, 262001 (2003)
  [hep-ex/0309032].

\bibitem{Ebert:2009ua} 
  D.~Ebert, R.~N.~Faustov and V.~O.~Galkin,
  ``Heavy-light meson spectroscopy and Regge trajectories in the relativistic quark model,''
  Eur.\ Phys.\ J.\ C {\bf 66}, 197 (2010)
  [arXiv:0910.5612 [hep-ph]].

\bibitem{Eshraim:2014eka} 
  W.~I.~Eshraim, F.~Giacosa and D.~H.~Rischke,
  ``Phenomenology of charmed mesons in the extended Linear Sigma Model,''
  Eur.\ Phys.\ J.\ A {\bf 51}, 112 (2015)
  [arXiv:1405.5861 [hep-ph]].

\bibitem{Fischer:2014cfa} 
  C.~S.~Fischer, S.~Kubrak and R.~Williams,
  ``Spectra of heavy mesons in the Bethe-Salpeter approach,''
  Eur.\ Phys.\ J.\ A {\bf 51}, 10 (2015)
  [arXiv:1409.5076 [hep-ph]].
  
  





\bibitem{Mohler:2015zsa}
  D.~Mohler,
  ``Recent progress in lattice calculations of properties of open-charm mesons,''
  arXiv:1508.02753 [hep-lat].

\bibitem{DeTar:2011nn} 
  C.~DeTar,
  ``Charmonium spectroscopy from Lattice QCD,''
  Int.\ J.\ Mod.\ Phys.\ Conf.\ Ser.\  {\bf 02}, 31 (2011)
  [arXiv:1101.0212 [hep-lat]].

\bibitem{Prelovsek:2015fra} 
  S.~Prelovsek,
  ``Lattice studies of charmonia and exotics,''
  arXiv:1508.07322 [hep-lat].

\bibitem{Weber:2013eba} 
  M.~Wagner, S.~Diehl, T.~Kuske and J.~Weber,
  ``An introduction to lattice hadron spectroscopy for students without quantum field theoretical background,''
  arXiv:1310.1760 [hep-lat].

\bibitem{Dudek:2007wv} 
  J.~J.~Dudek, R.~G.~Edwards, N.~Mathur and D.~G.~Richards,
  ``Charmonium excited state spectrum in lattice QCD,''
  Phys.\ Rev.\ D {\bf 77}, 034501 (2008)
  [arXiv:0707.4162 [hep-lat]].

\bibitem{Dong:2009wk} 
  S.~J.~Dong {\it et al.},
  ``The charmed-strange meson spectrum from overlap fermions on domain wall dynamical fermion configurations,''
  PoS Lattice {\bf 2009}, 090 (2009)
  [arXiv:0911.0868 [hep-ph]].

\bibitem{Burch:2009az} 
  T.~Burch {\it et al.},
  ``Quarkonium mass splittings in three-flavor lattice QCD,''
  Phys.\ Rev.\ D {\bf 81}, 034508 (2010)
  [arXiv:0912.2701 [hep-lat]].

\bibitem{Dudek:2010wm} 
  J.~J.~Dudek, R.~G.~Edwards, M.~J.~Peardon, D.~G.~Richards and C.~E.~Thomas,
  ``Toward the excited meson spectrum of dynamical QCD,''
  Phys.\ Rev.\ D {\bf 82}, 034508 (2010)
  [arXiv:1004.4930 [hep-ph]].

\bibitem{Mohler:2011ke} 
  D.~Mohler and R.~M.~Woloshyn,
  ``$D$ and $D_s$ meson spectroscopy,''
  Phys.\ Rev.\ D {\bf 84}, 054505 (2011)
  [arXiv:1103.5506 [hep-lat]].

\bibitem{Namekawa:2011wt} 
  Y.~Namekawa {\it et al.} [PACS-CS Collaboration],
  ``Charm quark system at the physical point of 2+1 flavor lattice QCD,''
  Phys.\ Rev.\ D {\bf 84}, 074505 (2011)
  [arXiv:1104.4600 [hep-lat]].

\bibitem{Bali:2011dc} 
  G.~Bali {\it et al.},
  ``Spectra of heavy-light and heavy-heavy mesons containing charm quarks, including higher spin states for $N_f=2+1$,''
  PoS LATTICE {\bf 2011}, 135 (2011)
  [arXiv:1108.6147 [hep-lat]].

\bibitem{Bali:2011rd} 
  G.~S.~Bali, S.~Collins and C.~Ehmann,
  ``Charmonium spectroscopy and mixing with light quark and open charm states from $n_F=2$ lattice QCD,''
  Phys.\ Rev.\ D {\bf 84}, 094506 (2011)
  [arXiv:1110.2381 [hep-lat]].

\bibitem{Liu:2012ze} 
  L.~Liu {\it et al.} [Hadron Spectrum Collaboration],
  ``Excited and exotic charmonium spectroscopy from lattice QCD,''
  JHEP {\bf 1207}, 126 (2012)
  [arXiv:1204.5425 [hep-ph]].

\bibitem{Yang:2012mya} 
  Y.~B.~Yang {\it et al.} [CLQCD Collaboration],
  ``Lattice study on $\eta_{c2}$ and $X(3872)$,''
  Phys.\ Rev.\ D {\bf 87}, 014501 (2013)
  [arXiv:1206.2086 [hep-lat]].

\bibitem{Dowdall:2012ab} 
  R.~J.~Dowdall {\it et al.},
  ``Precise heavy-light meson masses and hyperfine splittings from lattice QCD including charm quarks in the sea,''
  Phys.\ Rev.\ D {\bf 86}, 094510 (2012)
  [arXiv:1207.5149 [hep-lat]].

\bibitem{Bali:2012ua} 
  G.~Bali, S.~Collins and P.~Perez-Rubio,
  ``Charmed hadron spectroscopy on the lattice for $N_f=2+1$ flavours,''
  J.\ Phys.\ Conf.\ Ser.\  {\bf 426}, 012017 (2013)
  [arXiv:1212.0565 [hep-lat]].

\bibitem{Moir:2013ub} 
  G.~Moir {\it et al.},
  ``Excited spectroscopy of charmed mesons from lattice QCD,''
  JHEP {\bf 1305}, 021 (2013)
  [arXiv:1301.7670 [hep-ph]].

\bibitem{Galloway:2014tta} 
  B.~A.~Galloway {\it et al.} [HPQCD collaboration],
  ``Radial and orbital excitation energies of charmonium,''
  PoS LATTICE {\bf 2014}, 092 (2014)
  [arXiv:1411.1318 [hep-lat]].

\bibitem{Bali:2015lka} 
  P.~Perez-Rubio, S.~Collins and G.~S.~Bali,
  ``Charmed baryon spectroscopy and light flavor symmetry from lattice QCD,''
  Phys.\ Rev.\ D {\bf 92}, 034504 (2015)
  [arXiv:1503.08440 [hep-lat]].

\bibitem{Cheung:2016bym} 
  G.~K.~C.~Cheung, C.~O'Hara, G.~Moir, M.~Peardon, S.~M.~Ryan, C.~E.~Thomas and D.~Tims,
  ``Excited and exotic charmonium, $D_s$ and $D$ meson spectra for two light quark masses from lattice QCD,''
  arXiv:1610.01073 [hep-lat].  
  
\bibitem{Prelovsek:2011nk} 
  S.~Prelovsek, C.~B.~Lang and D.~Mohler,
  ``Scattering phase shift and resonance properties on the lattice: an introduction,''
  Bled Workshops in Physics, vol.\ 12, no.\ 1
  [arXiv:1110.4520 [hep-ph]].

\bibitem{Gong:2011nr} 
  M.~Gong {\it et al.},
  ``Study of the scalar charmed-strange meson $D_{s0}^*(2317)$ with chiral fermions,''
  PoS Lattice {\bf 2010}, 106 (2014)
  [arXiv:1103.0589 [hep-lat]].

\bibitem{Mohler:2012na} 
  D.~Mohler, S.~Prelovsek and R.~M.~Woloshyn,
  ``$D \pi$ scattering and $D$ meson resonances from lattice QCD,''
  Phys.\ Rev.\ D {\bf 87}, 034501 (2013)
  [arXiv:1208.4059 [hep-lat]].

\bibitem{Liu:2012zya} 
  L.~Liu {\it et al.},
  ``Interactions of charmed mesons with light pseudoscalar mesons from lattice QCD and implications on the nature of the $D_{s0}^\ast(2317)$,''
  Phys.\ Rev.\ D {\bf 87}, 014508 (2013)
  [arXiv:1208.4535 [hep-lat]].

\bibitem{Prelovsek:2013cra} 
  S.~Prelovsek and L.~Leskovec,
  ``Evidence for $X(3872)$ from $D D^\ast$ scattering on the lattice,''
  Phys.\ Rev.\ Lett.\ {\bf 111}, 192001 (2013)
  [arXiv:1307.5172 [hep-lat]].

\bibitem{Prelovsek:2013xba} 
  S.~Prelovsek and L.~Leskovec,
  ``Search for $Z_c^+(3900)$ in the $1^{+ -}$ channel on the lattice,''
  Phys.\ Lett.\ B {\bf 727}, 172 (2013)
  [arXiv:1308.2097 [hep-lat]].

  

  
  
  
\bibitem{Mohler:2013rwa} 
  D.~Mohler {\it et al.},
  ``$D_{s0}^\ast(2317)$ meson and $D$ meson-kaon scattering from lattice QCD,''
  Phys.\ Rev.\ Lett.\ {\bf 111}, 222001 (2013)
  [arXiv:1308.3175 [hep-lat]].

\bibitem{Ikeda:2013vwa}
  Y.~Ikeda {\it et al.} [HAL QCD Collaboration],
  ``Charmed tetraquarks $T_{cc}$ and $T_{cs}$ from dynamical lattice QCD simulations,''
  Phys.\ Lett.\ B {\bf 729}, 85 (2014)
  [arXiv:1311.6214 [hep-lat]].

\bibitem{Lang:2014yfa} 
  C.~B.~Lang {\it et al.},
  ``$D_s$ mesons with $D K$ and $D^\ast K$ scattering near threshold,''
  Phys.\ Rev.\ D {\bf 90}, 034510 (2014)
  [arXiv:1403.8103 [hep-lat]].

\bibitem{Prelovsek:2014swa} 
  S.~Prelovsek, C.~B.~Lang, L.~Leskovec and D.~Mohler,
  ``Study of the $Z_c^+$ channel using lattice QCD,''
  Phys.\ Rev.\ D {\bf 91}, 014504 (2015)
  [arXiv:1405.7623 [hep-lat]].

\bibitem{Guerrieri:2014nxa} 
  A.~L.~Guerrieri {\it et al.},
  ``Flavored tetraquark spectroscopy,''
  PoS LATTICE {\bf 2014}, 106 (2015)
  [arXiv:1411.2247 [hep-lat]].

\bibitem{Padmanath:2015era} 
  M.~Padmanath, C.~B.~Lang and S.~Prelovsek,
  ``X(3872) and Y(4140) using diquark-antidiquark operators with lattice QCD,''
  Phys.\ Rev.\ D {\bf 92}, 034501 (2015)
  [arXiv:1503.03257 [hep-lat]].  
  
\bibitem{Lang:2015sba} 
  C.~B.~Lang, L.~Leskovec, D.~Mohler and S.~Prelovsek,
  ``Vector and scalar charmonium resonances with lattice QCD,''
  JHEP {\bf 1509}, 089 (2015)
  [arXiv:1503.05363 [hep-lat]].
  
  
\bibitem{Moir:2016srx} 
  G.~Moir, M.~Peardon, S.~M.~Ryan, C.~E.~Thomas and D.~J.~Wilson,
  ``Coupled-Channel $D\pi$, $D\eta$ and $D_{s}\bar{K}$ Scattering from Lattice QCD,''
  arXiv:1607.07093 [hep-lat].  
  
  
\bibitem{Carrasco:2014cwa} 
  N.~Carrasco {\it et al.} [ETM Collaboration],
  ``Up, down, strange and charm quark masses with $N_f = 2+1+1$ twisted mass lattice QCD,''
  Nucl.\ Phys.\ B {\bf 887}, 19 (2014)
  [arXiv:1403.4504 [hep-lat]].
 
  
\bibitem{Alexandrou:2012rm} 
  C.~Alexandrou {\it et al.} [ETM Collaboration],
  ``Lattice investigation of the scalar mesons $a_0(980)$ and $\kappa$ using four-quark operators,''
  JHEP {\bf 1304}, 137 (2013)
  [arXiv:1212.1418 [hep-lat]].

\bibitem{Abdel-Rehim:2014zwa} 
  A.~Abdel-Rehim {\it et al.},
  ``Investigation of the tetraquark candidate $a_0(980)$: technical aspects and preliminary results,''
  PoS LATTICE {\bf 2014}, 104 (2014)
  [arXiv:1410.8757 [hep-lat]].

\bibitem{Berlin:2015faa} 
  J.~Berlin, A.~Abdel-Rehim, C.~Alexandrou, M.~D.~Brida, M.~Gravina and M.~Wagner,
  ``Computation of correlation matrices for tetraquark candidates with $J^P = 0^+$ and flavor structure $q_1 \bar{q}_2 q_3 \bar{q}_3$,''
  PoS LATTICE {\bf 2015}, 096 (2015)
  [arXiv:1508.04685 [hep-lat]].
  
  
\bibitem{Kalinowski:2012re} 
  M.~Kalinowski and M.~Wagner [ETM Collaboration],
  ``Strange and charm meson masses from twisted mass lattice QCD,''
  PoS ConfinementX, 303 (2012)
  [arXiv:1212.0403 [hep-lat]].

\bibitem{Kalinowski:2013wsa} 
  M.~Kalinowski and M.~Wagner [ETM Collaboration],
  ``Masses of mesons with charm valence quarks from 2+1+1 flavor twisted mass lattice QCD,''
  Acta Phys.\ Polon.\ Supp.\ {\bf 6}, 991 (2013)
  [arXiv:1304.7974 [hep-lat]].

\bibitem{Wagner:2013laa} 
  M.~Kalinowski and M.~Wagner [ETM Collaboration],
  ``Twisted mass lattice computation of charmed mesons with focus on $D^{**}$,''
  PoS LATTICE {\bf 2013}, 241 (2014)
  [arXiv:1310.5513 [hep-lat]].

\bibitem{Cichy:2015tma} 
  K.~Cichy, M.~Kalinowski and M.~Wagner,
  ``Mass spectra of mesons containing charm quarks -- continuum limit results from twisted mass fermions,''
  PoS LATTICE {\bf 2015}, 093 (2015)
  [arXiv:1510.07862 [hep-lat]].

\bibitem{Kalinowski:2015bwa} 
  M.~Kalinowski and M.~Wagner,
  ``Masses of $D$ mesons, $D_s$ mesons and charmonium states from twisted mass lattice QCD,''
  Phys.\ Rev.\ D {\bf 92}, 094508 (2015)
  [arXiv:1509.02396 [hep-lat]].

\bibitem{Baron:2008xa}
  R.~Baron {\it et al.} [ETM Collaboration],
  ``Status of ETMC simulations with $N_f = 2+1+1$ twisted mass fermions,''
  PoS {\bf LATTICE2008}, 094 (2008)
  [arXiv:0810.3807 [hep-lat]].

\bibitem{Jansen:2009xp} 
  K.~Jansen and C.~Urbach,
  ``tmLQCD: a program suite to simulate Wilson twisted mass lattice QCD,''
  Comput.\ Phys.\ Commun.\ {\bf 180}, 2717 (2009)
  [arXiv:0905.3331 [hep-lat]].

\bibitem{Baron:2009zq} 
  R.~Baron {\it et al.} [ETM Collaboration],
  ``First results of ETMC simulations with $N_f = 2+1+1$ maximally twisted mass fermions,''
  PoS {\bf LATTICE2009}, 104 (2009)
  [arXiv:0911.5244 [hep-lat]].

  
\bibitem{Baron:2010bv} 
  R.~Baron {\it et al.} [ETM Collaboration],
  ``Light hadrons from lattice QCD with light $(u,d)$, strange and charm dynamical quarks,''
  JHEP {\bf 1006}, 111 (2010)
  [arXiv:1004.5284 [hep-lat]].

\bibitem{Baron:2011sf} 
  R.~Baron {\it et al.} [ETM Collaboration],
  ``Light hadrons from $N_f = 2+1+1$ dynamical twisted mass fermions,''
  PoS {\bf LATTICE2010}, 123 (2010)
  [arXiv:1101.0518 [hep-lat]].

\bibitem{Iwasaki:1985we}
  Y.~Iwasaki,
  ``Renormalization group analysis of lattice theories and improved lattice action: two-dimensional non-linear $\mathcal{O}(N)$ sigma model,''
  Nucl.\ Phys.\ B {\bf 258}, 141 (1985).

\bibitem{Frezzotti:2000nk}
  R.~Frezzotti, P.~A.~Grassi, S.~Sint and P.~Weisz [ALPHA Collaboration],
  ``Lattice QCD with a chirally twisted mass term,''
  JHEP {\bf 0108}, 058 (2001)
  [arXiv:hep-lat/0101001].

\bibitem{Frezzotti:2003xj}
  R.~Frezzotti and G.~C.~Rossi,
  ``Twisted-mass lattice QCD with mass non-degenerate quarks,''
  Nucl.\ Phys.\ Proc.\ Suppl.\ {\bf 128}, 193 (2004)
  [arXiv:hep-lat/0311008].

\bibitem{Frezzotti:2003ni}
  R.~Frezzotti and G.~C.~Rossi,
  ``Chirally improving Wilson fermions. I: $\mathcal{O}(a)$ improvement,''
  JHEP {\bf 0408}, 007 (2004)
  [arXiv:hep-lat/0306014].

\bibitem{Frezzotti:2004wz}
  R.~Frezzotti and G.~C.~Rossi,
  ``Chirally improving Wilson fermions. II. Four-quark operators,''
  JHEP {\bf 0410}, 070 (2004)
  [arXiv:hep-lat/0407002].

\bibitem{Frezzotti:2005gi}
  R.~Frezzotti, G.~Martinelli, M.~Papinutto and G.~C.~Rossi,
  ``Reducing cutoff effects in maximally twisted lattice QCD close to the chiral limit,''
  JHEP {\bf 0604}, 038 (2006)
  [arXiv:hep-lat/0503034].

\bibitem{Chiarappa:2006ae}
  T.~Chiarappa, F.~Farchioni, K.~Jansen, I.~Montvay, E.~E.~Scholz, L.~Scorzato, T.~Sudmann and C.~Urbach,
  ``Numerical simulation of QCD with u, d, s and c quarks in the twisted-mass Wilson formulation,''
  Eur.\ Phys.\ J.\ C {\bf 50}, 373 (2007)
  [arXiv:hep-lat/0606011].
  
\bibitem{Shindler:2007vp} 
  A.~Shindler,
  ``Twisted mass lattice QCD,''
  Phys.\ Rept.\ {\bf 461}, 37 (2008)
  [arXiv:0707.4093 [hep-lat]].

\bibitem{Baron:2010th} 
  R.~Baron {\it et al.} [ETM Collaboration],
  ``Computing $K$ and $D$ meson masses with $N_f = 2+1+1$ twisted mass lattice QCD,''
  Comput.\ Phys.\ Commun.\ {\bf 182}, 299 (2011)
  [arXiv:1005.2042 [hep-lat]].

\bibitem{Baron:2010vp} 
  R.~Baron {\it et al.} [ETM Collaboration],
  ``Kaon and $D$ meson masses with $N_f = 2+1+1$ twisted mass lattice QCD,''
  PoS {\bf LATTICE2010}, 130 (2010)
  [arXiv:1009.2074 [hep-lat]].

\bibitem{Albanese:1987ds}
  M.~Albanese {\it et al.} [APE Collaboration],
  ``Glueball Masses and String Tension in Lattice QCD,''
  Phys.\ Lett.\ B {\bf 192}, 163 (1987).
  
\bibitem{Jansen:2008si} 
  K.~Jansen, C.~Michael, A.~Shindler and M.~Wagner [ETM Collaboration],
  ``The static-light meson spectrum from twisted mass lattice QCD,''
  JHEP {\bf 0812}, 058 (2008)
  [arXiv:0810.1843 [hep-lat]].

\bibitem{Blossier:2009kd} 
  B.~Blossier, M.~Della Morte, G.~von Hippel, T.~Mendes and R.~Sommer,
  ``On the generalized eigenvalue method for energies and matrix elements in lattice field theory,''
  JHEP {\bf 0904}, 094 (2009)
  [arXiv:0902.1265 [hep-lat]].

\bibitem{Urbach:2007rt} 
  C.~Urbach [ETM Collaboration],
  ``Lattice QCD with two light Wilson quarks and maximally twisted mass,''
  PoS LATTICE {\bf 2007}, 022 (2007)
  [arXiv:0710.1517 [hep-lat]].

\bibitem{Frezzotti:2007qv} 
  R.~Frezzotti and G.~Rossi,
  ``$\mathcal{O}(a^2)$ cutoff effects in Wilson fermion simulations,''
  PoS LATTICE {\bf 2007}, 277 (2007)
  [arXiv:0710.2492 [hep-lat]].
  
\bibitem{Michael:2010aa} 
  C.~Michael, A.~Shindler and M.~Wagner [ETM Collaboration],
  ``The continuum limit of the static-light meson spectrum,''
  JHEP {\bf 1008}, 009 (2010)
  [arXiv:1004.4235 [hep-lat]].

\bibitem{Levkova:2010ft} 
  L.~Levkova and C.~DeTar,
  ``Charm annihilation effects on the hyperfine splitting in charmonium,''
  Phys.\ Rev.\ D {\bf 83}, 074504 (2011)
  [arXiv:1012.1837 [hep-lat]].

\bibitem{Davies:2010ip} 
  C.~T.~H.~Davies {\it et al.} [HPQCD collaboration],
  ``Update: precision $D_s$ decay constant from full lattice QCD using very fine lattices,''
  Phys.\ Rev.\ D {\bf 82}, 114504 (2010)
  [arXiv:1008.4018 [hep-lat]].

\bibitem{Gregory:2010gm} 
  E.~B.~Gregory {\it et al.} [HPQCD collaboration],
  ``Precise $B$, $B_s$ and $B_c$ meson spectroscopy from full lattice QCD,''
  Phys.\ Rev.\ D {\bf 83}, 014506 (2011)
  [arXiv:1010.3848 [hep-lat]].

\bibitem{Donald:2012ga}
  G.~C.~Donald {\it et al.} [HPQCD collaboration],
  ``Precision tests of the $J/\psi$ from full lattice QCD: mass, leptonic width and radiative decay rate to $\eta_c$,''
  Phys.\ Rev.\ D {\bf 86}, 094501 (2012)
  [arXiv:1208.2855 [hep-lat]].


\bibitem{Alexandrou:2014sha} 
  C.~Alexandrou, V.~Drach, K.~Jansen, C.~Kallidonis and G.~Koutsou,
  Phys.\ Rev.\ D {\bf 90}, 074501 (2014)
  [arXiv:1406.4310 [hep-lat]].  
  
\bibitem{deDivitiis:2013xla}
  G.~M.~de Divitiis {\it et al.} [RM123 Collaboration],
  ``Leading isospin breaking effects on the lattice,''
  Phys.\ Rev.\ D {\bf 87},  114505 (2013)
  [arXiv:1303.4896 [hep-lat]].

\bibitem{Borsanyi:2013lga}
  S.~Borsanyi {\it et al.} [Budapest-Marseille-Wuppertal Collaboration],
  ``Isospin splittings in the light baryon octet from lattice QCD and QED,''
  Phys.\ Rev.\ Lett.\  {\bf 111},  252001 (2013)
  [arXiv:1306.2287 [hep-lat]].
  
\bibitem{Dmitrasinovic:2005gc} 
  V.~Dmitrasinovic,
  ``$D_{s0}^+(2317)$-$D_0(2308)$ mass difference as evidence for tetraquarks,''
  Phys.\ Rev.\ Lett.\ {\bf 94}, 162002 (2005).

\bibitem{Cleven:2014oka}
  M.~Cleven {\it et al.},
  ``Strong and radiative decays of the $D_{s0}^\ast(2317)$ and $D_{s1}(2460)$,''
  Eur.\ Phys.\ J.\ A {\bf 50}, 149 (2014)
  [arXiv:1405.2242 [hep-ph]].
  
\bibitem{Bigi:2007qp}
  I.~I.~Bigi {\it et al.},
  ``Memorino on the `$1/2$ vs.\ $3/2$ puzzle'' in $\bar{B} \, \rightarrow \, l \, \bar{\nu} \, X_c$ -- a year later and a bit wiser,''
  Eur.\ Phys.\ J.\ C {\bf 52}, 975 (2007)
  [arXiv:0708.1621 [hep-ph]].
  
\bibitem{Atoui:2013sca} 
  M.~Atoui,
  ``Lattice computation of $B \rightarrow D^\ast , D^{\ast \ast} l \nu$ form factors at finite heavy masses,''
  arXiv:1305.0462 [hep-lat].

\bibitem{Atoui:2013ksa} 
  M.~Atoui {\it et al.},
  ``Semileptonic $B \rightarrow D^{\ast \ast}$ decays in Lattice QCD : a feasibility study and first results,''
  Eur.\ Phys.\ J.\ C {\bf 75}, 376 (2015)
  [arXiv:1312.2914 [hep-lat]].


  
\end{thebibliography}
\end{document}